\documentclass[prd,nofootinbib,preprint,superscriptaddress]{revtex4}
\pdfoutput=1
\usepackage[T1]{fontenc}
\usepackage{amsmath,amssymb}
\usepackage{epsfig}
\usepackage{subfigure}
\usepackage{float}
\usepackage{graphicx}
\usepackage[usenames,dvipsnames]{color}
\usepackage{slashed}
\usepackage{multirow}
\usepackage{epstopdf}
\usepackage[colorlinks,citecolor=blue]{hyperref}
\usepackage{pdfpages}
\usepackage{color}
\usepackage{comment}
\usepackage{cancel}
\newcommand{\be}{\begin{equation}}
\newcommand{\ee}{\end{equation}}
\newcommand{\bea}{\begin{eqnarray}}
\newcommand{\eea}{\end{eqnarray}}

\newcommand{\umt}{{\rm U(1)}_{L_{\mu}-L_{\tau}}}
\newcommand{\vmt}{v_{\mu \tau}}
\newcommand{\gmt}{g_{\mu \tau}}
\newcommand{\zmt}{Z_{\mu \tau}}

\newcommand{\nn}{\nonumber}
\newcommand{\chiL}{\chi_{{}_L}}
\newcommand{\chiR}{\chi_{{}_R}}

\begin{document}
\title{$(g-2)_{e,\,\mu}$ and strongly interacting dark matter with collider implications}
\author{Anirban Biswas}
\email{anirban.biswas.sinp@gmail.com}
\affiliation{Department of Physics, Sogang University, Seoul 121-742, South Korea}
\affiliation{Center for Quantum Spacetime, Sogang University, Seoul 121-742, South Korea}
\author{Sarif Khan}
\email{sarif.khan@uni-goettingen.de}
\affiliation{Institut f\"{u}r Theoretische Physik,
Georg-August-Universit\"{a}t G\"{o}ttingen,
Friedrich-Hund-Platz 1, 37077 G\"{o}ttingen, Germany}
\begin{abstract}
The quest for new physics beyond the Standard Model is boosted
by the recently observed deviation in the anomalous magnetic moments of
muon and electron from their respective theoretical prediction.
In the present work, we have proposed a suitable
extension of the minimal $L_{\mu}-L_{\tau}$ model to address
these two experimental results as the minimal
model is unable to provide any realistic solution. In our model,
a new Yukawa interaction involving first generation of leptons, a
singlet vector like fermion ($\chi^{\pm}$) and a scalar (either
an SU(2)$_{L}$ doublet $\Phi^\prime_2$ or a complex singlet
$\Phi^\prime_4$) provides the additional one loop contribution to
$a_{e}$ only on top of the usual contribution coming from the
$L_{\mu}-L_{\tau}$ gauge boson ($Z_{\mu\tau}$) to both electron
and muon. The judicious choice of $L_{\mu}-L_{\tau}$
charges to these new fields results in a strongly
interacting scalar dark matter in $\mathcal{O}({\rm MeV})$ range
after taking into account the bounds from relic density,
unitarity and self interaction. The freeze-out dynamics of dark matter
is greatly influenced by $3\rightarrow2$ scatterings
while the kinetic equilibrium with the SM bath is ensured by $2\rightarrow2$
scatterings with neutrinos where $Z_{\mu\tau}$ plays a pivotal role.
The detection of dark matter is possible directly through scatterings
with nuclei mediated by the SM $Z$ bosons. Moreover, our proposed model can
also be tested in the upcoming $e^+e^-$ colliders by searching
opposite sign di-electron and missing energy signal i.e. $e^{+} e^{-}
\rightarrow \chi^{+} \chi^{-} \rightarrow e^{+} e^{-} \cancel{E}_T$
at the final state.
\end{abstract}
\maketitle
\section{Introduction}
\label{intro}
The Standard Model (SM) is a very well established theory of nature
and is confirmed fully with the discovery of the Higgs boson. But with time,
we have understood from various astrophysical phenomena
\cite{Sofue:2000jx, Bartelmann:1999yn, Clowe:2003tk}
that the matter content of the Universe is not only made up of
with the elementary particles described by the SM but more than
80$\%$ matter content of the Universe is unknown to us. This mysterious
part is the so called dark matter (DM), and its presence has been confirmed
by many pieces of evidence from large scale to small scale observations.
The most precise determination of dark matter abundance at the present epoch
is $\Omega_{\rm DM} h^{2}=0.120\pm 0.001$ by the Planck satellite \cite{Planck:2018vyg}. 
Therefore, in order to have a viable dark matter candidate(s), it is essential
to extend the particle spectrum of the SM. 
Moreover, a long standing discrepancy exits over the last two decades
between the theoretical prediction of the SM and the experimental
measurement of the anomalous magnetic moment of muon \cite{Muong-2:2001kxu,
Muong-2:2006rrc, Jegerlehner:2009ry, Davier:2019can, Aoyama:2020ynm}.
Recently, Fermilab has announced a $4.2 \sigma$ discrepancy between the experimental and 
theoretical value \cite{Muong-2:2021ojo},
\begin{eqnarray}
\Delta a_{\mu} = a^{\rm exp}_{\mu} - a^{\rm SM}_{\mu} = 
\left(2.51 \pm 0.59 \right) \times 10^{-9}\,. 
\end{eqnarray}
The uncertainty in $\Delta{a_{\mu}}$\footnote{The anomalous magnetic moment
$a_{\ell}$ of a lepton $\ell$ is defined as $a_{\ell}\equiv\dfrac{(g-2)_{\ell}}{2}$,
where $g$ is the Lande $g$ factor.} will go down in future when more
data will be available from the ongoing experiment at Fermilab \cite{Muong-2:2021vma}
as well as the future experiment at JPARC \cite{Abe:2019thb}. Besides 
$(g-2)_{\mu}$, there is also an inconsistency in $(g-2)$ for electron
between theoretical and experimental values. {However,
the magnitude of the deviation is four orders smaller than that of muon
and depending on measurement of the fine structure constant $\alpha^{-1}_{em}$
using $^{137}$Cs atom at Berkeley \cite{Parker:2018vye} ($\alpha^{-1}_{em}=$137.035999046(27))
and ${}^{87}$Rb atom at LKB \cite{Morel:2020dww}($\alpha^{-1}_{em} = 137.035999206(11)$),
we have deviations both in negative and positive directions respectively from the SM
expectation, as described below,
\begin{eqnarray}
\Delta a_{e} &=& a^{\rm exp}_{e} - a^{\rm B(LKB)}_{e} \nonumber \\ &=& 
\left(-8.8 \pm 3.6 \right) \times 10^{-13}\,\, 
{\rm using}\,\,^{137}{\rm Cs}\,\,
{\rm at}\,\,{\rm Berkeley}\,\,{\rm with}\,\,
2.4\sigma\,\,{\rm discrepancy}\,, \nonumber \\
&=&  \left(+4.8 \pm 3.0 \right) \times 10^{-13}\,\,{\rm using}\,\,
^{87}{\rm Rb}\,\,{\rm at}\,\,{\rm LKB}\,\,{\rm with}\,\,1.6\sigma\,\,{\rm discrepancy}\,,
\end{eqnarray} 
and we need further investigations of the electron anomalous magnetic
moment in future by using different techniques 
\cite{Bell:1982qr, Anghel:2012zz} to confirm the deviation in one particular direction.}

Keeping in view of the above discussions, in this work, we have
considered an extension of the minimal $L_{\mu}-L_{\tau}$ model \cite{He:1990pn, He:1991qd}
to address both the anomalies in $(g-2)$ of muon and electron on the basis
of the experimental data available to us so far. The
$L_{\mu}-L_{\tau}$ gauge extension of the SM is not only an anomaly free theory but
also is very well motivated from the neutrino mass generation and produces correct
mixing angles as measured by several experiments over the last two decades \cite{Ma:2001md,
Grimus:2005jk, Rodejohann:2005ru, Aizawa:2005yy, Xing:2006xa, Adhikary:2006rf, Fuki:2006ag,
Haba:2006hc, Joshipura:2009tg, Adhikary:2009kz, Joshipura:2009fu, Xing:2010ez, Araki:2010kq,
He:2011kn, Heeck:2011wj, Grimus:2012hu, Altmannshofer:2014pba, Xing:2015fdg, Asai:2017ryy,
Dev:2017fdz, Chen:2017gvf, Nomura:2018vfz, Nomura:2018cle, Banerjee:2018eaf}.
Moreover, both thermal as well as nonthermal dark matter have also 
been {studied} earlier in
$L_{\mu}-L_{\tau}$ model by several authors \cite{Altmannshofer:2016jzy, Patra:2016shz,
Biswas:2016yan, Biswas:2016yjr, Biswas:2017ait, Arcadi:2018tly, Singirala:2018mio,
Foldenauer:2018zrz, Escudero:2019gzq, Biswas:2019twf, Okada:2019sbb, Borah:2020jzi,
Asai:2020qlp, Borah:2021khc, Drees:2021rsg, Hapitas:2021ilr, Tapadar:2021kgw}. 
{
First, we have considered the kinetic mixing 
between  $U(1)_{L_{\mu} - L_{\tau}}$ and $U(1)_{Y}$ and discussed
its relevance in the present work. Due to
the kinetic mixing, the detection prospects of the present model 
increase at the different ongoing and proposed experiments namely
Borexino \cite{Borexino:2017rsf,Altmannshofer:2019zhy} and 
ShiP \cite{SHiP:2015vad}. We have found that to explain 
electron and muon ($g-2$) anomalies together we need larger kinetic mixing 
which is already ruled out by the Borexino experiment 
\cite{Borexino:2017rsf,Altmannshofer:2019zhy}. 
In Fig. \ref{Fig:parameter_space_Min_model}, we have shown in detail the regions which are already
ruled out by different experiments in
$g_{\mu\tau}-M_{Z_{\mu\tau}}$ plane and the regions which will be 
accessed at the different proposed experiments.   
} 

{We have extended the minimal model by
a singlet scalar $\Phi^\prime_4$, a SU(2)$_{\rm L}$ scalar doublet
$\Phi^\prime_2$ and a ``vector-like'' charged fermion $\chi^{\pm}$ which
is singlet under both SU(2)$_{\rm L}$ and SU(3)$_{\rm c}$.
We have assigned suitable $L_{\mu}-L_{\tau}$ charges
to all these new fields which allow us to incorporate
two additional Yukawa terms. This results in an
additional one loop contribution to $\Delta{a}_e$ over the contribution
due to $\zmt$ through $Z-\zmt$ mixing. 
On the other hand, $\Delta{a}_{\mu}$ gets only
the $\zmt$ induced one loop contribution and it does
not depend significantly on the $Z-\zmt$ mixing as unlike
electron, muon has nonzero $L_{\mu}-L_{\tau}$ charge.
Due to this additional contribution to $\Delta{a_e}$, we now have the
freedom to choose the kinetic mixing parameter
respecting the current bounds \cite{Bauer:2018onh}.
There are earlier works where authors have explored both electron
and muon anomalous magnetic moments which can be found in
\cite{Davoudiasl:2018fbb, Crivellin:2018qmi, Liu:2018xkx, Han:2018znu, 
Bauer:2019gfk, Endo:2019bcj, Badziak:2019gaf, Hiller:2019mou, Bigaran:2020jil,Endo:2020mev,
Dorsner:2020aaz, Haba:2020gkr, Calibbi:2020emz, Chen:2020jvl, Dutta:2020scq,
Abdallah:2020biq, Chen:2020tfr, Botella:2020xzf, Jana:2020joi, Hati:2020fzp, Chun:2020uzw,
Li:2020dbg, Banerjee:2020zvi, Cao:2021lmj, DelleRose:2020oaa, Escribano:2021css,
Frank:2021nkq, Borah:2021khc, Bharadwaj:2021tgp}.}

Apart from addressing the anomalous magnetic moments,
we have also studied the phenomenology of a viable dark matter candidate
which is an admixture of two neutral complex scalars namely,
$\Phi_4^\prime$ and $\phi^\prime_2$ (neutral component of the
scalar doublet $\Phi_2^\prime$). The dynamics of the dark sector
especially our dark matter candidate $\phi_4$ is greatly influenced
by the choice of $L_{\mu}-L_{\tau}$ charges of
the fields involving in the new Yukawa interaction necessary for
$(g-2)_e$. This results in the strongly interacting dark matter
scenario as we get a cubic self interaction term for
$\phi_4$ when $L_{\mu}-L_{\tau}$ symmetry is broken by
the vacuum expectation value (VEV) of $\Phi^\prime_3$. Therefore,
the freeze-out era of $\phi_4$ is predominantly determined
by the competition between $3\rightarrow2$ interaction rates
with the Hubble expansion rate. Moreover, the kinetic equilibrium
of the dark matter with the SM bath, as required for the Strongly
Interacting Massive Particle (SIMP) scenario \cite{Hochberg:2014dra}, is achieved
by the elastic scattering between $\phi_4$ and $\nu_\alpha$ ($\alpha=\mu$,\,$\tau$)   
where $\zmt$ plays the role of the dominant mediator. Therefore,
in this way parameters of the new gauge interaction such as $\gmt$ and
$M_{\zmt}$ have large impact on the cosmic evolution of dark matter and
at the same time they are tightly constrained by the
precise experimental measurement of $(g-2)_{\mu}$. We have
shown that the $2\sigma$ parameter space for addressing $(g-2)_{\mu}$
has some overlap with the region in $\gmt-M_{\zmt}$ plane that keeps
an MeV scale dark matter in kinetic equilibrium till freeze-out.
For earlier works focusing on the SIMP scenario see Refs.\,\,\cite{Hochberg:2014kqa,
Hochberg:2015vrg, Daci:2015hca, Bernal:2015xba, Bernal:2015bla, Lee:2015gsa, Choi:2015bya,
Choi:2016tkj, Choi:2017zww, Choi:2017mkk, Ho:2017fte, Davis:2017noy, Hochberg:2018rjs, Bhattacharya:2019mmy,
Smirnov:2020zwf, Katz:2020ywn, Choi:2021yps}.
Finally, we have looked for the prospect of collider signature of 
the charged fermion ($\chi^\pm$) at the $e^{+} e^{-}$ linear collider
for the centre of mass (c.o.m) energies $\sqrt{s} = 1000$ GeV
and $3000$ GeV respectively. Here we have investigated the opposite
sign di-electron and missing energy signal at the final state i.e.
$e^{+} e^{-} \rightarrow \chi^{+} \chi^{-} \rightarrow e^{+} e^{-} \cancel{E}_T$.
We have shown that the signal strength of the charged fermion significantly
improved for the presence of the $t$-channel process mediated by SIMP dark 
matter $\phi_4$ which remains absent at the hadron collider. This
enhancement in the cross section will ensure the detection of
the present model in the early run of $e^{+} e^{-}$ collider. 

The rest of the paper is organised in the following way.
In the Section\,\,\ref{Sec:minimal_model} we have described the 
minimal $L_{\mu}-L_{\tau}$ model and have shown that it is
not possible to address both the anomalies simultaneously in
the minimal model. The extended model has been described in detail
in the Section\,\,\ref{Sec:Model_extn}. A brief discussion on neutrino
masses via Type-I seesaw mechanism in the context of the present
model is presented in the Section\,\,\ref{Sec:nu_mass}. The Section\,\,\ref{Sec:simp}
is devoted to a comprehensive study on the SIMP dark matter and
related numerical analyses. The signature of the new charged
fermion $\chi^\pm$ has been studied in the Section\,\,\ref{Sec:collider}.
Finally, we summarise in the Section\,\,\ref{Sec:conclu}.
The contributions in the anomalous magnetic moment of
electron due to new scalars are given in the Appendix\,\,\ref{Sec:g-2_ext}.
The couplings necessary for calculating all the Feynman diagrams are listed in
the Appendix\,\,\ref{App:Vertices}. 
\section{The minimal $L_{\mu}-L_{\tau}$ Model and anomalous magnetic moment}
\label{Sec:minimal_model}
As discussed in the previous section, one of our prime motivations is to address
both the anomalies reported in the anomalous magnetic moments of $\mu$ and
$e$ within a single framework. It is well known for quite a while that
the minimal U(1)$_{L_{\mu}-L_{\tau}}$ model can resolve the enduring
discrepancy between the experimentally measured value of
$(g-2)_{\mu}$ and the SM prediction efficiently, where an MeV scale
($\mathcal{O}(10\,\,{\rm MeV}\sim100\,\,{\rm MeV})$)
new gauge boson ($Z_{\mu\tau}$) provides the
require deficit on top of the SM contribution to match
the experimental prediction \cite{Biswas:2016yan, Biswas:2019twf}.
Keeping this in mind, we have investigated the possibility
of addressing $(g-2)$ of electron in
the minimal $L_{\mu}-L_{\tau}$ model alongside $(g-2)_{\mu}$. Since
the electrons are not charged under U(1)$_{L_{\mu}-L_{\tau}}$ symmetry,
the effect of $L_{\mu}-L_{\tau}$ gauge boson on the anomalous magnetic
moment comes only through the kinetic mixing between U(1)$_{L_{\mu}-L_{\tau}}$
and U(1)$_{Y}$ of the SM. Before going into the details of anomalous magnetic
moments of $e$ and $\mu$ we would first like to describe the minimal $L_{\mu}-L_{\tau}$
model briefly.

In the minimal $L_{\mu}-L_{\tau}$ model, in addition to the SM gauge
symmetry, we demand another local U(1)$_{L_{\mu}-L_{\tau}}$ gauge invariance,
where $L_{\ell}$ represents the lepton number corresponding to the lepton $\ell$.
Therefore, the ${L_{\mu}-L_{\tau}}$ charges for three generations of the
SM leptons are $Q^{{}^{e,\,\nu_e}}_{\mu\tau}=0$,
$Q^{{}^{\mu,\,\nu_\mu}}_{\mu\tau}=+1$ and
$Q^{{}^{\tau,\,\nu_\tau}}_{\mu\tau}=-1$ respectively
while all the quarks possess zero ${L_{\mu}-L_{\tau}}$ charge.
One of the biggest advantages of ${L_{\mu}-L_{\tau}}$ 
gauge extension is that it does not introduce any axial vector anomaly
\cite{Bilal:2008qx, Adler:1969gk, Bardeen:1969md}
and gauge-gravitational anomaly \cite{Delbourgo:1972xb, Eguchi:1976db}
since they cancel automatically
between second and third generations of the SM leptons. In addition to
the usual SM fields, we only need a scalar field having nonzero ${L_{\mu}-L_{\tau}}$
charge to break this local U(1) symmetry spontaneously and thereby generating
a massive neutral gauge boson $Z_{\mu\tau}$. The ${L_{\mu}-L_{\tau}}$ symmetric
Lagrangian for the minimal model is given by
\begin{eqnarray}
\mathcal{L}&=&\mathcal{L}_{\rm SM} 
+ \left(D_{\alpha}{\Phi^\prime_3}\right)^{\dagger} \left(D^{\alpha}{\Phi^\prime_3}\right) 
- \frac{1}{4} \hat{X}_{\alpha \beta} \hat{X}^{\alpha \beta} +
\frac{\epsilon}{2} \hat{X}_{\alpha \beta} \hat{B}^{\alpha \beta}
- \gmt\sum_{\substack{\ell = \mu, \nu_{\mu}, \\\,\,\, \tau, \nu_{\tau}}}
Q^{\ell}_{\mu\tau} \bar{\ell} \gamma^{\alpha} \ell
\hat{X}_{\alpha}
\nn\\&&-
\lambda_{13} \left({\Phi^\prime_1}^\dagger \Phi^\prime_1\right)\left({\Phi^\prime_3}^\dagger\Phi^\prime_3\right),
\label{Lag_mu_tao}
\end{eqnarray}  
where $\mathcal{L}_{\rm SM}$ denotes the SM Lagrangian. Here  we have considered
a SM singlet scalar field ($\Phi^\prime_3$) having ${L_{\mu}-L_{\tau}}$
charge equal to unity which breaks the $L_{\mu}-L_{\tau}$ symmetry
spontaneously. As mentioned above, the fourth term represents
kinetic mixing between the hypercharge gauge boson ($\hat{B}_{\mu}$)
and the ${L_{\mu}-L_{\tau}}$ gauge boson ($\hat{X}_{\mu}$), where the respective
abelian field strength tensor is denoted by the same letter but with two Lorentz
indices. The last but one term corresponds to the interactions of
second and third generation leptons with $L_{\mu}-L_{\tau}$
gauge boson while the last term is the only gauge invariant interaction between the
SM Higgs doublet $\Phi^\prime_1$ and the singlet scalar $\Phi^\prime_3$.
 
To obtain the physical gauge boson $Z_{\mu\tau}$,
first we need a basis transformation
from the ``hat'' states ($\hat{B}_{\alpha}$, $\hat{X}_{\alpha}$) to the ``un-hat''
states (${B}_{\alpha}$, ${X}_{\alpha}$) so that the off-diagonal term proportional
to $\epsilon$ vanishes. This requires a transformation like
\begin{eqnarray}
\hat{B}_{\alpha} &=& B_{\alpha} - \epsilon \,X_{\alpha}\,, \nonumber \\
\hat{X}_{\alpha} &=& \,X_{\alpha} + \mathcal{O}(\epsilon^2)\,.
\label{hat-unhat-matrix}
\end{eqnarray}
 Where, we have considered terms up to linear order of $\epsilon$
as there exist various experimental constraints on the kinetic mixing
parameter ($\epsilon$), which forces us to consider $\epsilon \ll 1$ \cite{Bauer:2018onh}.
We would like to mention here that the above transformation is
neither an orthogonal transformation nor it is a unique one. Although,
the basis transformation given in Eq.\,\,(\ref{hat-unhat-matrix}) removes
the off-diagonal term (proportional to $\epsilon$), it reintroduces $\epsilon$
again in the mass matrix of neutral gauge bosons written in the ``un-hat''
basis ($W^3_{\alpha}, \,\,B_{\alpha},\,\,X_{\alpha}$) after both electroweak
symmetry breaking and $L_{\mu}-L_{\tau}$ breaking as
\begin{eqnarray}
\mathcal{M}^2_{\rm gauge} = \left(\begin{array}{ccc}
\frac{1}{4}g^2_2v^2 &-\frac{1}{4}g_2 g_1v^2 & -\frac{1}{4}g_2 g_1v^2\epsilon \\
-\frac{1}{4}g_2 g_1v^2 & \frac{1}{4}g^2_1v^2 & \frac{1}{4} g^2_1 v^2\epsilon \\
-\frac{1}{4}g_2 g_1v^2\epsilon  & \frac{1}{4} g^2_1 v^2\epsilon & g^2_{\mu\tau}v^2_{\mu\tau}
\end{array}\right)\,\,,
\label{gauge-mixing}
\end{eqnarray}
where $g_1$, $g_2$ and $g_{\mu\tau}$ are the gauge couplings of U(1)$_Y$, SU(2)$_L$
and U(1)$_{L_{\mu}-L_{\tau}}$ respectively while the VEVs of $\Phi^\prime_1$ and $\Phi^\prime_3$
are $v/\sqrt{2}$ and $v_{\mu\tau}/\sqrt{2}$ respectively. The above mass matrix
has a special symmetry that if we rotate the upper $2\times2$ block between $W^3_{\alpha}$
and $B_{\alpha}$ by the Weinberg angle $\theta_W$ ($\tan\theta_W=\frac{g_1}{g_2}$),
not only the $2\times2$ block becomes diagonal but also the entire mass matrix
reduces to a block diagonal form where a new $2\times2$ block is formed
between $\mathcal{Z}_{\alpha}$ ($\equiv\cos\theta_W W^3_\alpha -\sin \theta_W B_\alpha$)
and $X_\alpha$ while the state $A_\alpha$ ($\equiv\sin\theta_W W^3_\alpha
+\cos \theta_W B_\alpha$), orthogonal to the state $\mathcal{Z}_\alpha$, decouples completely
with a zero eigenvalue. It is then natural to identify the state ${A_{\alpha}}$ 
as the photon, the gauge boson corresponding to the unbroken U(1)$_{em}$ symmetry.   
This is the reason behind our choice of Eq.\,\,(\ref{hat-unhat-matrix})
among many other possibilities. 

Finally, to obtain the other two physical gauge bosons, we need
to diagonalise the $2\times2$ block between $\mathcal{Z}_\alpha$
and $X_\alpha$, the elements of which are given by
\begin{eqnarray}
\mathcal{M}^2_{\mathcal{Z}X}\,=\,
\dfrac{1}{4}\left(\begin{array}{cc}
(g_1^2+g_2^2)v^2 & -{\epsilon} g_1
\sqrt{g_1^2+g_2^2}\,v^2 \\
-{\epsilon} g_1 \sqrt{g_1^2+g_2^2}\,v^2
& 4\,g^2_{\mu\tau} v^2_{\mu\tau} 
\end{array}
\right)\,.
\end{eqnarray}
After diagonalisation, the physical gauge bosons are given by 
\begin{eqnarray}
Z^\alpha &=& \cos\theta_{\mu\tau} \mathcal{Z}^\alpha -\sin \theta_{\mu\tau} X^\alpha
= \cos\theta_{\mu\tau} \left(\cos\theta_W W^3_\alpha -\sin \theta_W B_\alpha\right)
- \sin \theta_{\mu\tau} X^\alpha\,,
\label{Z-boson}\\
Z_{\mu\tau}^\alpha &=& \sin\theta_{\mu\tau} \mathcal{Z}^\alpha +\cos \theta_{\mu\tau} X^\alpha
=\sin\theta_{\mu\tau}\left(\cos\theta_W W^3_\alpha -\sin \theta_W B_\alpha\right)
+\cos \theta_{\mu\tau} X^\alpha\,
\label{Zmt-boson}
\end{eqnarray}
having masses as follows
\begin{eqnarray}
M_{Z} &=& \sqrt{\frac{(g^2_1+g^2_2)v^2}{4}\cos^2\theta_{\mu\tau} +
 g^2_{\mu\tau}v^2_{\mu\tau}\sin^2\theta_{\mu\tau} + 
 \epsilon \,\frac{g_1\sqrt{g^2_1+g^2_2}\,v^2}{4}\sin 2\theta_{\mu\tau}}\;, 
\label{Zmass} \\
M_{Z_{\mu\tau}} &=& \sqrt{\frac{(g^2_1+g^2_2)v^2}{4}\sin^2\theta_{\mu\tau} 
+ g^2_{\mu\tau}v^2_{\mu\tau}\cos^2\theta_{\mu\tau} - 
\epsilon \,\frac{g_1\sqrt{g^2_1+g^2_2}\,v^2}{4}\sin 2\theta_{\mu\tau}}\;,
\label{Zmt-mass}
\end{eqnarray}
and the $Z-Z_{\mu\tau}$ mixing angle $\theta_{\mu\tau}$ has the following expression
\begin{eqnarray}
\theta_{\mu\tau} = \frac{1}{2}\,\tan^{-1}\left(\dfrac{\dfrac{2\, g_1\,\epsilon}
{\sqrt{g^2_1 + g^2_2}}}{1 - \dfrac{4g^2_{\mu\tau}}{g^2_1 + g^2_2}
\dfrac{v^2_{\mu\tau}}{v^2}}\right) \,\, .
\label{theta_mt}
\end{eqnarray}
As expected the mixing angle is proportional to the kinetic mixing parameter
$\epsilon$. The Eq.\,\,(\ref{Z-boson}) represents the neutral gauge
boson of weak interaction namely, the $Z$ boson. If we set $\epsilon=0$,
Eq.\,\,(\ref{Z-boson}) reduces to the well known SM $Z$ boson with
mass $M_Z=\frac{1}{2}\sqrt{g_1^2+g_2^2}\,v$ (see Eq.\,\,(\ref{Zmass}))
as given in the SM. Moreover, one can notice that
in spite of the kinetic mixing between U(1)$_Y$
and U(1)$_{L_{\mu}-L_{\tau}}$, the state representing photon ($A_\alpha$) remains unaltered. 
The effect of $\epsilon$ enters only into the expressions of $Z$ and $Z_{\mu\tau}$.  

Due to this $Z-Z_{\mu\tau}$ mixing all the SM fermions, particularly
the first generation of leptons which do not possess any $L_{\mu}-L_{\tau}$ charge,
will now be able to interact with $Z_{\mu\tau}$. Consequently, we have an
additional contribution in the anomalous magnetic moment of electron, analogous to
muon, besides the usual SM contribution involving photon. However, the only difference
is that the magnitude of such BSM effect in the context of electron will be much less
compared to that of $\mu$ as the $e^+e^-Z_{\mu\tau}$ vertex is suppressed by tiny
$Z-Z_{\mu\tau}$ mixing angle $\theta_{\mu\tau}$. The general structure of
$\overline{\ell}\ell Z_{\mu\tau}(Z)$ interaction is
$\overline{\ell}\gamma_{\alpha}(g^{Z_{\mu\tau}(Z)}_{{}_V}
+ g^{Z_{\mu\tau}(Z)}_{{}_A} \gamma_{{}_5})\ell\,Z^\alpha_{\mu\tau}(Z^\alpha)$
for any lepton $\ell$. The expressions of $g_{{}_V}$ and $g_{{}_A}$ for both
$e$ and $\mu$ are given in Table \ref{Tab:llZ_coupling}. One can easily
recover the familiar $\overline{\ell}\ell Z$ vertex of the SM in the
limit $\epsilon \rightarrow 0$. Moreover, we can see that
in the $\epsilon\rightarrow0$ limit, both vector and axial vector
couplings of the $\overline{e}eZ_{\mu\tau}$ vertex disappear while
the $\overline{\mu}\mu Z_{\mu\tau}$ vertex becomes purely
vectorial with $g^{Z_{\mu\tau}}_{{}_V}=-g_{\mu\tau}$. 
\begin{table}
\begin{center}
\begin{tabular}{||c|c|c||}
\hline
Field & \multicolumn{2}{|c||}{$Z_{\mu\tau}$}\\   
\hline
& $g^{\zmt}_{{}_V}$ & $g^{\zmt}_{{}_A}$ \\
\hline
$\ell$ &
{\scriptsize $-\dfrac{g_2\left(T^{\ell}_3 - 2\,Q^{\ell}_{em}\sin^2\theta_W\right)
}{2\,\cos\theta_W}\sin\theta_{\mu\tau}$
$-\left[g_{\mu\tau}\,Q^{\ell}_{\mu\tau}-
\dfrac{g_2\,\epsilon\left(T^{\ell}_3 - 2\,Q^{\ell}_{em}\right)
}{2\,\cot\theta_W}\right]\cos\theta_{\mu\tau}$} 
&
{\scriptsize $\dfrac{g_2}{2\,\cos\theta_W}\Big(\sin\theta_{\mu\tau}
$
$-\,\epsilon\sin\theta_W\cos\theta_{\mu\tau}\Big)\,T^{\ell}_3$}
\\
\hline
$e$ & 
{\footnotesize$-\dfrac{g_2\left(-1 + 4\,\sin^2\theta_W\right)
}{4\,\cos\theta_W}\sin\theta_{\mu\tau}$
$+\left(\dfrac{3\,g_2\,\epsilon
}{4\,\cot\theta_W}\right)\cos\theta_{\mu\tau}$}  
&
{\footnotesize$-\dfrac{g_2}{4\,\cos\theta_W}\Big(\sin\theta_{\mu\tau}
$
$-\,\epsilon\sin\theta_W\cos\theta_{\mu\tau}\Big)$}
\\
\hline
$\mu$ & 
{\footnotesize$-\dfrac{g_2\left(-1 + 4\,\sin^2\theta_W\right)
}{4\,\cos\theta_W}\sin\theta_{\mu\tau}$
$-\left(g_{\mu\tau}-\dfrac{3\,g_2\,\epsilon
}{4\,\cot\theta_W}\right)\cos\theta_{\mu\tau}$}  
&
{\footnotesize $-\dfrac{g_2}{4\,\cos\theta_W}\Big(\sin\theta_{\mu\tau}
$
$-\,\epsilon\sin\theta_W\cos\theta_{\mu\tau}\Big)$}
\\
\hline
\hline
 & \multicolumn{2}{|c||}{$Z$}\\  
\hline
& $g^{Z}_{{}_V}$ & $g^{Z}_{{}_A}$ \\
\hline
$\ell$
&
{\scriptsize$-\dfrac{g_2\left(T^{\ell}_3 - 2\,Q^{\ell}_{em}\sin^2\theta_W\right)
}{2\,\cos\theta_W}\cos\theta_{\mu\tau}$ 
$+\,\left[g_{\mu\tau}\,Q^{\ell}_{\mu\tau}-
\dfrac{g_2\,\epsilon\left(T^{\ell}_3 - 2\,Q^{\ell}_{em}\right)
}{2\,\cot\theta_W}\right]\sin\theta_{\mu\tau}$} 
&
{\scriptsize$\dfrac{g_2}{2\,\cos\theta_W}\Big(\cos\theta_{\mu\tau}$
$+\,\epsilon\sin\theta_W\sin\theta_{\mu\tau}\Big)\,T^{\ell}_3$} \\
\hline
$e$ 
&
{\footnotesize$-\dfrac{g_2\left(-1 + 4\,\sin^2\theta_W\right)
}{4\,\cos\theta_W}\cos\theta_{\mu\tau}$
$-\left(\dfrac{3\,g_2\,\epsilon
}{4\,\cot\theta_W}\right)\sin\theta_{\mu\tau}$} 
&
{\footnotesize$-\dfrac{g_2}{4\,\cos\theta_W}\Big(\cos\theta_{\mu\tau}
$
$+\,\epsilon\sin\theta_W\sin\theta_{\mu\tau}\Big)$} \\
\hline  
$\mu$
&
{\footnotesize$-\dfrac{g_2\left(-1 + 4\,\sin^2\theta_W\right)
}{4\,\cos\theta_W}\cos\theta_{\mu\tau}$ 
$+\left(g_{\mu\tau} -\dfrac{3\,g_2\,\epsilon
}{4\,\cot\theta_W}\right)\sin\theta_{\mu\tau}$} 
&
{\footnotesize$-\dfrac{g_2}{4\,\cos\theta_W}\Big(\cos\theta_{\mu\tau}
$
$+\,\epsilon\sin\theta_W\sin\theta_{\mu\tau}\Big)$} 
\\
\hline
\end{tabular}
\end{center}
\caption{$\overline{\ell}\ell Z_{\mu\tau}(Z)$ vertex factors for a general
lepton $\ell$ with third component of weak isospin $T^\ell_3$, electric charge $Q^\ell_{em}$
and $L_{\mu}-L_{\tau}$ charge $Q^\ell_{\mu\tau}$.
The vertex factors for electron ($T^e_3 = -\frac{1}{2},\,Q^e_{em} =-1,\,Q^e_{\mu\tau}=0$)
and muon ($T^{\mu}_3 = -\frac{1}{2},\,Q^{\mu}_{em} =-1,\,Q^{\mu}_{\mu\tau}=+1$)
are also listed.}
\label{Tab:llZ_coupling}
\end{table}

\begin{figure}[h!]
\includegraphics[height=4cm,width=15cm]{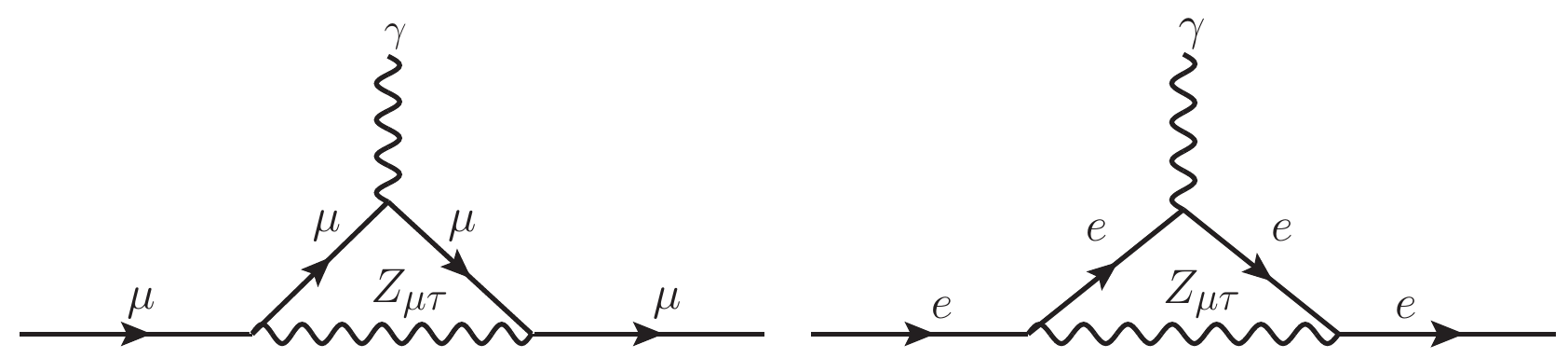}
\caption{Feynman diagrams for one loop contributions in the anomalous
magnetic moments of muon and electron.}
\label{Fig:Feyn_dia_g-2_Zmt}
\end{figure}

The Feynman diagrams contributing to the anomalous
magnetic moments of both $\mu$ and $e$ at one loop level
are shown in Fig.\,\,\ref{Fig:Feyn_dia_g-2_Zmt}. The expression
$\Delta{a}_{{}_\ell}$ ($\ell = e$, $\mu$) due to
the $L_{\mu}-L_{\tau}$ gauge boson $Z_{\mu\tau}$ is given by \cite{Biswas:2019twf}
\begin{eqnarray}
\Delta a_{{}_\ell} = \frac{1}{8\pi^2}~
\left((g^{Z_{\mu\tau}}_{{}_V})^2\,F^V_{Z_{\mu\tau}}(R_{Z_{\mu\tau}})
-(g^{Z_{\mu\tau}}_{{}_A})^2\,F^{A}_{Z_{\mu\tau}}(R_{Z_{\mu\tau}})\right)\,,
\label{del_al}
\end{eqnarray}
where $R_{Z_{\mu\tau}}= \dfrac{M^2_{Z_{\mu\tau}}}{m^2_{\ell}}$ and
\begin{eqnarray}
g^{Z_{\mu\tau}}_{{}_V} &=& -\dfrac{g_2\left(T^{\ell}_3 - 2\,Q^{\ell}_{em}\sin^2\theta_W\right)
}{2\,\cos\theta_W}\sin\theta_{\mu\tau} -\left[g_{\mu\tau}\,Q^{\ell}_{\mu\tau}-
\dfrac{g_2\,\epsilon\left(T^{\ell}_3 - 2\,Q^{\ell}_{em}\right)
}{2\,\cot\theta_W}\right]\cos\theta_{\mu\tau}\,,\\
\label{azz} 
g^{Z_{\mu\tau}}_{{}_A}&=-& \dfrac{g_2}{2\,\cos\theta_W}\Big(\sin\theta_{\mu\tau}
-\,\epsilon\sin\theta_W\cos\theta_{\mu\tau}\Big)\,T^{\ell}_3\;.
\label{bzz} 
\end{eqnarray}
The loop functions for the vectorial and axial vectorial interactions are
given by \cite{Biswas:2019twf}
\begin{eqnarray}
F^{V}_{Z_{\mu\tau}}(R_{Z_{\mu\tau}}) &=& \int_0^1 dx\, \frac{2x(1-x)^2}{(1
 -x)^2+R_{Z_{\mu\tau}} x}
 \;, \\
F^{A}_{Z_{\mu\tau}}(R_{Z_{\mu\tau}}) &=& \int_0^1 dx\,\frac{2x(1-x)(3+x)}{(1
 -x)^2+R_{Z_{\mu\tau}} x}\;.
\end{eqnarray}
From Eq.\,\,(\ref{del_al}), it is clearly seen that the vectorial part
and the axial vectorial part of the interaction between $Z_{\mu\tau}$ 
and leptons act oppositely (true for any gauge boson) in the anomalous
magnetic moment and the net effect due to a new gauge boson is
the difference between the contributions of both interactions. In the
limit of small kinetic mixing ($\epsilon\rightarrow0$), $\Delta{a}_e$
goes to zero while $\Delta{a}_{\mu}\propto g^2_{\mu\tau}$, gets contribution
from the vectorial part only.

\begin{figure}[h!]
\includegraphics[height=10cm,width=15cm]{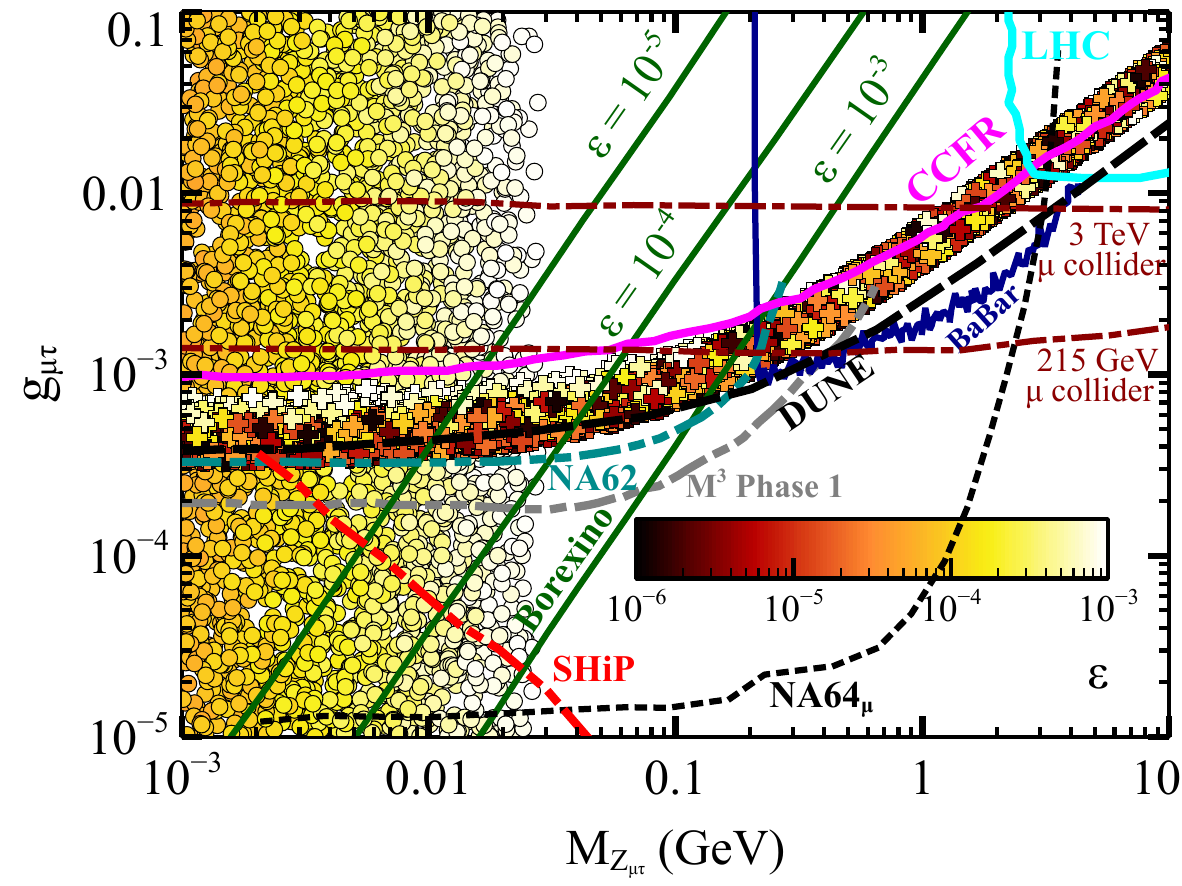}
\caption{Parameter space for $(g-2)_e$ (circular points) and $(g-2)_\mu$ (plus shaped points)
with relevant experimental constraints in $\gmt-M_{\zmt}$ plane. Solid lines
represent existing bounds while other lines are projected exclusion
limits from future experiments.}
\label{Fig:parameter_space_Min_model}
\end{figure}

The recent measurement of $(g-2)_{\mu}$ shows a $4.2\sigma$ deviation
from the SM prediction with a magnitude of $\Delta{a}_{\mu}=
a^{exp}_{\mu}-a^{SM}_{\mu}=(2.51\pm0.59)\times10^{-9}$ \cite{Muong-2:2021ojo}.
On the other hand, there are two existing values of $\Delta{a}_e$
due to two different measurements of the fine structure constant ($\alpha_{em}$)
at Berkeley with $^{137}{\rm Cs}$ atom \cite{Parker:2018vye}
and LKB with $^{87}{\rm Rb}$ atom \cite{Morel:2020dww}
respectively. Using these measurements, the SM prediction for
$a_e\equiv\dfrac{(g-2)_e}{2}$ is either higher ($2.4\sigma$ deviation) or
lower ($1.6\sigma$ deviation) than the experiment value \cite{Hanneke:2008tm}.
More importantly, the nature of new physics
is determined by the measurement of $\alpha_{em}$, where
the former case requires a destructive BSM contribution while an opposite
situation is need for the latter. In the case of muon we
always need a positive contribution to $a_{\mu}$ from the
BSM theory as the SM prediction is lower than the experimental measurement.
Therefore, depending upon the value of $\alpha_{em}$,
we either need positive values of $\Delta{a_{\ell}}$
for both $e$ and $\mu$ or we require a positive $\Delta{a_\mu}$
along with a negative $\Delta{a_e}$.  
Although, there is a relative sign difference between the contributions coming from
the vectorial and the axial vectorial parts of the interaction to $\Delta{a}_{\ell}$ as  
seen from Eq.\,(\ref{del_al}), it is not possible to achieve the second possibility
in the minimal $L_{\mu}-L_{\tau}$ model. The reason behind this is that for muon
we need dominance of the vectorial part of interaction while for $e$ the axial
vectorial dominance, which mainly comes from tiny $Z-Z_{\mu\tau}$ mixing, is required. 
On the other hand, the first possibility where we need same sign of $\Delta{a_{\ell}}$ can be
achieved easily in the minimal model. However, the parameter space addressing both
the experimental values for $\Delta{a_{\mu}}=(2.51\pm0.59)\times10^{-9}$ and
$\Delta{a_e}=(4.8\pm3.0)\times10^{-13}$ is already excluded by the
measurement of  $\nu_e\,e$ scattering at Borexino experiment.
In Fig.\,\,\ref{Fig:parameter_space_Min_model}
we have summarised all results in the familiar $g_{\mu\tau}-M_{Z_{\mu\tau}}$ plane.

In this figure, the $2\sigma$ contour of $(g-2)_{\mu}$ has been
shown by plus shaped points while the corresponding contour for
$(g-2)_{e}$ is denoted by circular points. The colour-bar indicates
variation of the kinetic mixing parameter in the following range
$10^{-6}\leq\epsilon\leq10^{-3}.$ It is clearly seen that to
obtain a sufficient contribution in $\Delta{a_e}$ from $\zmt$
we need $\epsilon\gtrsim10^{-4}$ while for $\mu$, the parameter $\epsilon$
can be as low as $10^{-6}$ or even smaller. This is mainly due
to the fact that the interaction of $e^+\,e^-$ with $\zmt$ is
entirely governed by the $Z-\zmt$ mixing since $e^\pm$ do not have any
$L_{\mu}-L_{\tau}$ charge. However, the vectorial part of interaction
for $\mu^\pm$, responsible for getting a positive $\Delta{a_{\mu}}$,
is dominated by a factor proportional to  the $L_{\mu}-L_{\tau}$ charge.
The overlapping region in Fig.\,\ref{Fig:parameter_space_Min_model}
satisfies both the experimental values of anomalous
magnetic moments and the corresponding parameters are 
$3\times 10^{-4}\lesssim\gmt\lesssim 10^{-3}$, $M_{\zmt}\lesssim0.3$ GeV
and $\epsilon\gtrsim10^{-4}$.  However, as one can see from Fig.\,\ref{Fig:parameter_space_Min_model}
that the above parameter space for $\epsilon\gtrsim10^{-4}$ is already ruled-out
from the observation of $e-\nu$ scatterings at Borexino experiment. We
have depicted excluded regions in $\gmt-M_{\zmt}$ plane for three different
values of the kinetic mixing parameters for which the $e-\nu$ scattering
cross section in the minimal $L_{\mu}-L_{\tau}$ model lies
outside the $2\sigma$ range obtained from Borexino experiment i.e.
$0.88< \dfrac{\sigma^{L_{\mu}-L_{\tau}}_{e\nu}}
{\sigma^{\rm SM}_{e\nu}}<1.24$ 
\cite{Borexino:2017rsf, Altmannshofer:2019zhy}. 
The other relevant existing/proposed experimental bounds from
CCFR, DUNE using neutrino tridents ($\nu_{\mu} N \rightarrow
\nu_{\mu}N\mu^+\mu^-$) \cite{Altmannshofer:2014pba}
and also from four $\mu$ production at LHC
($p\,p\rightarrow\zmt\mu^+\mu^-\rightarrow\mu^+\mu^-\mu^+\mu^-$) \cite{CMS:2018yxg},
BaBar ($e^{+}\,e^{-}\rightarrow\zmt\mu^+\mu^-\rightarrow\mu^+\mu^-\mu^+\mu^-$) \cite{BaBar:2016sci}
are also shown in Fig.\,\ref{Fig:parameter_space_Min_model}. Additionally,
projected bounds from the proposed muon collider \cite{MuonCollider:2022xlm} for two
different centre of mass energies are also presented
in the $\gmt-M_{\zmt}$ plane, where the corresponding
bounds are obtained from a combination of $\zmt$+$\gamma$ searches
and from variation in angular observables of the Bhabha scattering.
Moreover, proposed exclusion limits in the $\gmt-M_{\zmt}$ plane
from different beam-dump experiments namely,  Na62
(from charged kaon decays, $K\rightarrow \mu\nu\zmt \rightarrow \nu\mu\,\nu\bar{\nu}$)
\cite{Krnjaic:2019rsv}, Na64$_{\mu}$ (from large missing energy of muon beam
due to bremsstrahlung of muons in presence of target nuclei and
subsequent invisible decay of $Z_{\mu}\rightarrow\nu\bar{\nu}$) \cite{Gninenko:2014pea},
SHiP \cite{SHiP:2015vad} etc. are included to indicate the future detection
prospects of the $L_{\mu}-L_{\tau}$ scenario. Finally, for completeness,
we have also demonstrated the proposed sensitivity region
from a new muon missing momentum experiment M$^3$ at
Fermi lab \cite{Kahn:2018cqs}. Therefore, it is quiet evident that
there are lots of experimental efforts to detect a possible light $\zmt$ and within
a next few years the entire $\gmt-M_{\zmt}$ parameter space that satisfies
$(g-2)_{\mu}$ in $2\sigma$ range will be probed.
\section{Extended $L_{\mu}-L_{\tau}$ Model}
\label{Sec:Model_extn}

In the previous section, we have tried to expound
both the anomalies reported in $g-2$ of muon and electron in a common
framework. While doing so we have noticed that the minimal
model is not sufficient and we need some extension of the
minimal model. Moreover, we would also like to see that the
extended model is good enough to address issues related to
neutrino mass and dark matter. Therefore, in order to accomplish these unresolved
issues, we extend particle contents of the minimal $L_{\mu}-L_{\tau}$
model. In particular, we introduce a vector like fermion
$\chi$ with nonzero U(1)$_{Y}$ and U(1)$_{L_{\mu}-L_{\tau}}$
quantum numbers and has no colour and weak-isospin charges.
Additionally, in the fermionic sector we include three right handed neutrinos,
having transformation properties similar to the SM leptons
under U(1)$_{L_{\mu}-L_{\tau}}$, for neutrino mass generation
via Type-I seesaw mechanism. In the scalar sector, besides the
SM Higgs doublet $\Phi^\prime_1$ and previously introduced $L_{\mu}-L_{\tau}$ breaking
scalar $\Phi^\prime_3$, we have one SU(2)$_{L}$ doublet $\Phi^\prime_2$ and a
SU(2)$_{L}$ singlet $\Phi^\prime_4$ with suitable $L_{\mu}-L_{\tau}$ charges
to construct new Yukawa interactions between the SM Leptons
and $\chi$. As we will see later, both these scalars
along with the charged fermion $\chi$ have played
a pivotal role in generating $(g-2)_e$ in the ballpark
of experimental measurements.  
In Tables (\ref{tab1-SM-Gauge}, \ref{tab2-U(1)-LmLt}), we have shown the
complete particle spectrum and associated charges under the complete
gauge group $SU(3)_{C} \times SU(2)_{L} \times U(1)_{Y} 
\times U(1)_{L_{\mu} - L_{\tau}}$ of the present model. Here we would like to note
that since all fermions are vector like under the $L_{\mu}-L_{\tau}$ symmetry
the extended model is also anomaly free as it was for the minimal model.
\begin{center}
\begin{table}[h!]
\begin{tabular}{||c|c|c|c||}
\hline
\hline
\begin{tabular}{c}
    Gauge\\
    Group\\ 
    \hline
    
    ${\rm SU(2)}_{\rm L}$\\ 
    \hline
    ${\rm U(1)}_{\rm Y}$\\ 
\end{tabular}
&

\begin{tabular}{c|c|c}
    \multicolumn{3}{c}{Baryon Fields}\\ 
    \hline
    $Q_{L}^{i}=(u_{L}^{i},d_{L}^{i})^{T}$&$u_{R}^{i}$&$d_{R}^{i}$\\ 
    \hline
    $2$&$1$&$1$\\ 
    \hline
    $1/6$&$2/3$&$-1/3$\\ 
\end{tabular}
&
\begin{tabular}{c|c|c|c|c}
    \multicolumn{5}{c}{Lepton Fields}\\
    \hline
    $L_{L}^{i}=(\nu_{L}^{i},e_{L}^{i})^{T}$ & $e_{R}^{i}$ & $N_{R}^{i}$ & $\chiL$ & $\chiR$\\
    \hline
    $2$&$1$&$1$&$1$ &$1$\\
    \hline
    $-1/2$&$-1$&$0$&$-1$ &$-1$\\
\end{tabular}
&
\begin{tabular}{c|c|c|c}
    \multicolumn{4}{c}{Scalar Fields}\\
    \hline
    $\Phi^\prime_{1}$&$\Phi^\prime_{2}$&$\Phi^\prime_{3}$&$\Phi^\prime_4$\\
    \hline
    $2$&$2$&$1$&$1$\\
    \hline
    $1/2$&$1/2$&$0$&$0$\\
\end{tabular}\\
\hline
\hline
\end{tabular}
\caption{Particle contents and their corresponding
charges under the SM gauge group.}
\label{tab1-SM-Gauge}
\end{table}
\end{center}

\begin{center}
\begin{table}[h!]
\begin{tabular}{||c|c|c|c||}
\hline
\hline
\begin{tabular}{c}
    Gauge\\
    Group\\ 
    \hline
    $\umt$\\ 
    
\end{tabular}
&
\begin{tabular}{c}
    \multicolumn{1}{c}{Baryonic Fields}\\ 
    \hline
    $(Q^{i}_{L}, u^{i}_{R}, d^{i}_{R})$\\ 
    \hline
    $0$ \\ 
    
\end{tabular}
&
\begin{tabular}{c|c|c|c|c}
    \multicolumn{5}{c}{Lepton Fields}\\ 
    \hline
    $(L_{L}^{e}, e_{R}, N_{R}^{e})$ & $(L_{L}^{\mu}, \mu_{R},
    N_{R}^{\mu})$ & $(L_{L}^{\tau}, \tau_{R}, N_{R}^{\tau})$ & $\chiL$ & $\chiR$\\ 
    \hline
    $0$ & $1$ & $-1$ & $1/3$ & $1/3$\\ 
    
\end{tabular}
&
\begin{tabular}{c|c|c|c}
    \multicolumn{4}{c}{Scalar Fields}\\
    \hline
    $\Phi^\prime_{1}$ & $\Phi^\prime_{2}$ & $\Phi^\prime_{3}$&$\Phi^\prime_4$ \\
    \hline
    $0$ & $-1/3$ & $1$&$-1/3$\\
\end{tabular}\\
\hline
\hline
\end{tabular}
\caption{$L_{\mu}-L_{\tau}$ charges for fermions and scalars of the present model.}
\label{tab2-U(1)-LmLt}
\end{table}
\end{center}   
The full Lagrangian of the present model is given by
\begin{eqnarray}
&&\hspace{-0.6cm}
\mathcal{L} = \mathcal{L}_{\rm SM} 
- \frac{1}{4} \hat{X}_{\alpha \beta} \hat{X}^{\alpha \beta} +
\frac{\epsilon}{2} \hat{X}_{\alpha \beta} \hat{B}^{\alpha \beta}
- \gmt\sum_{\substack{\ell = \mu, \nu_\mu, \\\,\,\, \tau, \nu_{\tau}}}
Q^{\ell}_{\mu\tau} \bar{\ell} \gamma^{\alpha} \ell
\hat{X}_{\alpha}
+\bigg[
i\,\bar{\chi}\slashed{\partial}\chi 
- \bar{\chi} \gamma^{\alpha} \chi
\left(\dfrac{\gmt}{3} \hat{X}_{\alpha} - {g_1}\hat{B}_{\alpha}\right) 
\nn \\ && \hspace{-0.6cm}
- M_{\chi} \overline{\chi}\chi
- \biggl( \beta^{R}_{e\chi} \overline{L_{e}} \Phi^\prime_2 \chiR 
+ \beta^{L}_{e\chi} \overline{e_R}\,\chiL \Phi^\prime_4 + {\it h.c.} \biggr)\bigg]
+ \mathcal{L}_{N}
+\sum_{i=2}^{4} \left(D_{\alpha}{\Phi^\prime_i}\right)^{\dagger} \left(D^{\alpha}{\Phi^\prime_i}\right) 
 - \mathcal{V}(\Phi^\prime_1, \Phi^\prime_2, \Phi^\prime_3, \Phi^\prime_4) \,.\nn\\
\label{lagrangian}
\end{eqnarray}
Apart from the terms which have already defined in the previous section, the terms
within the square brackets represent the Lagrangian of vector like fermion $\chi$
including the new Yukawa interactions with couplings
$\beta^{R}_{e\chi}$ and $\beta^{L}_{e\chi}$
respectively. The Lagrangian for three right handed neutrinos are denoted by  
$\mathcal{L}_{N}$. We have written the exact form of $\mathcal{L}_{N}$ below.
\begin{eqnarray}
\mathcal{L}_{N} &=& \frac{i}{2} \sum_{\substack{i = e, \mu,\\ \tau}} \overline{N_{i}} \slashed{D} N_{i}
+ \sum_{\substack{i = e, \mu,\\\tau}} \big( y _{ii} \,\overline{L_{i}} \Phi^\prime_1 N_i  + {\it h.c.} \big) 
+ \big( M_{ee} N_{e} N_{e} + h_{e\mu} N_{e} N_{\mu} {\Phi^{\prime}_{3}}^\dagger
 \nn \\
&& + h_{e\tau} N_{e} N_{\tau} \Phi^\prime_{3} 
+ M_{\mu\tau} e^{i \theta} N_{\mu}N_{\tau}  + {\it h.c.}\big)\,,
\label{rh-neutrino-lagrangian}
\end{eqnarray} 
where, the first term is the kinetic term of $N_i$ while
the rest are interaction terms responsible for  
the light neutrino mass generation via Type-I seesaw mechanism.
We refrain further discussion on this topic and will discuss
again when we talk about neutrino mass in Section \ref{Sec:nu_mass}. Finally,
the last two terms in Eq.\,(\ref{lagrangian}) are the kinetic
and interaction terms for the BSM scalars ($\Phi^\prime_i,\,i=2,\,3,\,4$).
In the kinetic term, $D_{\alpha}$ being the usual covariant derivative
involving gauge boson(s) and generator(s) of each group
under which $\Phi^{\prime}_i$ transforms non-trivially.
The explicit form of $\mathcal{V}$, invariant under the full
gauge group, is given below  
\begin{eqnarray}
\mathcal{V}(\Phi^\prime_1, \Phi^\prime_2, \Phi^\prime_3, \Phi^\prime_4)
&=&  \sum^{4}_{i=2} \left[ \mu^2_{i} \left({\Phi^{\prime}}^\dagger_i
\Phi^\prime_i \right) 
+  \lambda_i \left({\Phi^{\prime}}^\dagger_i\Phi^\prime_i\right)^2 \right] 
+ \sum_{i,j, j > i} \lambda_{ij} \left({\Phi^{\prime}}^\dagger_i\Phi^\prime_i\right)
\left({\Phi^{\prime}}^\dagger_j\Phi^\prime_j\right) \nn \\
&& \hspace{-0.7cm}
+ \lambda^\prime_{12} \left({\Phi^{\prime}}^\dagger_1\Phi^\prime_2\right)
\left({\Phi^{\prime}}^\dagger_2\Phi^\prime_1\right)
+ \left[\mu \left({\Phi^{\prime}_1}^\dagger \Phi^\prime_2 \right) {\Phi^{\prime}_{4}}^\dagger 
+ \xi \left( \Phi^\prime_{3} {\Phi^{\prime}_{4}}^{3} \right)
+ {\it h.c.} \right].
\label{potential}
\end{eqnarray}
Here for necessary vacuum alignment we need $\mu^2_{1,3} < 0$, $\mu^2_{2,4} > 0$
and $\lambda_{i} > 0$. The second condition, $\mu^2_{2,4} > 0$, 
ensures that two new $L_{\mu}-L_{\tau}$ charged scalars $\Phi^\prime_2$ and $\Phi^\prime_4$
do not have any VEV. Apart from the usual self-conjugate terms, we have two
non-self-conjugate terms also in the Lagrangian which have utmost
importance in the present context. The trilinear term with coefficient
$\mu$ introduces mixing between the neutral component of $\Phi^\prime_2$
and $\Phi^\prime_4$. Since the VEVs of both these scalars are zero, the lightest
one is automatically stable and can be a viable dark matter candidate. Therefore,
for a singlet like dark matter candidate (dominated by $\Phi^\prime_4$),
which is precisely the case we are considering, this mixing opens
up a direct detection prospect through exchange of the SM $Z$
boson. On the other hand, the quartic term proportional to $\xi$
is responsible for cubic interaction among the dark matter particles,
which results in some higher order number changing processes ($3\rightarrow 2$).
After both EWSB and $L_{\mu}-L_{\tau}$ breaking, the $2\times2$ mass matrix for the
neutral component of $\Phi^\prime_2$ ($\phi^\prime_2$) and $\Phi^\prime_4$ is
given by 
 \begin{eqnarray}
 M^2_{\phi^\prime_{2} - \Phi^\prime_{4}} = \begin{pmatrix}
 a & b \\
 b & c
 \end{pmatrix}\,.
 \end{eqnarray}
 where $a = \mu^2_2 + (\lambda_{12} +\lambda^\prime_{12})\frac{v^2}{2} + \lambda_{23} \frac{v_{\mu\tau}}{2}$,
 $c = \mu^2_4 + \lambda_{14} \frac{v^2}{2} + \lambda_{34} \frac{v_{\mu\tau}}{2}$
 and $b = \frac{\mu v}{\sqrt{2}}$. One can
 easily diagonalise the above mass matrix using an orthogonal transformation by
 an angle $\theta_{D}$ and the resultant eigenstates are related to
 the old basis sates $(\phi^\prime_2,\,\,\Phi^\prime_4)$ in the following way
 \begin{eqnarray}
 \begin{pmatrix}
 \phi_2 \\
 \phi_4
 \end{pmatrix}
 =
 \begin{pmatrix}
 \cos \theta_{D} & - \sin \theta_{D} \\
 \sin \theta_{D} & \cos \theta_{D}
 \end{pmatrix}
 \,\,\begin{pmatrix}
 \phi^{\prime}_2\\
 \Phi^{\prime}_4
 \end{pmatrix}\,,
 \label{thetaD}
 \end{eqnarray}
 where, the mixing angle $\theta_{D}$ can be expressed in terms of the parameters
 of the Lagrangian as,
 \begin{eqnarray}
 \theta_{D} &=& \dfrac{1}{2} \tan^{-1}\left(\dfrac{2b}{c-a}\right)\,,\nn\\
&=& \dfrac{1}{2} \tan^{-1}\left[\dfrac{\sqrt{2}\,{\mu v}}
{\mu^2_4 -\mu^2_2 + (\lambda_{14}-\lambda_{12} -\lambda^\prime_{12})\frac{v^2}{2} 
+ (\lambda_{34}-\lambda_{23}) \frac{v_{\mu\tau}}{2}
}
\right]
\end{eqnarray} 
 and the masses corresponding to the physical states $\phi_2$ and $\phi_4$ are
 \begin{eqnarray}
 M^2_{\phi_2} &=& 
\left[\mu^2_2 + (\lambda_{12} +\lambda^\prime_{12})\frac{v^2}{2} + \lambda_{23} \frac{v_{\mu\tau}}{2}\right]
\cos^{2}\theta_{D}
+\left[\mu^2_4 + \lambda_{14} \frac{v^2}{2} + \lambda_{34} \frac{v_{\mu\tau}}{2}\right]
\sin^{2}\theta_{D}
-\frac{\mu v}{\sqrt{2}} \sin 2\theta_{D}\,,
\nn \\
M^2_{\phi_4} &=& \left[\mu^2_2 + (\lambda_{12} +\lambda^\prime_{12})\frac{v^2}{2} + \lambda_{23} \frac{v_{\mu\tau}}{2}\right]
 \sin^{2}\theta_{D}
+\left[\mu^2_4 + \lambda_{14} \frac{v^2}{2} + \lambda_{34} \frac{v_{\mu\tau}}{2}\right]
 \cos^{2}\theta_{D}
+\frac{\mu v}{\sqrt{2}} \sin 2\theta_{D}\,.\nn\\
 \end{eqnarray}
In our work, we have considered that $\phi_4$ is the lightest state and 
is a suitable for dark matter candidate.  Similar to the $\phi^\prime_2-\Phi^\prime_4$
mixing, there is another mass mixing between the
CP even neutral components $(\phi^\prime_1,\,\,\phi^\prime_3)$
of the Higgs doublet $\Phi^\prime_1$ and the $L_{\mu}-L_{\tau}$ breaking
scalar $\Phi^\prime_3$ respectively. This mixing is due to the presence
of a quartic interaction term with coefficient $\lambda_{13}$. Therefore,
after spontaneous symmetry breaking this quartic interaction generates off-diagonal
terms in the $2\times2$ mass matrix which is given by
\begin{eqnarray}
M^2_{\phi^\prime_{1} - \phi^\prime_{3}} = \begin{pmatrix}
2 \lambda_{1} v^2 & \lambda_{13} v v_{\mu\tau} \\
\lambda_{13} v v_{\mu\tau} & 2 \lambda_{3} v^2_{\mu\tau}
\end{pmatrix}\,.
\end{eqnarray}          
This symmetric mass matrix can be diagonalised in a similar manner as
above and as a result, we get two physical states $h_1$ and $h_3$ with
a mixing angle $\theta$ which can be expressed as,
\begin{eqnarray}
\theta = \dfrac{1}{2}\tan^{-1}\left[\frac{\lambda_{13} v v_{\mu\tau}}
{\lambda_3 \vmt^2 - 
\lambda^2_1 v^2}\right].
\end{eqnarray}
In this work we have considered $h_1$ as the SM like Higgs boson which
was discovered by the ATLAS \cite{ATLAS:2012yve}
and the CMS \cite{CMS:2012qbp} collaborations
in 2012 and having mass $M_{h_1}=125.5$ GeV. We would like to mention
in passing that the CP odd neutral components of $\Phi^\prime_1$
and $\Phi^\prime_3$ turn into the Goldstone bosons after both
$Z$ and $\zmt$ become massive. Therefore, in the scalar sector of the
extended model, apart from the SM like Higgs boson $h_1$, we have one CP even
scalar ($h_2$), one charged scalar (part of the doublet $\Phi^\prime_2$) and
two complex scalars ($\phi_2$, $\phi_4$). The latter take part in
one loop diagram contributing to $(g-2)_e$ while
$\phi_4$ plays the role of dark matter with enhanced detection
possibilities directly due to $\phi_2$.    
{As mentioned in the previous section, besides the contribution coming
from $\zmt$ through kinetic mixing, the new Yukawa
interactions (defined in Eq.\,(\ref{lagrangian}) involving $e$, $\chi$
and $\phi_i$ ($i=2,\,4$) provide additional contribution to $(g-2)_{e}$.
The details about $(g-2)_e$ have been discussed in 
Appendix \ref{Sec:g-2_ext}. 
In Fig.\,\ref{beta-mchi-sigma}, we have shown the allowed parameter space
in $\beta^{L}_{e\chi}-M_{\chi}$ plane after demanding $\Delta{a}_e$ in $2\sigma$ range of experimental measurement. Moreover, for the
same parameter space we have also
shown the $3\sigma$ and $5 \sigma$ statistical significance of the 
charged fermion at the $e^{+}e^{-}$ linear collider.}
  
\section{Neutrino mass}
\label{Sec:nu_mass}
As described in Eq.\,(\ref{rh-neutrino-lagrangian}), the Lagrangian
associated with the right handed neutrinos generate light neutrino masses
by the Type-I seesaw mechanism \cite{Minkowski:1977sc, Yanagida:1979as,
Mohapatra:1979ia, Schechter:1980gr}. A similar technique for
generating neutrino masses in the context of ${\rm U(1)}_{L_{\mu}-L_{\tau}}$
symmetry has already been explored in detail
by the present authors in \cite{Biswas:2016yan}.
We get both Dirac and Majorana masses when the SM Higgs doublet $\Phi^\prime_1$
and the singlet scalar $\Phi^\prime_3$ acquire VEVs.
The Dirac  mass matrix $M_{D}$ 
has the following form once the electroweak symmetry
breaks,
\begin{eqnarray}
M_{D} = \left(\begin{array}{ccc}
 \frac{y_{ee} v}{\sqrt{2}}~~&~~ 0 ~~&~~ 0 \\
~~&~~\\
0 ~~&~~  \frac{y_{\mu\mu} v}{\sqrt{2}} ~~&~~ 0 \\
~~&~~\\
0 ~~&~~ 0 ~~&~~  \frac{y_{\tau\tau} v}{\sqrt{2}} \\
\end{array}\right) \,,
\label{md}
\end{eqnarray}
and the Majorana mass matrix $M_{R}$ takes the following form when 
${\rm U(1)}_{L_{\mu}-L_{\tau}}$ symmetry gets broken,
\begin{eqnarray}
\mathcal{M}_{N} = \left(\begin{array}{ccc}
M_{ee} ~~&~~ \dfrac{ \vmt}{\sqrt{2}} h_{e \mu}
~~&~~\dfrac{\vmt}{\sqrt{2}} h_{e \tau} \\
~~&~~\\
\dfrac{\vmt}{\sqrt{2}} h_{e \mu} ~~&~~ 0
~~&~~ M_{\mu \tau} \,e^{i\xi}\\
~~&~~\\
\dfrac{\vmt}{\sqrt{2}} h_{e \tau} ~~&
~~ M_{\mu \tau}\,e^{i\xi} ~~&~~ 0 \\
\end{array}\right) \,.
\label{mncomplex}
\end{eqnarray}
In the Dirac mass matrix, we can rotate away the phases by redefining 
the left handed neutrinos on the other hand we cannot get rid of
all the phases associated with the Majorana mass matrix and 
one phase remains which we have considered in (2,3) position.
We can write down the neutrino mass matrix in the basis 
$({\nu_{L}}_\alpha, {N_{R}}^c_{\alpha})$ ($\alpha = e, \mu, \tau$)
as follows,
\begin{eqnarray}
M_{\nu} = \begin{pmatrix}
0 & M_{D} \\
M^{T}_{D} & M_{N}
\end{pmatrix}\,.
\end{eqnarray}
As demanded by the oscillation experiments and cosmology, we have the
following allowed range of the oscillation parameters,
\begin{itemize}
\item bound on the sum of all three light neutrinos from cosmology,
$\sum_i m_{i} < 0.23$ eV at $2\sigma$ C.L. \cite{Ade:2015xua},
\item $2\sigma$ range of  mass squared 
differences, $7.20<\dfrac{\Delta m^2_{21}}
{10^{-5}}\,{\text{eV}^2} < 7.94$ and \\$2.44(2.34)<\dfrac{\Delta m^2_{31}}
{10^{-3}}\,{\text{eV}^2} < 2.57(2.47)$ for NO(IO) \cite{deSalas:2017kay},
\item $2\sigma$ bound on the  mixing angles $32.5^{\circ}<\,\theta_{12}\,<36.8^{\circ}$,
$43.1^{\circ}(44.5^{\circ})<\,\theta_{23}\,<49.8^{\circ}(48.9^{\circ})$
and $8.2^{\circ} (8.3^{\circ}) <\,\theta_{13}\,<8.8^{\circ}$ for NO(IO)
\cite{deSalas:2017kay}.
\end{itemize}
Moreover, we also have the bound on the effective number of relativistic
d.o.f ($N_{eff}$) allowed from cosmology which is $N_{eff} = 2.99\pm 0.17$ \cite{Planck:2018vyg}.
Therefore, to be consistent with all these observations and experimental
measurements we can have three neutrinos in the sub-eV scale and other three
are in the higher scale. This is indeed possible if we go to the seesaw 
regime where $M_{N} \gg M_{D}$. Then we can diagonalize the $M_{\nu}$ matrix
as follows,
\begin{eqnarray}
m^{light}_{\nu} \simeq -M^{T}_{D} M^{-1}_{N} M_{D}\,,\,\,\,
{\rm and}\,\,\,m^{heavy}_{N} \simeq M_{N}
\end{eqnarray}  
where $m^{light}_{\nu}$ corresponds to three sub-eV scale Majorana neutrinos while 
$m^{heavy}_{N}$ corresponds to three heavier Majorana neutrinos. The detailed
numerical analysis and the ranges of associated parameters are given in \cite{Biswas:2016yan}.
\section{SIMP dark matter}
\label{Sec:simp}
\begin{figure}[h!]
\includegraphics[height=12cm,width=16cm,angle=0]{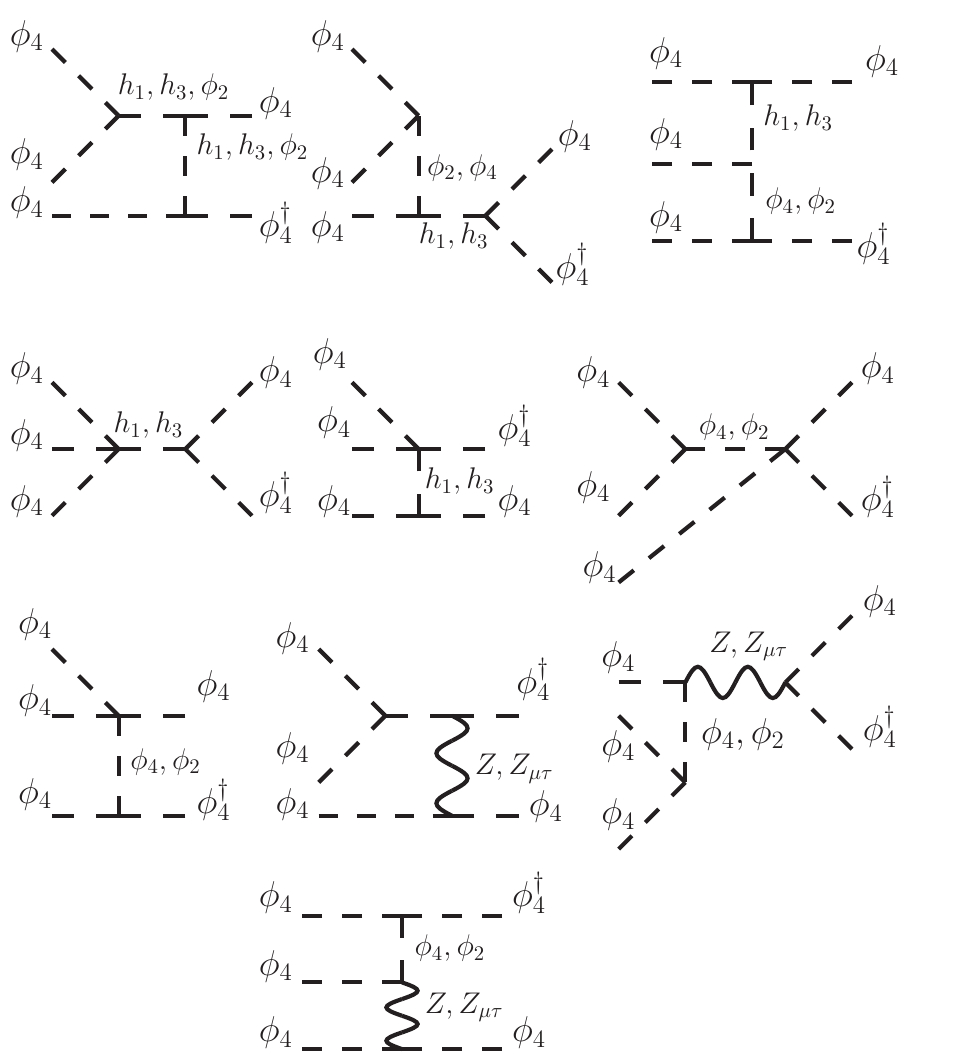}
\caption{Feynman diagrams for the $3\rightarrow2$ scattering $\phi_4\phi_4\phi_4
\rightarrow \phi_4 \phi^\dagger_4$, where comoving number density of
$\phi_4$ decreases by 2 unit per scattering.}
\label{Fig:simp_dia}
\end{figure}

We have seen in Section \ref{Sec:Model_extn} that there are two neutral scalars
$\phi_{2}$, $\phi_4$. The $L_{\mu}-L_{\tau}$ charge assignment
(see Table \ref{tab2-U(1)-LmLt}) among the scalar fields is extremely
crucial not only to get a stable dark matter candidate but also
it dictates the nature of dark matter As a result, the lightest scalar $\phi_4$
is naturally stable and has a cubic self-interaction term when $\Phi^\prime_3$
gets a VEV. The latter is responsible for number changing interactions
occurring through higher order scattering like  $3\rightarrow2$ processes.  
When the coupling $\xi$ of the cubic term is large enough so that
the main number changing processes are $3\rightarrow2$ scatterings
rather than the usual $2\rightarrow2$ pair annihilations of dark matter
into the SM fields, the freeze-out of $\phi_4$ is primarily determined
by the condition $\Gamma_{3\rightarrow2}\lesssim\mathcal{H}$. Moreover, the dark matter
$\phi_4$ maintains kinetic equilibrium with the SM bath by virtue of
elastic scattering with $\zmt$ where the latter remains thermally connected
with the SM leptons. This is known as the
SIMP paradigm \cite{Carlson:1992fn, Hochberg:2014dra}. 
In this work, we have explored the phenomenology of SIMP dark matter
in the context of U(1)$_{L_{\mu}-L_{\tau}}$ gauge extension. 
The Boltzmann equation expressing the evolution of comoving
number density of dark matter is given by\footnote{see
\cite{Biswas:2020ubd} for a detailed derivation of the Boltzmann equation}
\begin{eqnarray}
\frac{d Y_{\rm DM}}{d x} &=& \dfrac{{\rm s}^2}{\mathcal{H}\,x} 
\langle \sigma_{3 \rightarrow 2}^{\rm tot} {\rm v}^{2} \rangle\,
Y^{2}_{\rm DM}\left(Y_{\rm DM} - Y^{\rm eq}_{\rm DM} \right) 
-\dfrac{\rm s}{\mathcal{H}\,x} 
\langle \sigma_{2\rightarrow 2}^{\rm tot} {\rm v} \rangle 
\left(Y^{2}_{\rm DM} - (Y^{\rm eq}_{\rm DM})^2 \right). \nonumber \\ 
\label{Boltz}
\end{eqnarray}
Where $Y_{\rm DM}$ is the total comoving number densities of
both $\phi_4$ and $\phi^\dagger_4$ respectively while $x=M_{\phi_4}/T$ with
$T$ being the photon temperature. Here we have assumed that there is
no asymmetry between the number densities of particle and anti-particle
of dark matter. The entropy density and the Hubble
parameters are denoted by ${\rm s}$ and $\mathcal{H}$.
The quantity $\langle \sigma_{3 \rightarrow 2}^{\rm tot} {\rm v}^{2} \rangle$
is the thermal average of total scattering cross section for all relevant
number changing $3\rightarrow 2$ processes for $\phi_4$ taking into account
all symmetry factors 
\begin{eqnarray}
\sigma_{3 \rightarrow 2}^{\rm tot}
= -\dfrac{2}{3!}\,\sigma_{\phi_4\phi_4\phi_4 \rightarrow \phi^\dagger_4\phi_4}
-\dfrac{2}{2!}\,\sigma_{\phi^\dagger_4\phi_4\phi_4 \rightarrow \phi^\dagger_4\phi^\dagger_4}
+\dfrac{1}{3!}\,\sigma_{\phi^\dagger_4\phi^\dagger_4\phi^\dagger_4
\rightarrow \phi^\dagger_4\phi_4}
+\dfrac{1}{2!}\,\sigma_{\phi_4\phi^\dagger_4\phi^\dagger_4
\rightarrow \phi_4\phi_4}\,.
\label{sigma3to2}
\end{eqnarray}
Here, $+(-)$ represents increase(decrease) of $\phi_4$ due to a particular scattering,
e.g.\,\,the first term is the scattering cross section for a $3\rightarrow2$ process
like $\phi_4\phi_4\phi_4\rightarrow\phi_4\phi^\dagger_4$ which reduces the comoving
number density of $\phi_4$ by 2 unit per scattering.
Moreover, the factor $3!$ in the denominator
is due to three identical particles in the initial state. 
As we have computed these $3\rightarrow2$ processes in the non-relativistic limit
of dark matter following \cite{Berlin:2016gtr}, the thermal average
$\langle \sigma_{3 \rightarrow 2}^{\rm tot} {\rm v^2}\rangle$
is identical to $\sigma_{3 \rightarrow 2}^{\rm tot} {\rm v^2}$. The
matrix amplitude square ($|\mathcal{M}|^2$) for a particular $3\rightarrow2$ process is
related to $\sigma {\rm v^2}$, in the non-relativistic
limit, as \cite{Berlin:2016gtr}
\begin{eqnarray}
\langle \sigma 
{\rm v}^{2} \rangle_{3 \rightarrow 2} = 
\frac{\sqrt{5}}{S\times 384 \pi M^3_{\phi_4}} \int_{-1}^{+1}d\cos\theta\,
|\mathcal{M}_{3\rightarrow2}|^2\,, 
\end{eqnarray}
where initial and final state particles are either $\phi_4$ or $\phi_4^\dagger$ or both.
The symmetry factor $S$ depends on number of identical particles in the final state.
We have determined these amplitudes using \texttt{CalcHEP} \cite{Belyaev:2012qa} 
while the necessary model files have been generated by the Mathematica based
package \texttt{FeynRules} \cite{Alloul:2013bka}. Relevant diagrams for the scattering 
$\phi_4 \phi_4 \phi_4 \rightarrow \phi^\dagger_4 \phi_4$ are
shown in Fig.\,\ref{Fig:simp_dia} and the Feynman diagrams for other processes
can be generated easily following these diagrams. The necessary vertex factors
are listed in Appendix\,\,\ref{App:Vertices}.

The second term in Eq.\,\,(\ref{Boltz}) is coming
from $2\rightarrow2$ scatterings. These can
be proceed through mediation of scalars ($h_1$, $h_2$) and gauge bosons ($Z$, $\zmt$).
In this work, since we want to explore the phenomenon of freeze-out within the
dark sector only, we have chosen feeble scalar couplings. Therefore, the pair-annihilations
of $\phi_4$ and $\phi^\dagger_4$ into light SM fermions are possible through exchange of
$L_{\mu}-L_{\tau}$ gauge boson. Additionally, we have annihilation channel like
$\phi^\dagger_4\phi_4\rightarrow\zmt\zmt$. 
Once we obtain these scattering cross sections using \texttt{CalcHEP},
the thermal average $\langle\sigma^{\rm tot}_{2\rightarrow2} {\rm v}\rangle$
can be computed as
\begin{eqnarray}
\langle\sigma^{\rm tot}_{2\rightarrow2} {\rm v}\rangle =
\dfrac{1}{4} \dfrac{1}{r^4_{\phi_4}x^4\,{\rm K^2_2}(r_{\phi_4}x)}
\int_{2\,r_{\phi_4}x}^{\infty} dZ \,
\left(\sum_{Y}\sigma_{\phi_4\phi^\dagger_4\rightarrow Y\overline{Y}}\right)\,
Z^2 \left(Z^2-4r^2_{\phi_4}x^2\right){\rm K_1}(Z)\,.
\end{eqnarray}
The sum is over all possible final states $Y$ and $Z=\dfrac{\sqrt{s}}{T}$
where $s$ is one of the Mandelstam variables. The variables $r_{\phi_4}=
\dfrac{M_{\phi_4}}{M_0}$ and $x=\dfrac{M_0}{T}$ where $M_0$ is an arbitrary
mass scale. In this work we have chosen $M_0=M_{\phi_4}$, hence $r_{\phi_4}=1$.
We would like to note here that we need sufficient
elastic scatterings between dark matter and light $L_{\mu}-L_{\tau}$ charged leptons
to keep the dark sector thermally connected with the SM bath. We have discussed our
numerical results later and we now discuss the relevant bounds
that we have considered in the present work. These are listed below.
\begin{figure}[h!]
\includegraphics[height=9cm,width=17cm]{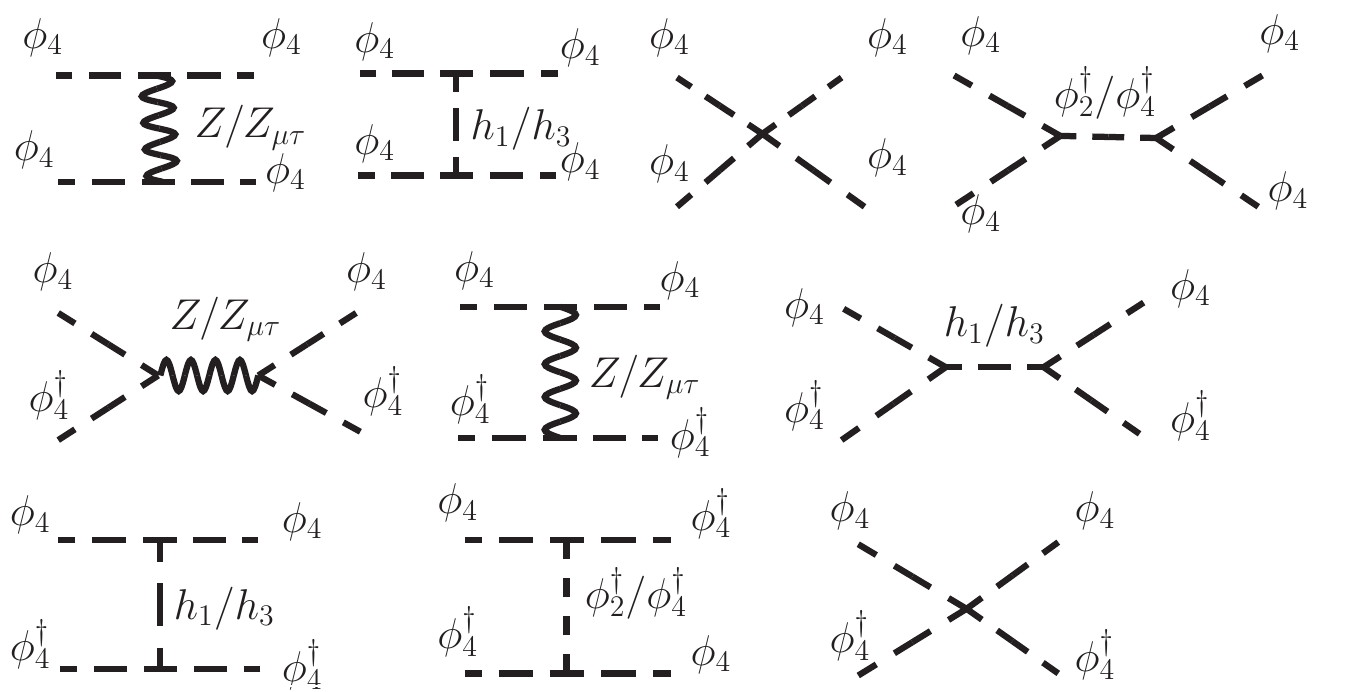}
\caption{Feynman diagrams for the self interactions between
$\phi_4 \phi_4$ and $\phi_4\phi^\dagger_4$. The diagrams for the
process $\phi^\dagger_4\phi^\dagger_4\rightarrow\phi^\dagger_4\phi^\dagger_4$
are same as those for $\phi_4\phi_4\rightarrow\phi_4\phi_4$.}
\label{Fig:self_int_Feyn_dia}
\end{figure}
\begin{itemize}
\item {\bf Self Interaction:}
Dark matter $\phi_4$ and $\phi^\dagger_4$ in our present model
have the following self scatterings which conserve their individual numbers
\begin{eqnarray}
\phi_4 \phi_4 \rightarrow \phi_4 \phi_4 \,,\,\,\,
\phi_4 \phi^{\dagger}_4 \rightarrow \phi_4 \phi^{\dagger}_4\,,\,\,\,
\phi^{\dagger}_4 \phi^{\dagger}_4 
\rightarrow \phi^{\dagger}_4 \phi^{\dagger}_4\,.
\end{eqnarray}
The effective self interaction cross section, considering
both $\phi_4$ and $\phi^\dagger_4$ contribute an equal amount
to the relic density, is given by
\begin{eqnarray}
\sigma_{\rm self} = \dfrac{1}{4} \left(\sigma_{\phi_4\phi_4}+
\sigma_{\phi_4\phi^\dagger_4} +
\sigma_{\phi^\dagger_4\phi^\dagger_4}
\right)\,.
\end{eqnarray}
The individual scattering cross sections are obtained using the
package \texttt{CalcHEP} and the corresponding Feynman diagrams are
shown in Fig.\,\,\ref{Fig:self_int_Feyn_dia}.
Thereafter we have enforced the non-relativistic
limit by putting $s\simeq4\,M^2_{\phi_4} + M^2_{\phi_4}\,{\rm v}^2$ and
taking ${\rm v} \rightarrow0$ where ${\rm v}$ is the relative velocity
in the centre of mass frame.
From the observation of the bullet cluster \cite{Clowe:2003tk, Markevitch:2003at} there is a
bound on the ratio $\frac{\sigma_{\rm self}}{M_{\phi_4}} \lesssim 1 \,\,
{\rm cm^{2}}/{\rm g}$. The bound depends on the relative velocity of
dark matter particles in a particular galaxy. Similarly, the Abell 520 cluster
merger predicts $\frac{\sigma_{\rm self}}{M_{\phi_4}} \sim 1 \,\,
{\rm cm^{2}}/{\rm g}$\,. Moreover, there is a wide range of the allowed
values of $\frac{\sigma_{\rm self}}{M_{\phi_4}}$ from various astrophysical
observations and N-body simulations \cite{Tulin:2017ara}. To be consistent with
maximum number of observations, in this work, we have
considered $0.1\,\,{\rm cm^2/g}\leq\frac{\sigma_{\rm self}}{M_{\phi_4}}
\leq 10\,\,{\rm cm^{2}}/{\rm g}$.
\item {\bf Perturbativity and Unitarity:}
We have considered perturbative limit ($<4\pi$) on all the quartic
couplings in Eq.\,\,(\ref{potential}) so that the vacuum does
not become unbounded from below for the large value of scalar fields.\,\,Moreover,
decomposing the matrix amplitude of a scattering process into partial
waves, the requirement of unitarity of S-matrix demands
\begin{eqnarray}
|\mathcal{M}| \leq 16 \pi\,.
\label{unitarity}
\end{eqnarray}   
\item {\bf Direct Detection bound:}
The dark matter candidates $\phi_4$ and $\phi^\dagger_4$
in the present model can be detected at the direct 
detection experiments by scattering with heavy nuclei and electrons as well.
Instead of being predominantly an SU(2)$_L$ singlet like state, the mixing
of $\Phi^\prime_4$ with the neutral component of inter doublet $\Phi^\prime_2$ generates
$\phi_4\phi_4^{\dagger}Z$ vertex. This is a vectorial interaction (proportional
to $\gamma_{\mu}$) only. On the other hand, the $\bar{f}fZ$ ($f$ is any SM fermion)
vertex factor has both the vectorial as well as the
axial vectorial (proportional to $\gamma_{\mu}\gamma_5$) parts.  
While the vectorial part is responsible for spin independent 
scattering, spin dependent scattering is possible 
due to the axial vectorial part. As a result, we have both spin independent
as well as spin dependent scatterings when dark matter scatters off through
$Z$ boson.\,\,The spin independent elastic scattering cross sections is given by 
\begin{eqnarray}
\sigma_{\rm SI}&=&\left(\dfrac{g_2\,\sin^2\theta_D}{2\,\cos\theta_W}\right)^2
\dfrac{\mu^2_{\phi_4\,N}}{\pi\,M^4_Z}\times \nn \\ && 
\left[\dfrac{\mathbb{Z}\left\{2a_q(\frac{1}{2},\frac{2}{3})+
a_q(-\frac{1}{2},-\frac{1}{3})\right\} +
(\mathbb{A-Z})\left\{a_q(\frac{1}{2},\frac{2}{3})+
2a_q(-\frac{1}{2},-\frac{1}{3})\right\}}{\mathbb{A}}
\right]^2,
\label{sigmaSI}
\end{eqnarray}
where,
\begin{eqnarray*}
a_{q}(T_3, Q^q_{em})=-\dfrac{g}{2\,\cos\theta_W}
\left\{\left(T_3-2\,Q^q_{em}\sin^2\theta_W\right)\cos\theta_{\mu\tau}
+\epsilon \sin\theta_W\left(T_3 - 2\,Q^q_{em}\right)\sin\theta_{\mu\tau}\right\}
\end{eqnarray*}
is the vectorial part
of $\bar{q}qZ$ coupling while $\mathbb{Z}$, $\mathbb{A}$ are
atomic number and mass number of the detector nucleus respectively.
The reduced mass between dark matter matter and nucleon $N$ is denoted
by $\mu_{\phi_4\,N}$. The spin dependent scattering cross section
is given by
\begin{eqnarray}
\sigma_{\rm SD} =\left(\dfrac{g_2\,\sin^2\theta_D}
{2\,\cos^2\theta_W}\right)^2 \dfrac{\mu^2_{\phi_4\,N}}{\pi\,M^4_Z}
{\rm v^2_{lab}}\,F_q^2\,,
\label{sigmaSD}
\end{eqnarray}
the quantity $F_q$ is given by
{\footnotesize
\begin{eqnarray}
F_q= \dfrac{b_q(\frac{1}{2})\left\{\Delta^{p}_u\,\langle S_p \rangle
+ \Delta^{n}_u\,\langle S_n\rangle\right\}
+ b_q(-\frac{1}{2})\left\{\Delta^{p}_d\,\langle S_p \rangle
+ \Delta^{n}_d\,\langle S_n\rangle\right\}
+b_q(-\frac{1}{2})\left\{\Delta^{p}_s\,\langle S_p \rangle
+ \Delta^{n}_s\,\langle S_n\rangle\right\}
}{\langle S_p\rangle + \langle S_n\rangle} \nn
\end{eqnarray}}
where, $\Delta^{p(n)}_q$ represents the spin content of quark $q$ in
proton(neutron). The recent values of $\Delta$'s are $\Delta^p_u=\Delta^n_d=0.84$,
$\Delta^p_d=\Delta^n_u=-0.43$ and $\Delta^p_s=\Delta^n_s=-0.09$ \cite{Cheng:2012qr}. 
The contributions of proton and neutron to nuclear spin are denoted
by $\langle S_p\rangle$ and $\langle S_n\rangle$ respectively.
For $^{129}$Xe isotope $\langle S_p\rangle=0.010$ and
$\langle S_n\rangle=0.329$ \cite{XENON:2019rxp}. The function
$b_q(T_3) = \dfrac{g_2}{2\,\cos\theta_W}\left(\cos\theta_{\mu\tau}+
\epsilon \sin\theta_W \sin\theta_{\mu\tau}\right)T_3$ is the axial
vectorial part of $\bar{q}qZ$ coupling and ${\rm v_{lab}}\simeq10^{-3}$ is
the local velocity of dark matter with respect to the laboratory frame.
Moreover, in the present work since the dark mass range is in sub-GeV range,
the elastic scatterings with electron also transfer energy efficiently
\cite{1108.5383}. As shown in \cite{Essig:2015cda},
MeV scale DM can excite electron from valence band to conduction
band and give rise to ionisation excitation. Therefore,
our dark matter can also be detected through elastic scatterings with electron. 
We have calculated $\phi_4-e$ elastic scattering for the range of parameters
we needed for the phenomenology and we have found that it is well
below the current bound. The cross section for $\phi_4-e$ elastic scattering
has the following form,
\begin{eqnarray}
\sigma_{elec}&=&\left(\dfrac{g_2\,\sin^2\theta_D}{2\,\cos\theta_W}\right)^2
\dfrac{\mu^2_{\phi_4\,e}}{\pi\,M^4_Z}\left\{a^2_e(-\frac{1}{2},-1)+
b^2_e(-\frac{1}{2}){\rm v^2_{lab}}\right\}\,,
\end{eqnarray}
where the functions $a_e(T_3, Q^e_{em})$ and $b_e(T_3)$ are
identical with $a_q$ and $b_q$ for the quarks. We can easily
notice that the axial vector part (proportional to $b_e$) of
$\bar{e}eZ$ interaction gives a velocity suppressed contribution
to $\sigma_{elec}$ as in the case for spin dependent scattering
with nuclei (Eq.\,\,(\ref{sigmaSD})).  
\item {\bf Kinetic equilibrium:}
In this work, although the freeze-out occurs after the chemical
imbalance for $3\rightarrow2$ scatterings is created within the dark sector,
the kinetic equilibrium between the two sectors continues and it is
primarily possible through elastic scatterings of dark matter with
$\nu_{\mu}$ and $\nu_{\tau}$ where light gauge boson $\zmt$ plays
an important role.
\begin{figure}[h!]
\includegraphics[height=7.5cm,width=10cm,angle=0]{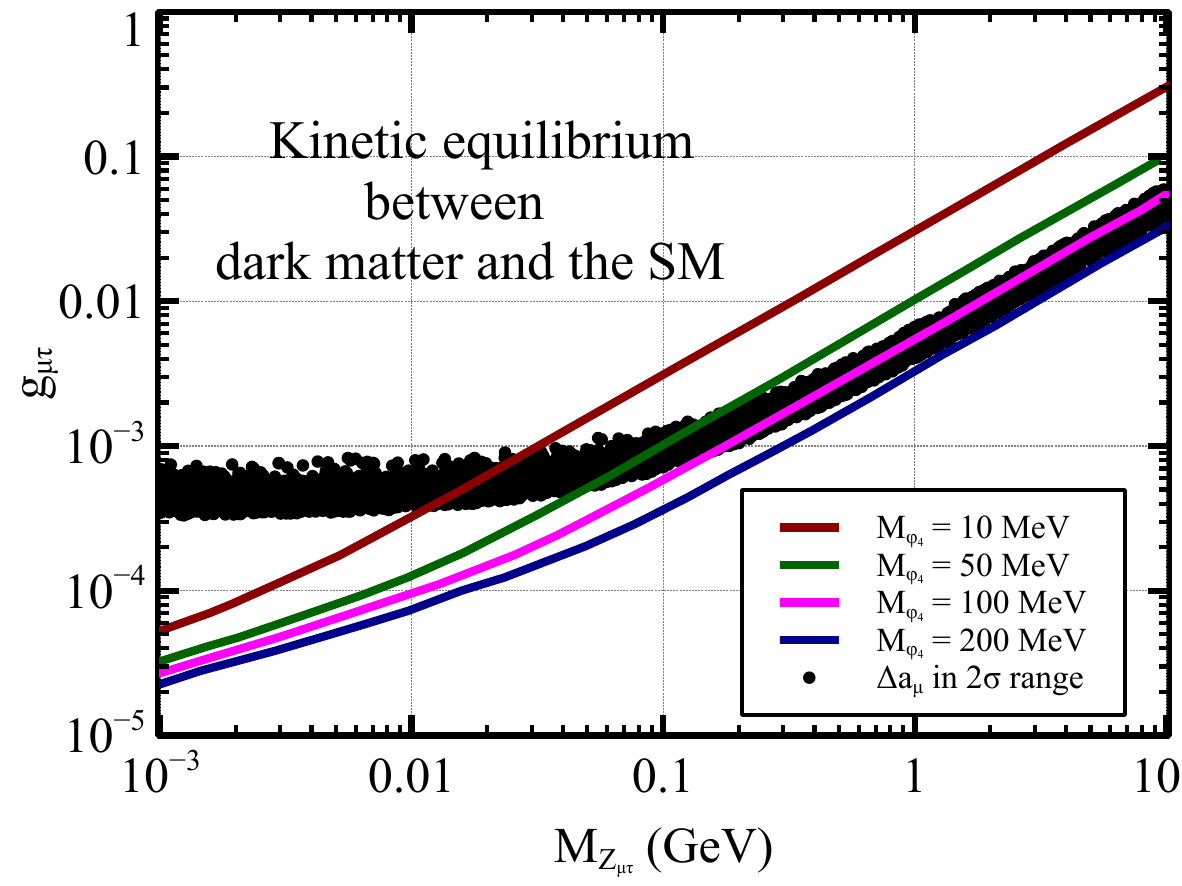}
\caption{The range of $\gmt$ and $M_{\zmt}$ for which dark matter
maintains kinetic equilibrium with the SM bath through elastic
scatterings with $\nu_{\mu}$ and $\nu_{\tau}$ respectively.}
\label{Fig:Kinetic_decup}
\end{figure}
In Fig\,\,\ref{Fig:Kinetic_decup}, we show the regions in
$g_{\mu\tau}-M_{\zmt}$ plane for four different values of
$M_{\phi_4}$, where the kinetic equilibrium is maintained
between the dark and the visible sectors. The allowed
regions are the upper portions of the solid lines. 
For that we have
used the condition $\dfrac{1}{n_{\rm scatt}}
\dfrac{\Gamma_{\rm el}}{\mathcal{H}}\bigg|_{T\simeq M_{\phi_4}/20}>1$
\cite{Gorbunov:2011zz, Gondolo:2012vh, Biswas:2021kio},
where $\Gamma_{\rm el}=\sum_{\alpha=\mu,\,\tau}n^{\rm eq}_{\nu_\alpha}
\langle\sigma_{\phi_4\nu_\alpha\rightarrow\phi_4\nu_\alpha}{\rm v}\rangle$ 
is the total scattering rate per dark matter and $n_{\rm scatt}=M_{\phi_4}/T$ is
the number of scatterings needed to transfer energy $\sim T$ between the
SM bath and $\phi_4$. We have compared the effective interaction rate 
$\Gamma_{\rm el}/n_{\rm scatt}$ with the Hubble parameter ($\mathcal{H}$)
around the freeze-out era of $\phi_4$ which is $\sim M_{\phi_4}/20$. For completeness,
in the same $g_{\mu\tau}-M_{\zmt}$ plane, we have shown the allowed parameter space 
satisfying $(g-2)_{\mu}$ in $2\sigma$ range by the black dots.
\item {\bf Relic density bound:} The abundance of dark matter has been
determined quite precisely by satellite borne CMB experiments particularly
the Planck experiment. The current value of dark
matter relic density is \cite{Planck:2018vyg}
\begin{eqnarray}
\Omega_{\rm DM} h^{2} = 0.120 \pm 0.001
\end{eqnarray}
\item {\bf Invisible Higgs Decay:}
We are considering sub-GeV dark matter in the present case. As a result, the SM like
Higgs boson $h_1$ can decay into a pair of $\phi_4$ and $\phi^\dagger_4$. Moreover,
$h_1$ can decay into a pair of light gauge boson $\zmt$ also. These additional
decay modes contribute to the invisible decay of $h_1$. The LHC has placed
an upper bound on the branching ratio of total invisible decay of the Higgs boson,
which is \cite{CMS:2018yfx},
\begin{eqnarray}
{\rm Br}(h_{1} \rightarrow {\rm invisible\,\,channels}) < 0.19\,.
\end{eqnarray}
However, the bound is easily satisfied in our model as we have considered
feeble scalar portal couplings while the other decay mode $h_1\rightarrow\zmt\zmt$
is suppressed by $\sin^4\theta_{\mu\tau}$. 
\end{itemize} 
\subsection{Numerical Results}
\label{Sec:Results}
\begin{figure}[h!]
\centering
\subfigure[$Y_{\rm DM}$ vs $x$ plot for two different values of $\xi$ ]
	{
      \includegraphics[height=3.5cm,width=7cm]
      {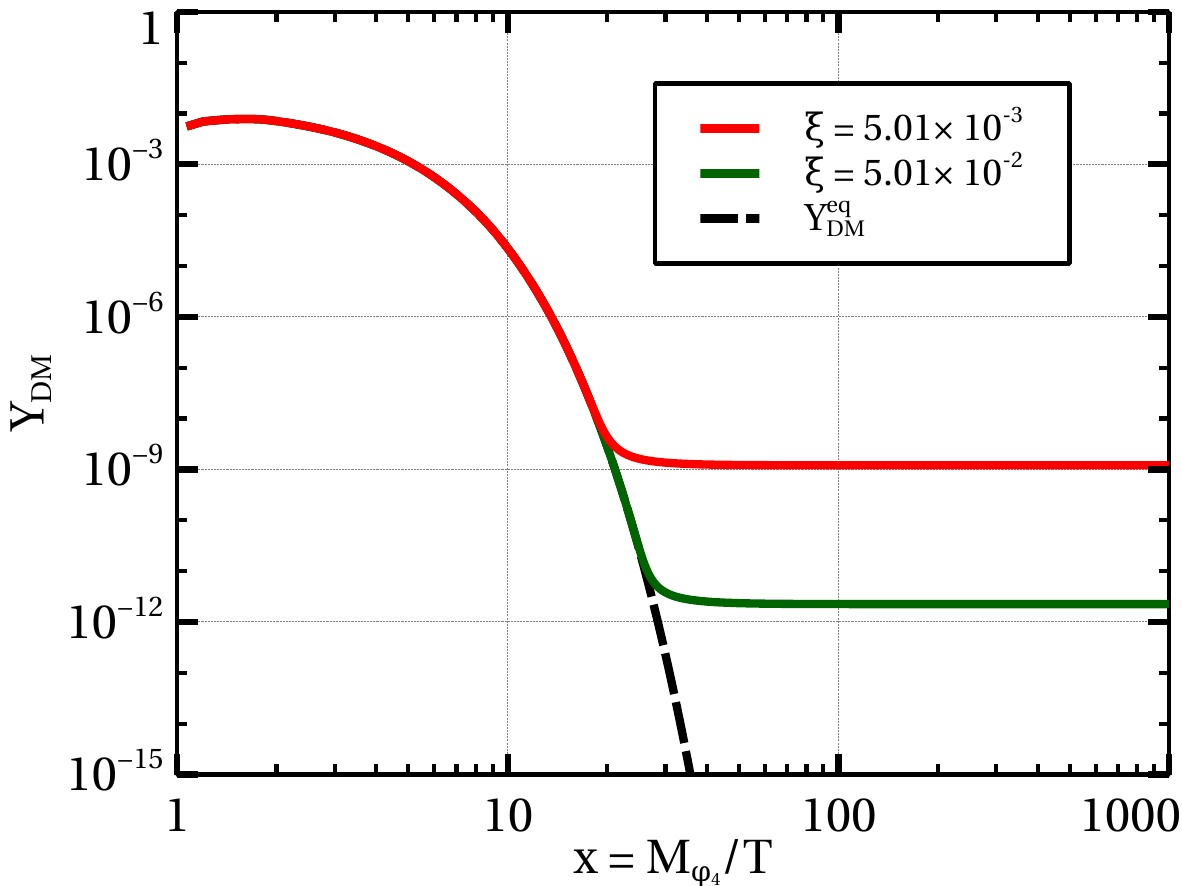}
     }
\subfigure[$Y_{\rm DM}$ vs $x$ plot for two different values of $\theta_D$]
	{
      \includegraphics[height=3.5cm,width=7cm]
      {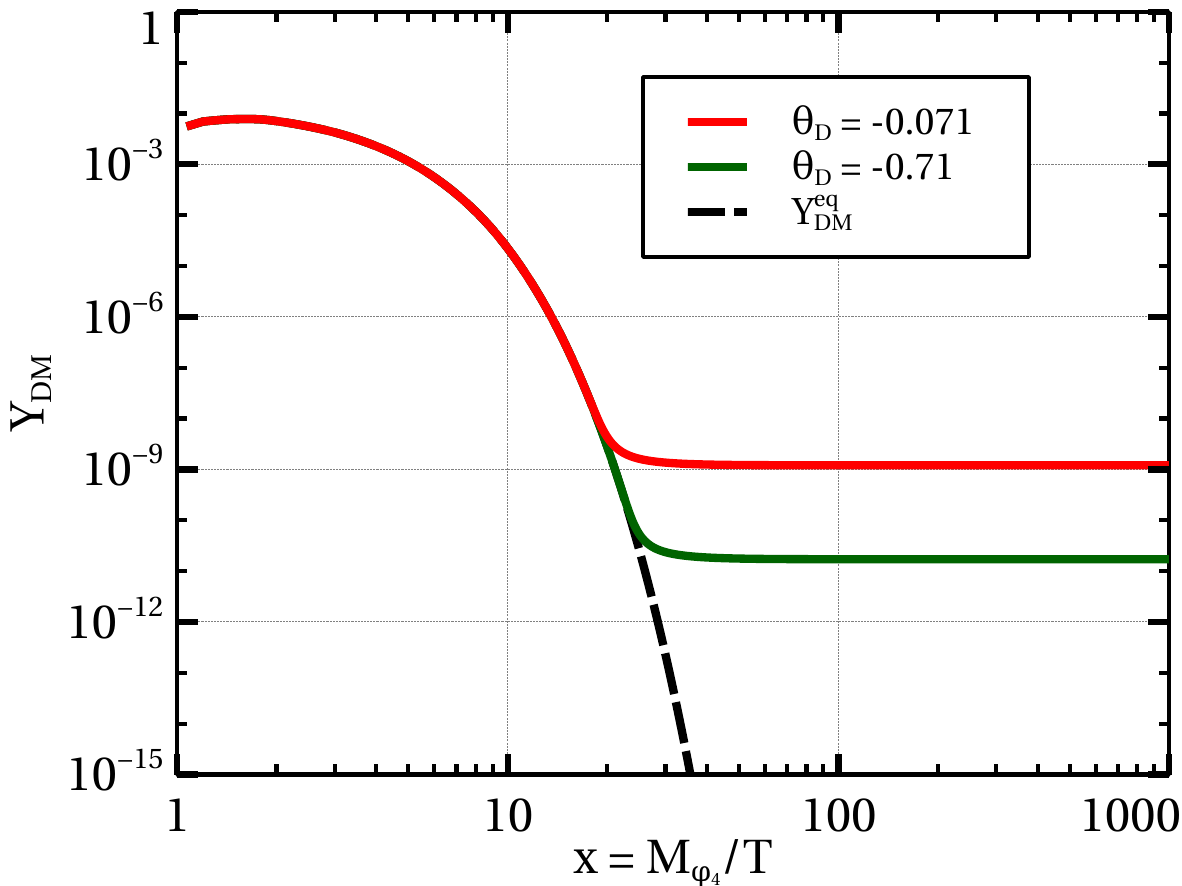}
    }
    \subfigure[$Y_{\rm DM}$ vs $x$ plot for two different $M_{\phi_2}$]
	{
      \includegraphics[height=3.5cm,width=7cm]
      {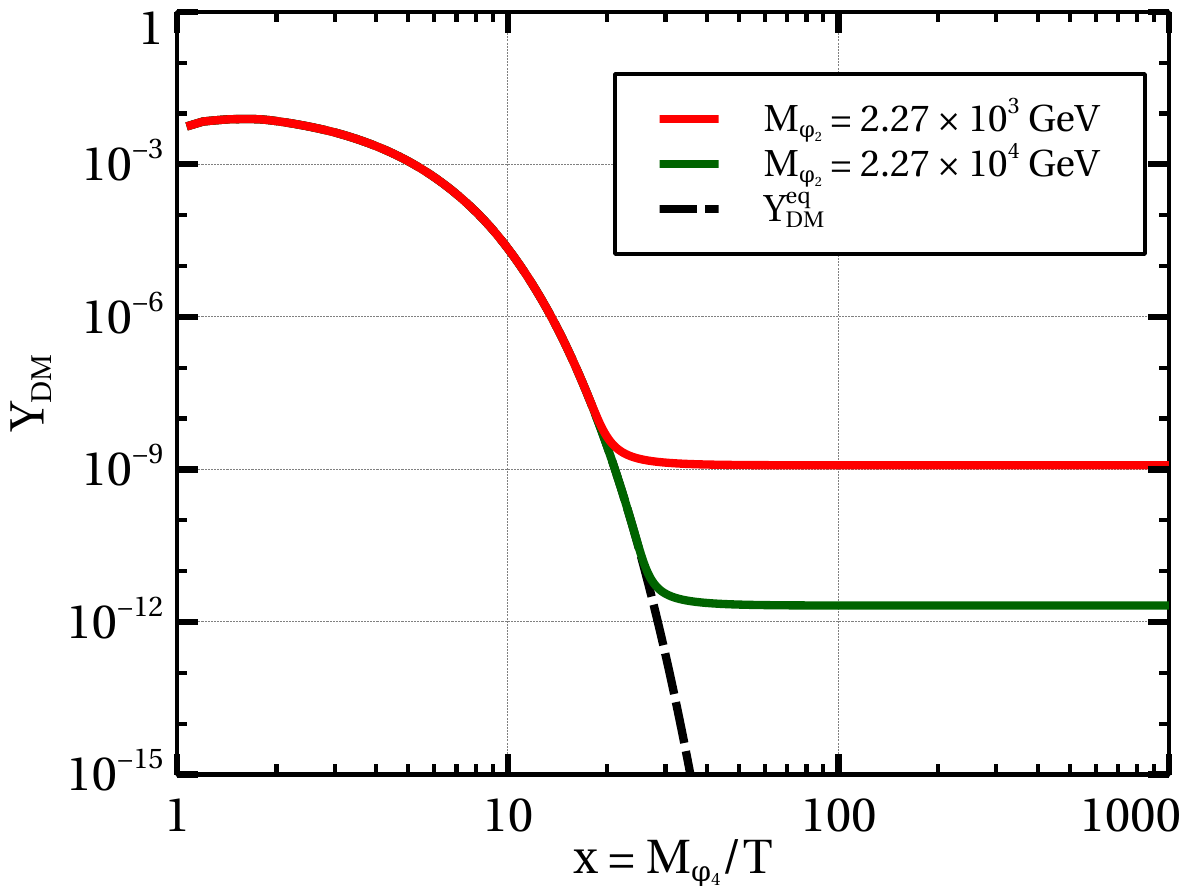}
    }
    \subfigure[$Y_{\rm DM}$ vs $x$ plot for two values of $M_{\phi_4}$]
	{
      \includegraphics[height=3.5cm,width=7cm]
      {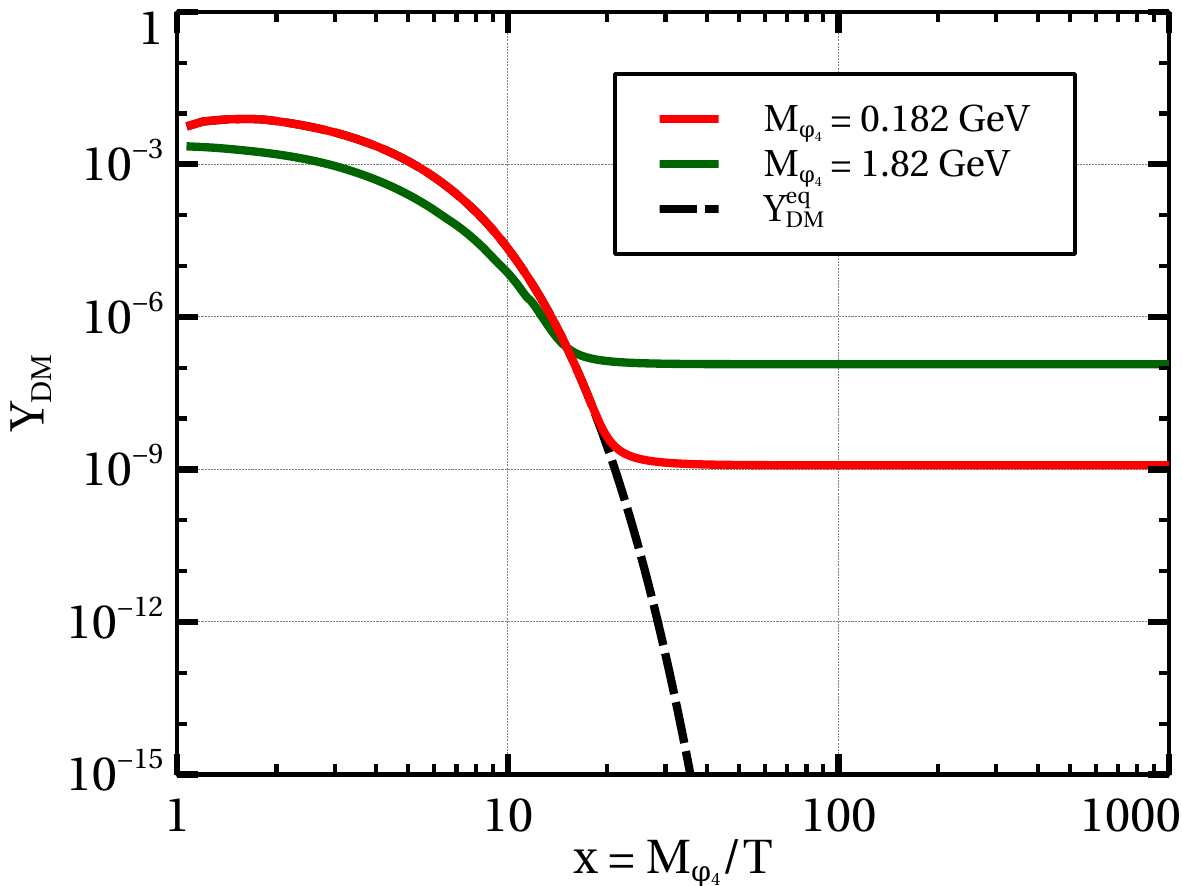}
     }
    \subfigure[$Y_{\rm DM}$ vs $x$ plot for two values of $M_{\zmt}$]
	{
      \includegraphics[height=3.5cm,width=7cm]
      {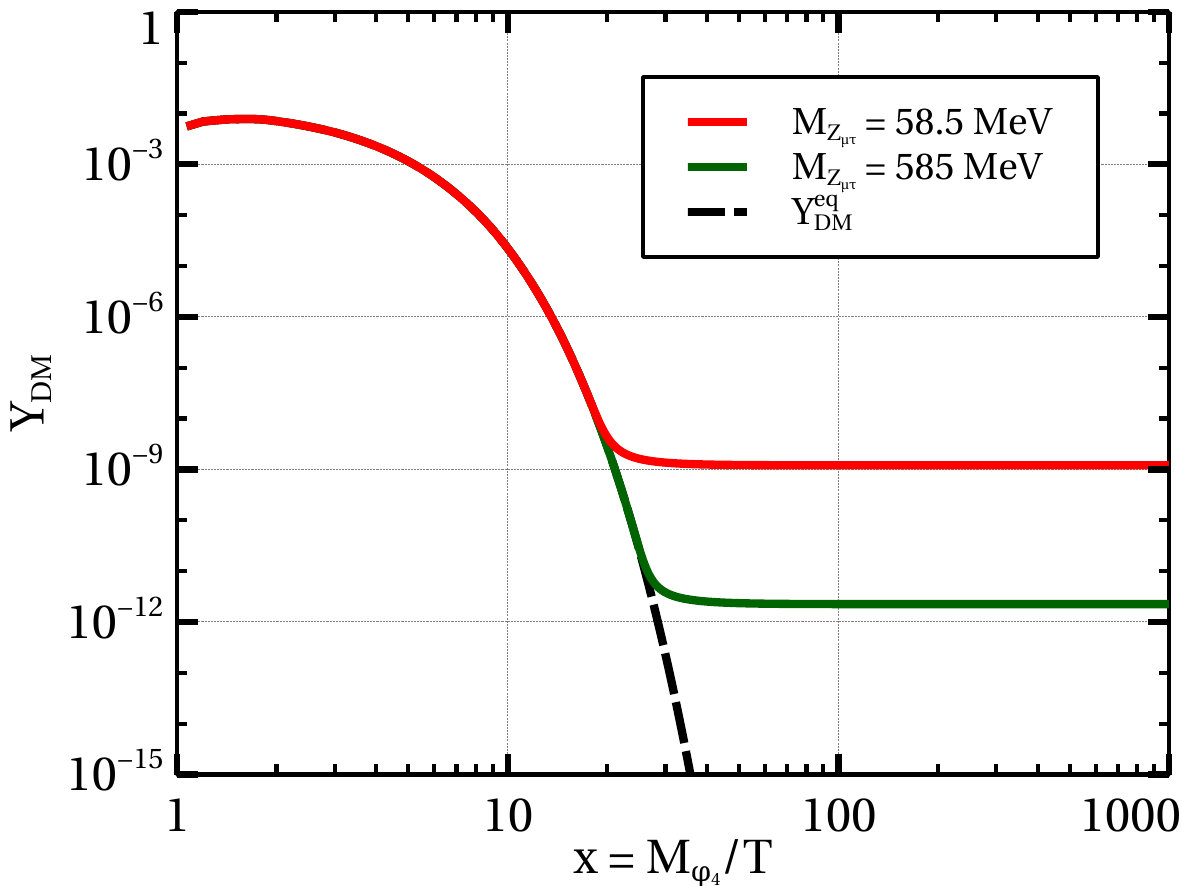}
    }
    \subfigure[$Y_{\rm DM}$ vs $x$ plot for two different values of $\gmt$]
	{
      \includegraphics[height=3.5cm,width=7cm]
      {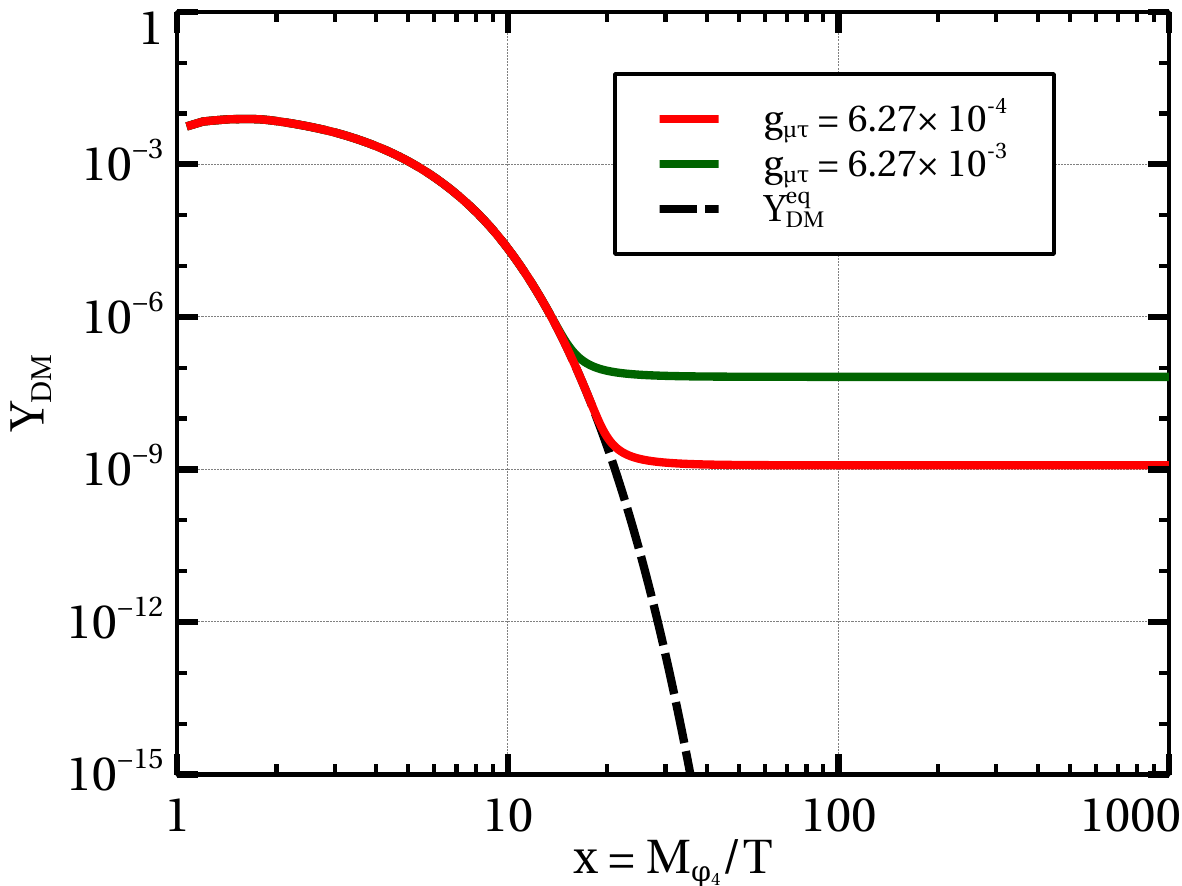}
    }
    \subfigure[Comparison on the effect of $3\rightarrow2$ and $2\rightarrow2$ scatterings in
    dark matter freeze-out]
	{
      \includegraphics[height=3.5cm,width=7cm]
      {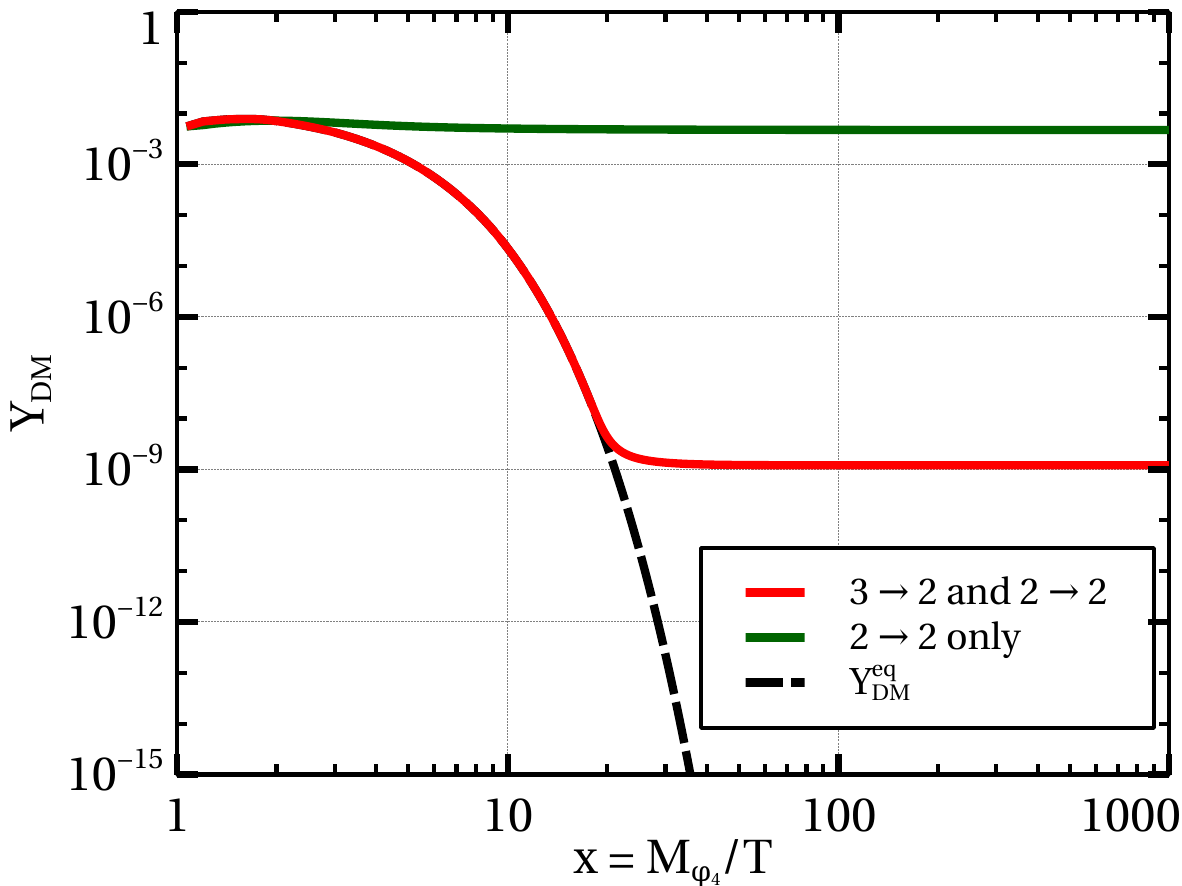}
    }
\caption{Numerical results: Evolution of $Y_{\rm DM}$ with $x$ for various model
parameters like $\xi$, $\theta_D$, $M_{\phi_2}$, $M_{\phi_4}$, $M_{\zmt}$ and $\gmt$.}
\label{Fig:Y-vs-x}
\end{figure}
In this section we have shown our results which we have obtained by solving
the Boltzmann equation (Eq\,\,(\ref{Boltz})) numerically. The solution
of Eq.\,\,(\ref{Boltz}) is shown in Fig.\,\,\ref{Fig:Y-vs-x} where we
have demonstrated the evolution of $Y_{\rm DM}$ with $x$ for different
model parameters. In all these plots of Fig.\,\,\ref{Fig:Y-vs-x}, the
red solid line is the solution of the Boltzmann equation for the following
set of model parameters $\xi=0.501\times10^{-2}$,
$\theta_D=-0.071$, $M_{\phi_2}=2.27$ TeV, $M_{\phi_4}=0.182$ GeV, $\gmt=6.27\times 10^{-4}$
and $M_{\zmt}=58.5$ MeV, which reproduces the correct relic density.
In plot (a) of Fig.\,\ref{Fig:Y-vs-x}, we have shown how the
era of freeze-out and the final abundance both change when we increase
$\xi$ from $0.501\times10^{-2}$ to 0.0501 and it is indicated by the
green solid line. As the $3\rightarrow2$ scattering cross section increases
with $\xi$ this results in a delayed freeze-out with a reduced final
abundance. In plot (b), we have demonstrated the effect of increasing $\theta_D$
on $Y_{\rm DM}$. We have found that the change in mixing angle $\theta_D$
has a similar effect on $Y_{\rm DM}$ as it is shown in plot (a) for the
parameter $\xi$. However, in this case $Y_{\rm DM}$ does not decrease as much
as it is for the parameter $\xi$ for one order increase in magnitude of $\theta_D$.

In plot (c) we have shown the impact of $M_{\phi_2}$ on $Y_{\rm DM}$.
Here we have increased the value of $M_{\phi_2}$ from $2.27$ TeV to
22.7 TeV and corresponding $Y_{\rm DM}$ has been indicated by the green
solid line. Any increase in $M_{\phi_2}$ enhances the magnitude of cubic
coupling\footnote{Since $\theta_D$ is -ve, which represents
a rotation in the reverse direction than in Eq.\,\,(\ref{thetaD}), 
the coefficient $\mu$ is a +ve number.} $\mu$
as $\mu = -\dfrac{\left(M^2_{\phi_2}-M^2_{\phi_4}\right)
\sin 2\theta_D}{\sqrt{2}v}$ with $M_{\phi_2}>>M_{\phi_4}$,
which eventually increases the trilinear interaction
among $\phi_1$, $\phi_1$ and $h_1(h_3)$ and hence the
scattering cross section $\sigma^{\rm tot}_{3\rightarrow2}$. The effect
of the mass of dark matter on $Y_{\rm DM}$ has been demonstrated in
plot (d) where we have considered $M_{\phi_4}=0.182$ GeV (red solid line)
and 1.82 GeV (green solid line) respectively.
Plots (e) and (f) show the dependence of $L_{\mu}-L_{\tau}$
gauge coupling $\gmt$ and gauge boson mass $M_{\zmt}$ on $Y_{\rm DM}$.
It is seen from both the plots that any increase in $\gmt$ and $M_{\zmt}$
has opposite effect on $Y_{\rm  DM}$ respectively and it is determined
by the corresponding change in $\vmt$ that appears in the couplings. Finally,
in plot (g) we have shown the effect of $2\rightarrow2$ and $3\rightarrow2$
scatterings on the freeze-out of dark matter. Here the red solid line
represents a situation when both $3\rightarrow2$ as well as $2\rightarrow2$
scatterings are present and the dark matter freezes-out around $x\simeq20$.
However, if we switch off the $3\rightarrow2$ interactions, the freeze-out of
dark matter occurs a lot earlier ($x\simeq 2$). It is due to the reason
that the cross sections of $2\rightarrow2$ scatterings are not as large as
that of the $3\rightarrow2$ scatterings which are predominantly responsible
for the number changing processes.

\begin{figure}[h!]
\centering
\subfigure[Variation of $\Omega_{\phi_4}h^2$ with $\xi$ for
three different values of $M_{\phi_4}$.]
	{
      \includegraphics[height=6cm,width=7.65cm]
      {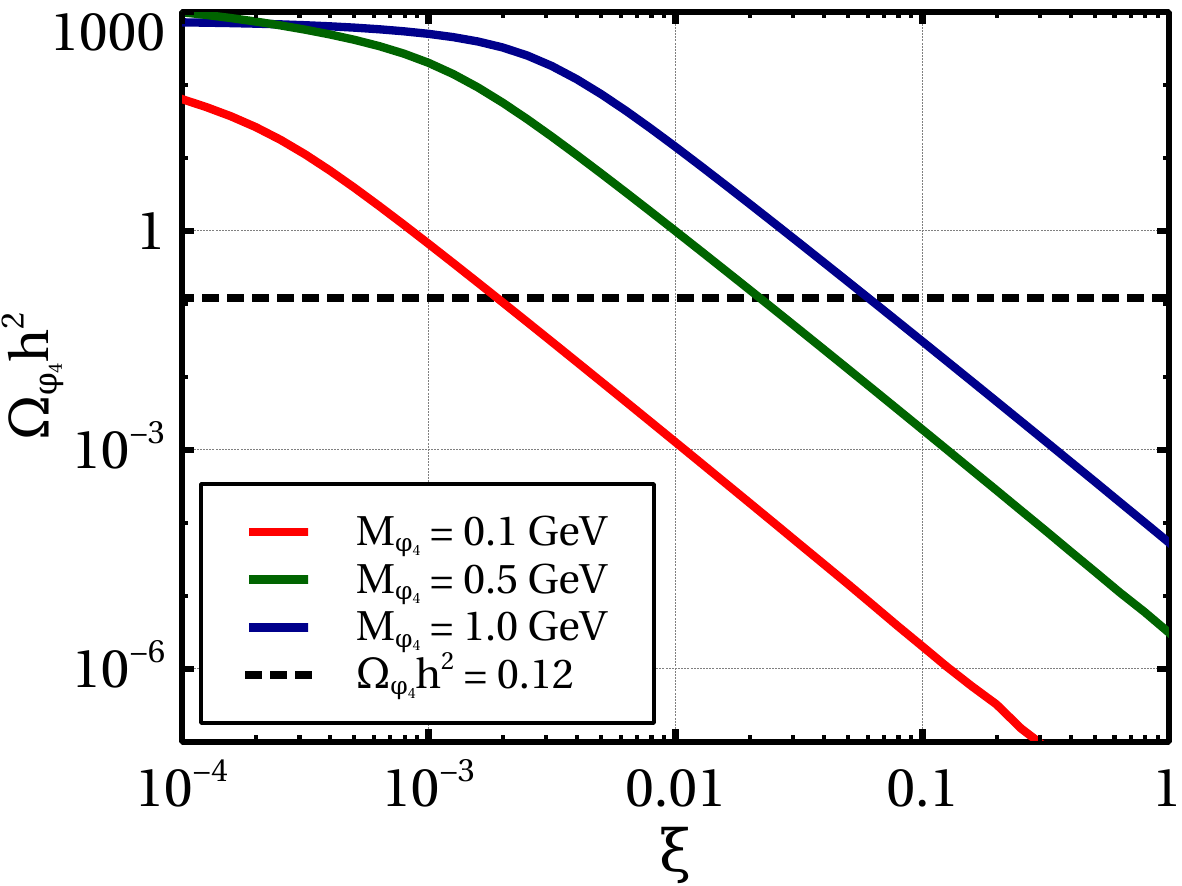}
     } 
\subfigure[Variation of $\Omega_{\phi_4}h^2$ with $M_{\phi_4}$ for
four different values of $\theta_D$.]
	{
      \includegraphics[height=6cm,width=7.65cm]
      {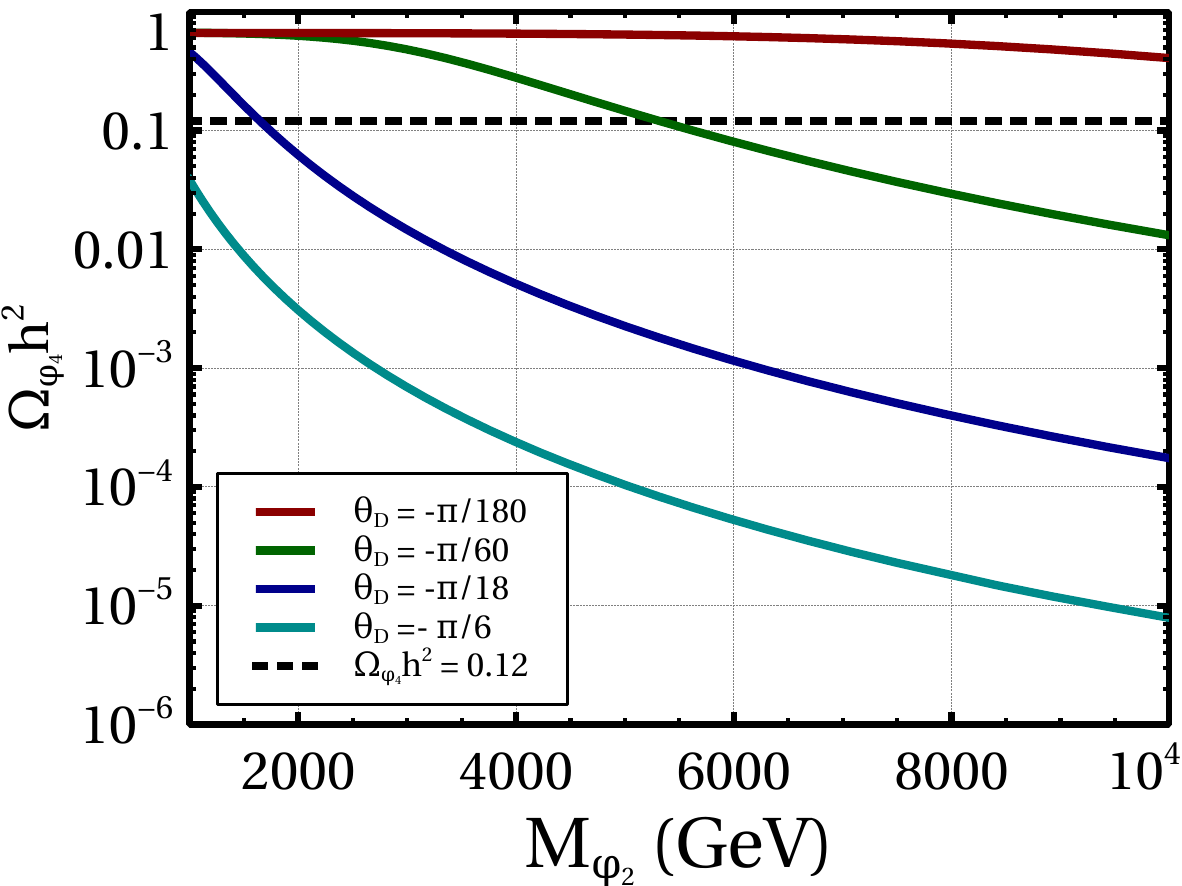} 
    }
\subfigure[$\Omega_{\phi_4}h^2$ vs $\theta_D$ for three different
values $M_{\phi_4}$.]
	{
      \includegraphics[height=6cm,width=7.65cm]
      {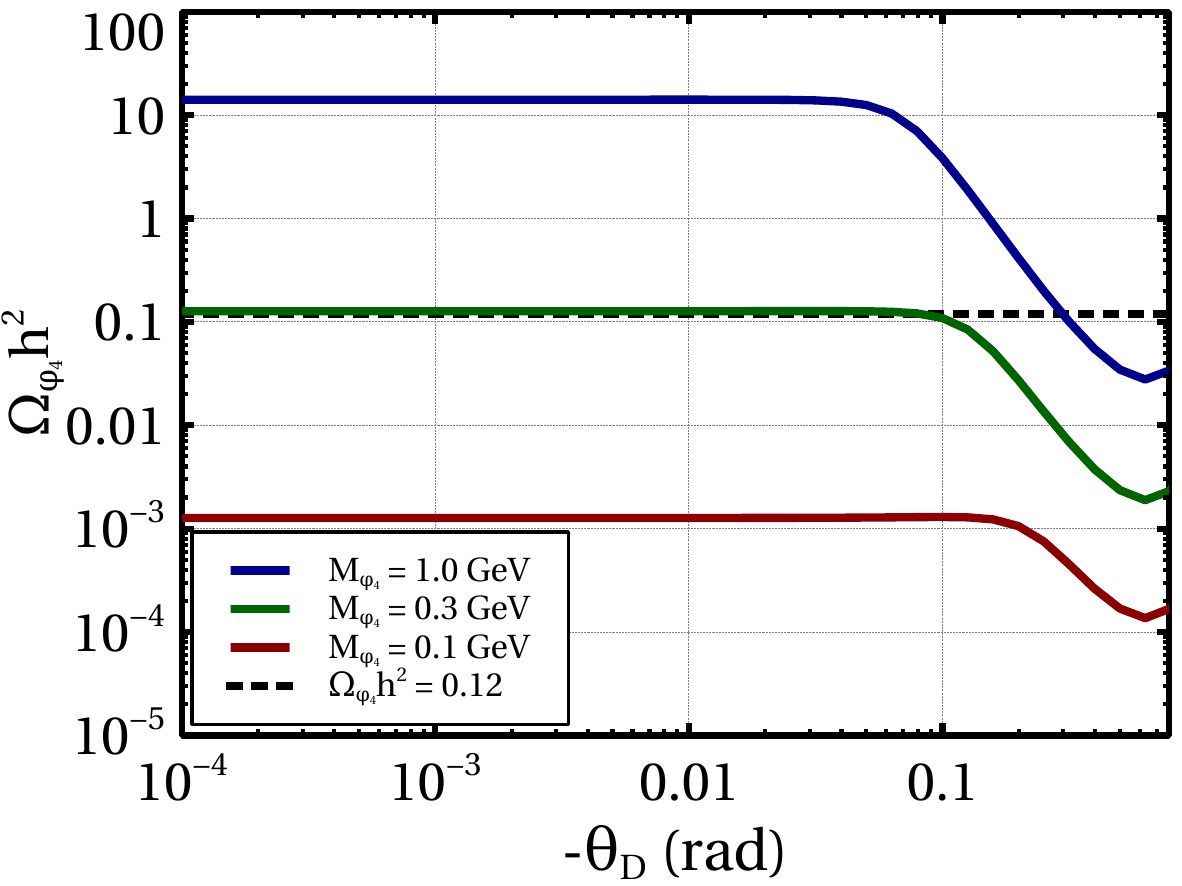} 
    }
\caption{Variation of relic density with relevant model parameters}
\label{Fig:omega_LP}
\end{figure}

The variation of relic density with three most relevant parameters
$\xi$, $M_{\phi_2}$ and $\theta_D$ is depicted in Fig.\,\,\ref{Fig:omega_LP}.
In plot (a), dependence of $\Omega_{\phi_4} h^2$ with $\xi$ has been shown
for three different values of $M_{\phi_4}$. Analogous to the previous plot
in Fig.\,\,\ref{Fig:Y-vs-x}(a), here also we have noticed a similar behaviour
except for $\xi<10^{-3}$, as the relic density decreases sharply with the
increase of quartic coupling $\xi$ due to enhancement of $\sigma^{\rm tot}_{3\rightarrow2}$. 
In plot (b), the effect of $M_{\phi_2}$ on $\Omega_{\phi_4}h^2$ is shown
for different values of $\theta_D$. We can see that for low value of mixing
angle ($|\theta_D|=\pi/180$ rad), the relic density is almost insensitive 
to the mass of $\phi_2$. However, as we increase the magnitude of $\theta_D$,
$\Omega_{\phi_4}h^2$ decreases with $M_{\phi_2}$ replicating the situation
shown in Fig.\,\,\ref{Fig:Y-vs-x}(c). The last figure in plot (c)
demonstrates $\Omega_{\phi_4} h^2$ as a function of $\theta_D$. Here, three
lines are for three different values of $M_{\phi_4}$ and the nature of
all three lines are exactly identical to each other, i.e. the relic density
is independent of the dark sector mixing angle for $|\theta_D|\lesssim 0.1$ rad
and thereafter it starts decreasing with the increase of magnitude of $\theta_D$.
The difference in magnitude of $\Omega_{\phi_4}h^2$ in these three
lines for different $M_{\phi_4}$ originates from two factors. The relic density
is proportional to both $M_{\phi_4}$ and $Y_{\rm DM}$ where the latter also
gets enhanced with $M_{\phi_4}$ as shown in Fig.\,\,\ref{Fig:Y-vs-x}(d).
\begin{figure}[h!]
\centering
\includegraphics[height=6cm,width=7.65cm]
      {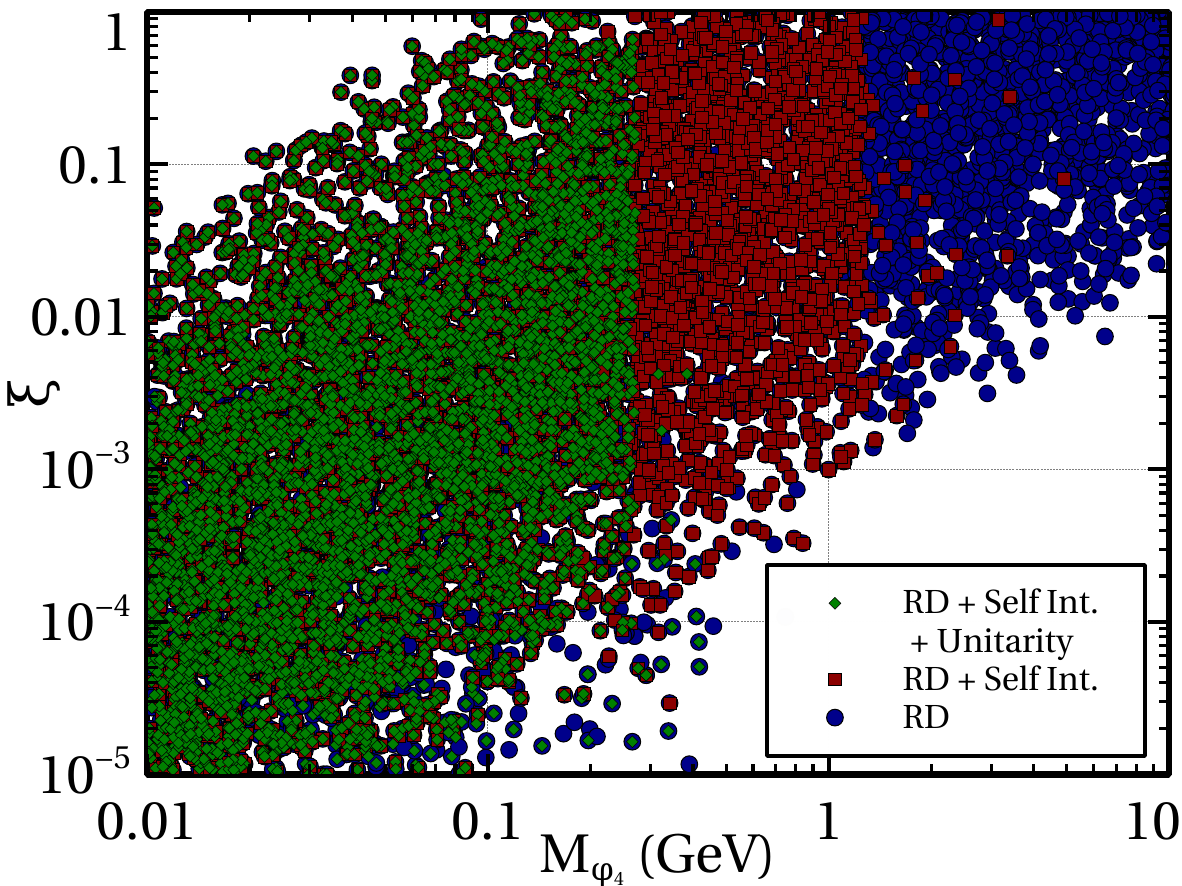}
      \includegraphics[height=6cm,width=7.65cm]
      {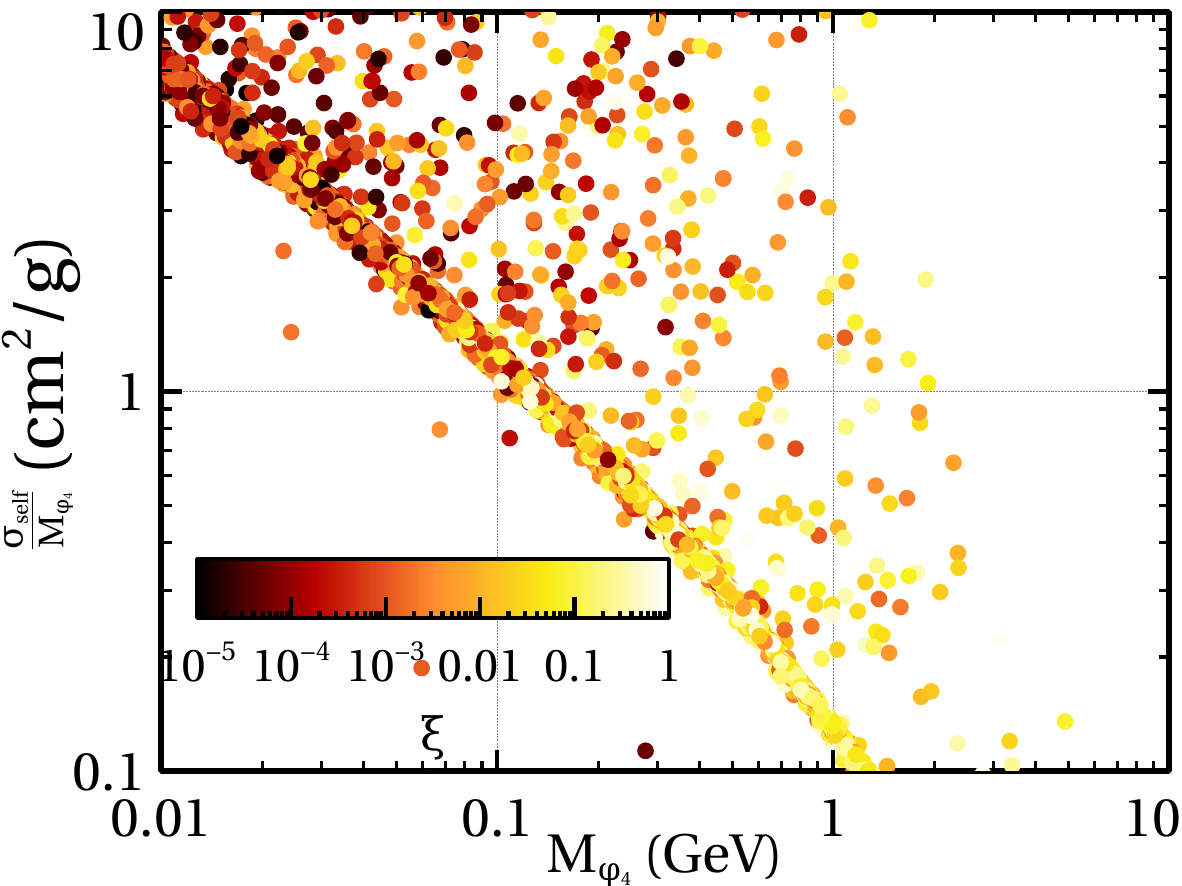} 
\caption{Left panel: Allowed parameter space is $\xi-M_{\phi_4}$ plane for
different constraints. Right pane: Variation of $\sigma_{\rm self}/M_{\phi_4}$
with $M_{\phi_4}$.}
\label{Fig:para_scan}
\end{figure}

In Fig.\,\,\ref{Fig:para_scan}, we show our allow parameter space in
$\xi-M_{\phi_4}$ plane. In order to obtain this we have scanned over
the parameters in the following range
\begin{eqnarray}
\begin{array}{cccccc}
10^{-5} &\leq & \xi & \leq & 1.0\,\,,\\ 
10^{-3}\,\,{\rm rad} &\leq &|\theta_{D}|& \leq & 0.1\,\,{\rm rad}\,\, ,\\
10^{-4} &\leq & \gmt & \leq & 10^{-2}\,\,,\\ 
10^{-3}\,\,{\rm GeV} &\leq & {M_{\zmt}}& \leq & 1.0\,\,{\rm  GeV}\,\,,\\
10^{3}\,\,{\rm GeV} &\leq & {M_{\phi_4}}& \leq & 10^{4}\,\,{\rm  GeV}\,\,,\\
10^{-2}\,\,{\rm GeV} &\leq & {M_{\phi_2}}& \leq & 10.0\,\,{\rm  GeV}\,\,,
\label{para-ranges}
\end{array}
\end{eqnarray}
and have imposed necessary constraints one by one. The result is shown
in the left panel of Fig.\,\,\ref{Fig:para_scan}. In this plot, the blue
dots describe a region in $\xi-M_{\phi_4}$ plane that reproduces the correct
dark matter relic density in $3\sigma$ range as determined by the
Planck experiment. On top of that, we have imposed bound from dark matter
self-interaction  $0.1\,\,{\rm cm^2/g}\leq\frac{\sigma_{\rm self}}{M_{\phi_4}}
\leq 10\,\,{\rm cm^{2}}/{\rm g}$. The resultant parameter space is indicated
by the red square shaped points. Finally, we have introduced another
constraint coming from the unitarity limit of scattering amplitudes as mentioned
in Eq.\,\,(\ref{unitarity}). The parameter space satisfying all three constraints
is shown by the green diamond shaped points. We can notice that in order to satisfy
these three constraints we need $M_{\phi_4}\lesssim 200$ MeV while the corresponding
quartic coupling is restricted to be $\lesssim 1$. The similar result has also been
presented in a different manner in the right panel of Fig.\,\,\ref{Fig:para_scan}. 
Here we have shown the variation of $\sigma_{\rm self}/M_{\phi_4}$ with
$M_{\phi_4}$ and the corresponding value of the parameter $\xi$
has been indicated by the colour bar. The only constraint applied
in this plot is that each and every point in $\sigma_{\rm self}/M_{\phi_4}-M_{\phi_4}$
plane satisfies the relic density bound i.e. $0.117\leq \Omega_{\phi_4}h^2\leq0.123$.
\begin{figure}[h!]
\centering
	{
      \includegraphics[height=7.5cm,width=10cm]
      {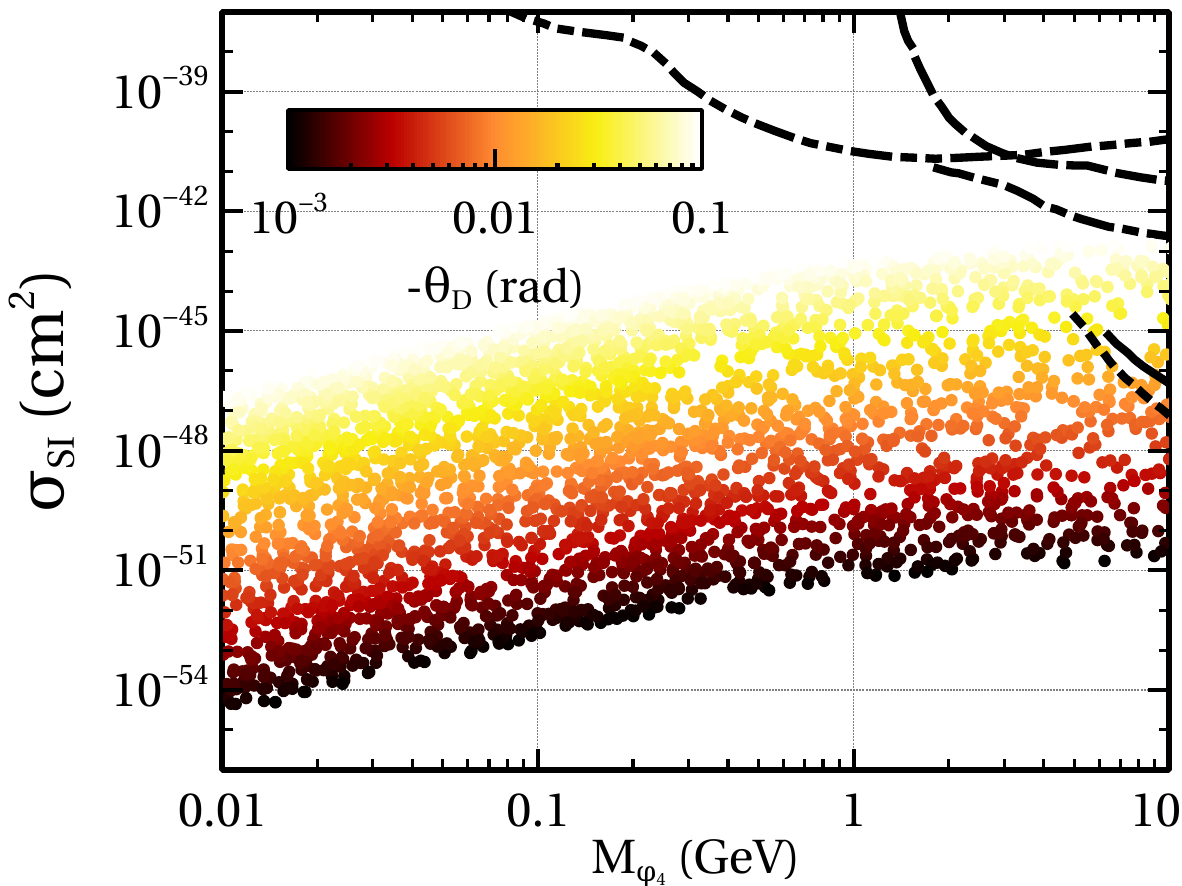}
     } 
\caption{Spin independent scattering cross sections with exclusion
limits on $\sigma_{\rm SI}$ from various ongoing as well as
future experiments.}
\label{Fig:DD}
\end{figure}

The spin independent and spin dependent elastic scattering cross sections are
calculated using Eqs.\,\,(\ref{sigmaSI},\,\ref{sigmaSD}) for the
parameter range given in Eq.\,\,(\ref{para-ranges}). We have found that
the spin dependent cross section is several orders of magnitude below the present
bound from XENON1T \cite{XENON:2019rxp}. Moreover, we also notice that
$\sigma_{\rm SD}$ for a particular value of $M_{\phi_4}$
is almost $10^{-6}$ times smaller than the corresponding $\sigma_{\rm SI}$ and
it is primarily due to the reason that $\sigma_{\rm SD}$ is suppressed by ${\rm v^2_{\rm lab}}$
(Eq.\,\,(\ref{sigmaSD})) where ${\rm v}_{\rm lab} \simeq 10^{-3}$ is the
local velocity of dark matter particles with respect to the laboratory frame.
Therefore, in Fig.\,\ref{Fig:DD}, we have shown the spin independent
scattering cross section only. In this figure, we demonstrate
$\sigma_{\rm SI}$ as a function of $M_{\phi_4}$ and the colour bar
provides the value of dark mixing angle $\theta_D$.
Moreover, we have shown existing and future bounds from various
direct detection experiments for comparison.\,\,In
the low mass region where $M_{\phi_4}\leq 1$ GeV,
we mainly have exclusion limit on $\sigma_{\rm SI}$ from NEWS-G \cite{Durnford:2021mzg}
and it has been indicated by the black dashed dot dot line. The current bounds
from two other low mass dark matter experiments
namely, CDMSlite \cite{SuperCDMS:2017nns} and DarkSide-50 \cite{DarkSide:2018bpj}
are shown by the dashed and the dashed dot lines respectively. The upper limit
on $\sigma_{\rm SI}$ from ``GeV-TeV'' scale experiment
like XENON1T \cite{XENON:2018voc}, which
has a very small overlap with our considered range of dark matter mass, has also
been depicted by the black solid line. Finally, the future
prediction from DARWIN \cite{DARWIN:2016hyl} is shown by the
dotted line.
\section{Collider Signature}
\label{Sec:collider}
\begin{figure}[h!]
\centering
\includegraphics[angle=0,height=4.5cm,width=15.0cm]{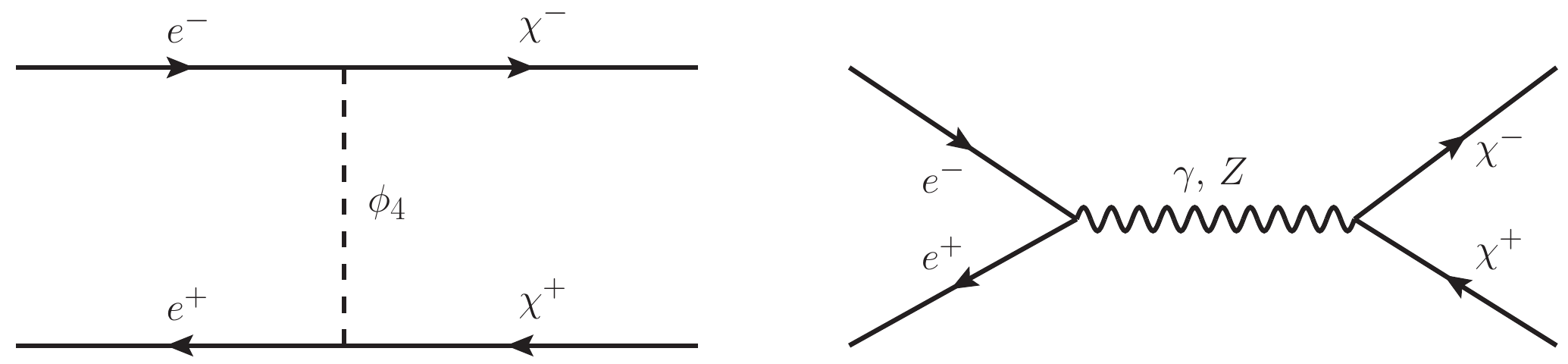}
\caption{Charged particle production at the $e^{+} e^{-}$ collider with 
dark matter $\phi_4$ as one of the intermediate states.}      
\label{collider-process}
\end{figure}
\begin{figure}[h!]
\centering
\includegraphics[angle=0,height=7cm,width=10.0cm]{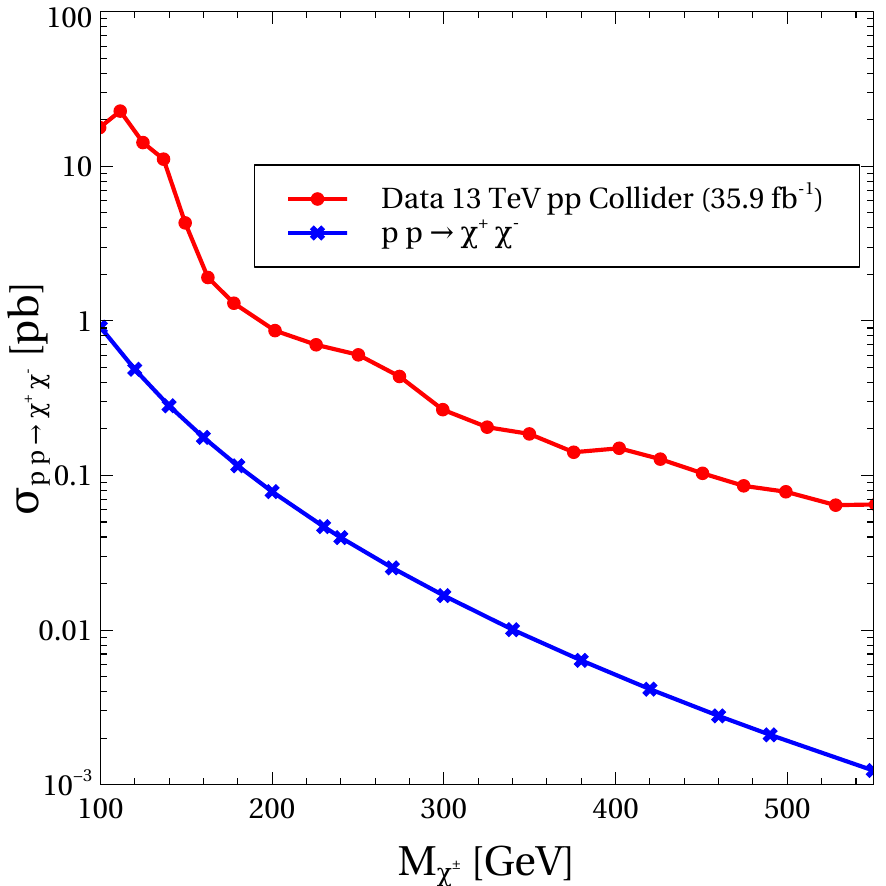}
\caption{Production of $\chi^{+} \chi^{-}$ at the 13 TeV pp collider. Red 
line corresponds the bound for the production of singly charged fermion when we
look for the $l^{+} l^{-} +$ MET signal at the final state. Any model
which predicts the production rate above the red line is already ruled out.}      
\label{pp-collider-CS}
\end{figure}

In this work, we have studied the pair production of charged particles 
($\chi^{+} \chi^{-}$) defined as $\chi = \chi_{L} \oplus \chi_R$. 
The produced $\chi^{\pm}$ subsequently decays into lepton and SIMP
dark matter {\it i.e.} $e^{+} e^{-} \rightarrow \chi^{+} (e^{+} \phi_{4})
\chi^{-} (e^{-} \phi_{4}) \rightarrow e^{+} e^{-} \cancel{E}_{T}$. The Feynman 
diagrams contributing in the signal are mediated
by $\gamma$, $Z$ and $\phi_4$ respectively and have been 
displayed in Fig.\,\,\ref{collider-process}. We have studied the signal
at two different $e^{+} e^{-}$ colliders namely, Compact Linear Collider (CLIC)
\cite{CLICPhysicsWorkingGroup:2004qvu,Dannheim:2012rn,Linssen:2012hp,
CLICDetector:2013tfe,CLICdp:2017vju,Aicheler:2018arh}\,\,and
International Linear Collider (ILC) \cite{Behnke:2013xla,Adolphsen:2013kya,
Behnke:2013lya,Baer:2013cma,Adolphsen:2013jya}. In the former case, 
we have considered centre of mass (c.o.m) energy $\sqrt{s} = 380$ GeV
and $3000$ GeV while $\sqrt{s} = 500$ GeV and $1000$ GeV
for the latter at the time of the pair production of vector like fermion. 
Depending on the c.o.m energy of the collider, we have an upper bound on
the mass of $\chi^{\pm}$ upto which it can be produced. 
In particular, in the present work we have investigated the signal
at the detector level for c.o.m energy $\sqrt{s} = 1000$ GeV 
and $3000$ GeV of the $e^{+}e^{-}$ linear collider.
Although there is no dedicated search for the present model at the CMS 
or ATLAS detector, still the same kind of signal can be produced at the 
hadron collider. We have produced the $\chi^{+} \chi^{-}$ final
state at the $p p$ collider using \texttt{MadGraph} \cite{Alwall:2014hca, Alwall:2011uj} for 
$\sqrt{s} = 13$ TeV and find that this is lower than the current 
exclusion limit given by the CMS collaboration for the 13 TeV run
of LHC with 35.9 $fb^{-1}$ integrated luminosity \cite{CMS:2018xqw}. 
This has been displayed in Fig.\,\,\ref{pp-collider-CS}
where the red points correspond to the upper limit on the
pair production cross section of the singly charged fermion coming
from the study of 13 TeV run of LHC by CMS whereas the blue cross points
correspond to the cross section for the present model at the $pp$
collider mediated by gauge bosons like $\gamma$ and $Z$.
Therefore, we conclude that the charge particle mass range
we have considered in the present model is safe from the LHC
bound. In contrary to the $pp$ collider, at $e^{+}e^{-}$ collider
the signal $e^{+} e^{-} \cancel{E}_T$ has an additional $t-$ channel
diagram mediated by the MeV scale SIMP dark matter $\phi_4$.
This $t-$channel diagram enhances the cross section
by an order of magnitude larger than the $s$-channel diagrams mediated
by $\gamma$ and $Z$ gauge bosons. 
In accomplishing the collider analysis for the present model, we have
considered cut based analysis using a series of sophisticated packages
available for the collider study. In particular, we have
used \texttt{FeynRules} \cite{Alloul:2013bka}
for implementing the present model and generated the UFO model files which fed 
into the \texttt{MadGraph} \cite{Alwall:2014hca, Alwall:2011uj} subsequently. 
We have then used \texttt{MadGraph} for generating the parton level process. For showering
we have used inbuilt \texttt{PYTHIA} \cite{Sjostrand:2006za} in the \texttt{MadGraph} 
and finally for the event analysis we use \texttt{DELPHES}
package \cite{deFavereau:2013fsa,Selvaggi:2014mya, Mertens:2015kba}.

\begin{figure}[h!]
\centering
\includegraphics[angle=0,height=6cm,width=7.5cm]{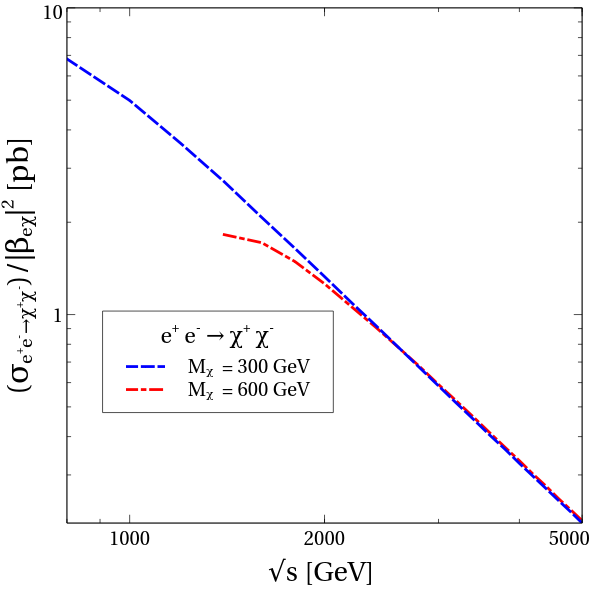}
\includegraphics[angle=0,height=6cm,width=7.5cm]{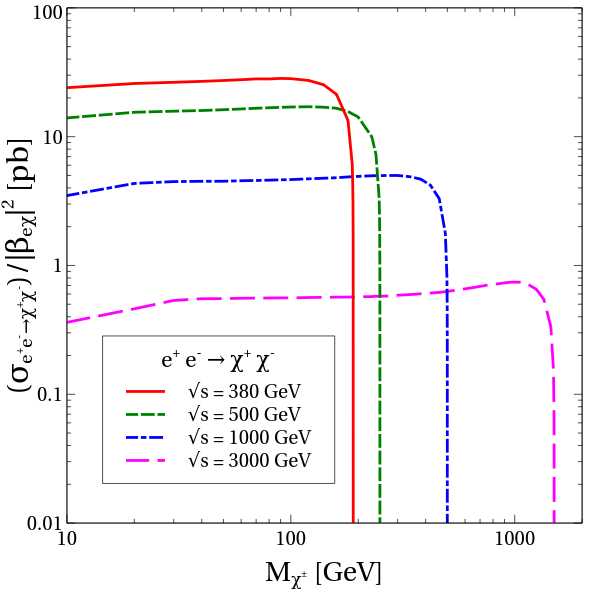}
\caption{Left panel shows the pair production of $\chi^{+}$ and $\chi^{-}$ 
for different c.o.m energy for two values of $\chi^{\pm}$ mass
at the $e^{+} e^{-}$ lepton collider.
Right panel shows the variation of $\chi^{+}$ and $\chi^{-}$ production rate with
$\chi^{\pm}$ mass $M_{\chi^{\pm}}$ for different values of c.o.m energies.}      
\label{ep-em-CS}
\end{figure}
In the left and right panels of Fig.\,(\ref{ep-em-CS}), we have shown the variation of 
$\chi^{+} \chi^{-}$ production cross section at $e^{+} e^{-}$ collider 
with the centre of mass energy of the collider for two different values
of $\chi^{\pm}$ mass and the mass of the charged particle for the
different centre of mass energies respectively.
In the left panel, blue dashed line corresponds to $M_{\chi} = 300$ GeV and
the red dashed-dot line corresponds to $M_{\chi} = 600$ GeV and both the 
line varies inversely with the center of mass energy $\sqrt{s}$. For
$M_{\chi^{+}} = 600$ GeV (red dash-dotted line), we can see that 
$\chi^{+} \chi^{-}$ starts producing when the c.o.m energy of the 
$e^{+} e^{-}$ reached a minimum
threshold energy which is $\sqrt{s} = 1200$ GeV. For the higher values of the
c.o.m energy, we can see that blue dashed and red dashed-dot lines coincide with
each other implies that the production cross section does not depend on
mass for higher c.o.m values. Whereas the right panel shows the variation of 
$\chi^{+}\chi^{-}$ production cross section with the mass $M_{\chi^{+}}$
for different values of the c.o.m energy. One can easily see that the
production cross section decrease with the increasing value of the c.o.m
energy which is consistent with the discussion of the left panel. 
The different c.o.m
energies are the proposed experimental set up for CLIC and ILC colliders. 
When the $\chi^{+}$ mass reached the threshold value {i.e.} 
$M_{\chi^{\pm}} \sim \frac{\sqrt{s}}{2}$, there is a sharp fall in 
the production cross section. In the later part of the draft, we have
done the collider analysis for $\sqrt{s} = 1000$ GeV and $\sqrt{s} = 3000$
GeV c.o.m energy.               
\subsection{Analysis}
In the present work, we are interested in the signal which contains
opposite sign di-electrons ($e^{+} e^{-}$) and 
transverse missing energy ($\cancel{E}_{T}$) in the final state. 
In discovering the 
signal from the present model, same kind of signal morphology can
appear in the final state from the known SM backgrounds. The dominant
backgrounds which can mimic the signal are as follows,

\begin{enumerate}
\item At the electron positron collider, the dominant background which can 
mimic the signal is $e^{+} e^{-} \rightarrow e^{+} e^{-} Z$, where 
$Z$ can decay to $\nu_{l} \bar{\nu}_{l}$. Therefore, finally it gives 
$e^{+} e^{-} \cancel{E}_{T}$ which exactly resembles the signal. This kind of 
background also includes $Z Z$ production channel which subsequently decays to
electrons ($e^{+} e^{-}$) and neutrinos ($\nu_{l} \bar{\nu}_{l}$).

\item Another dominant background can come from the pair production of
$W^{+} W^{-}$ mode at the $e^{+} e^{-}$ collider. The W-boson subsequently
decays leptonically to leptons and neutrinos and can replicate the signal as
$ e^{+} e^{-} \rightarrow W^{+} (\rightarrow l^{+} \nu_l)
W^{-} (\rightarrow l^{-} \bar{\nu}_l)
\rightarrow e^{+} e^{-} \cancel{E}_{T}$\,.

\item Another potentially relevant background is the pair production
of $t \bar{t}$ mode which can also mimic the signal when $t$ quark decays
leptonically associated with two b-quarks. This background can 
mock the signal at the electron positron collider as
$e^{+} e^{-} \rightarrow t (\rightarrow b\,l^{+} \nu_{l}) \bar{t} 
(\rightarrow \bar{b}\,l^{-} \bar{\nu}_{l}) \rightarrow b \bar{b}\,e^{+} e^{-}
\cancel{E}_{T}$. As we will see, this kind of signal is easy to avoid with the 
$b$-tagging. 
\end{enumerate}

\begin{figure}[h!]
\centering
\includegraphics[angle=0,height=6cm,width=7.5cm]{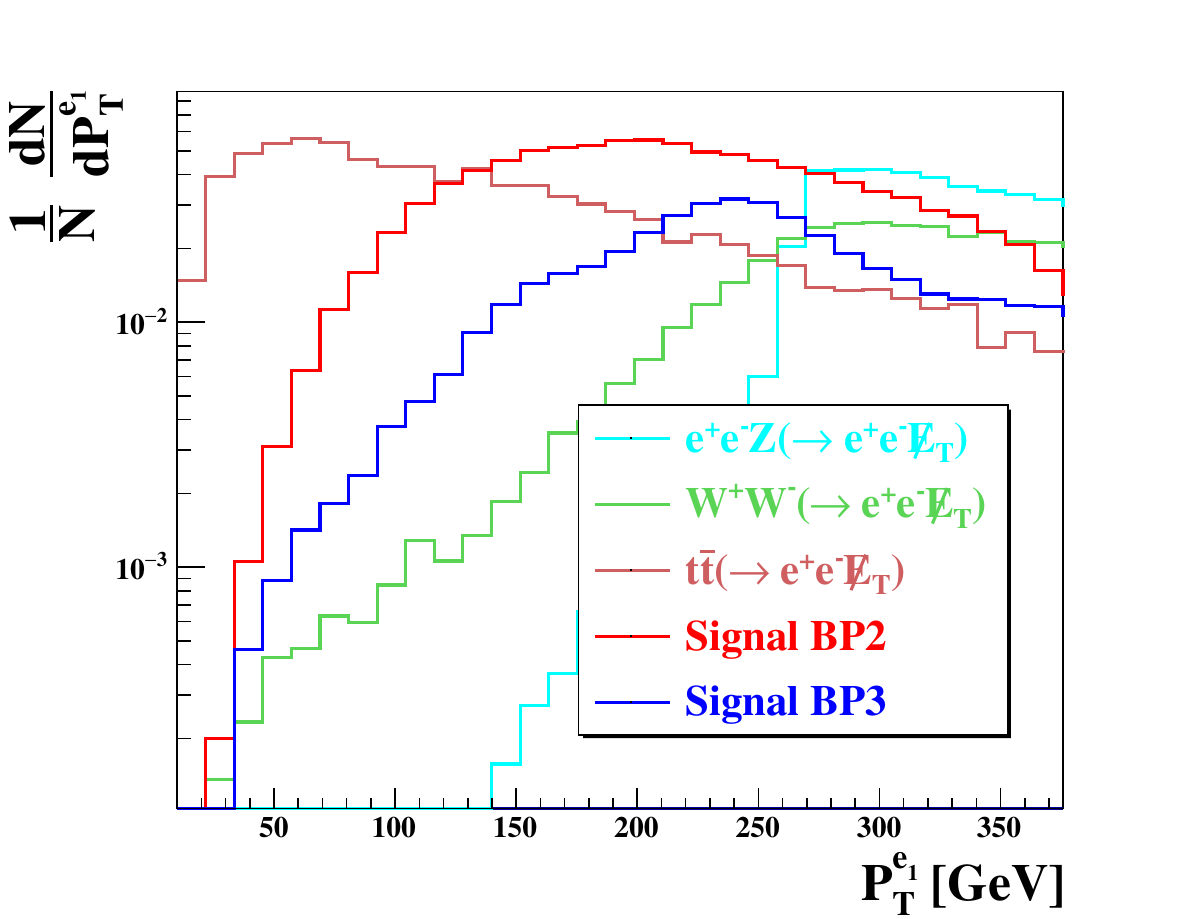}
\includegraphics[angle=0,height=6cm,width=7.5cm]{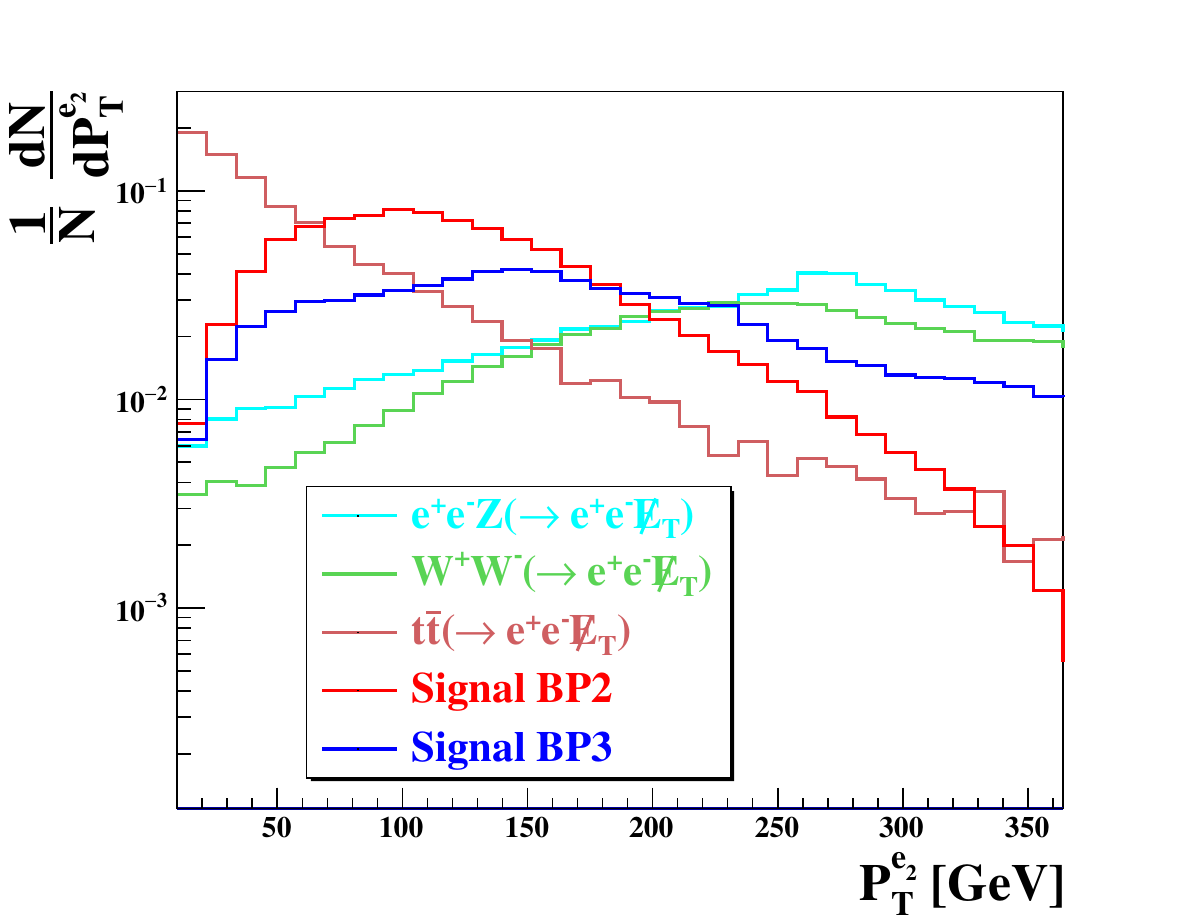}
\caption{Left panel shows the variation of fraction of events distribution
for backgrounds and signal with the transverse momentum of leading electron 
whereas the right panel shows the variation with respect to the rapidity
of the leading electron.}      
\label{histogram-1}
\end{figure}

\begin{figure}[h!]
\centering
\includegraphics[angle=0,height=6cm,width=7.5cm]{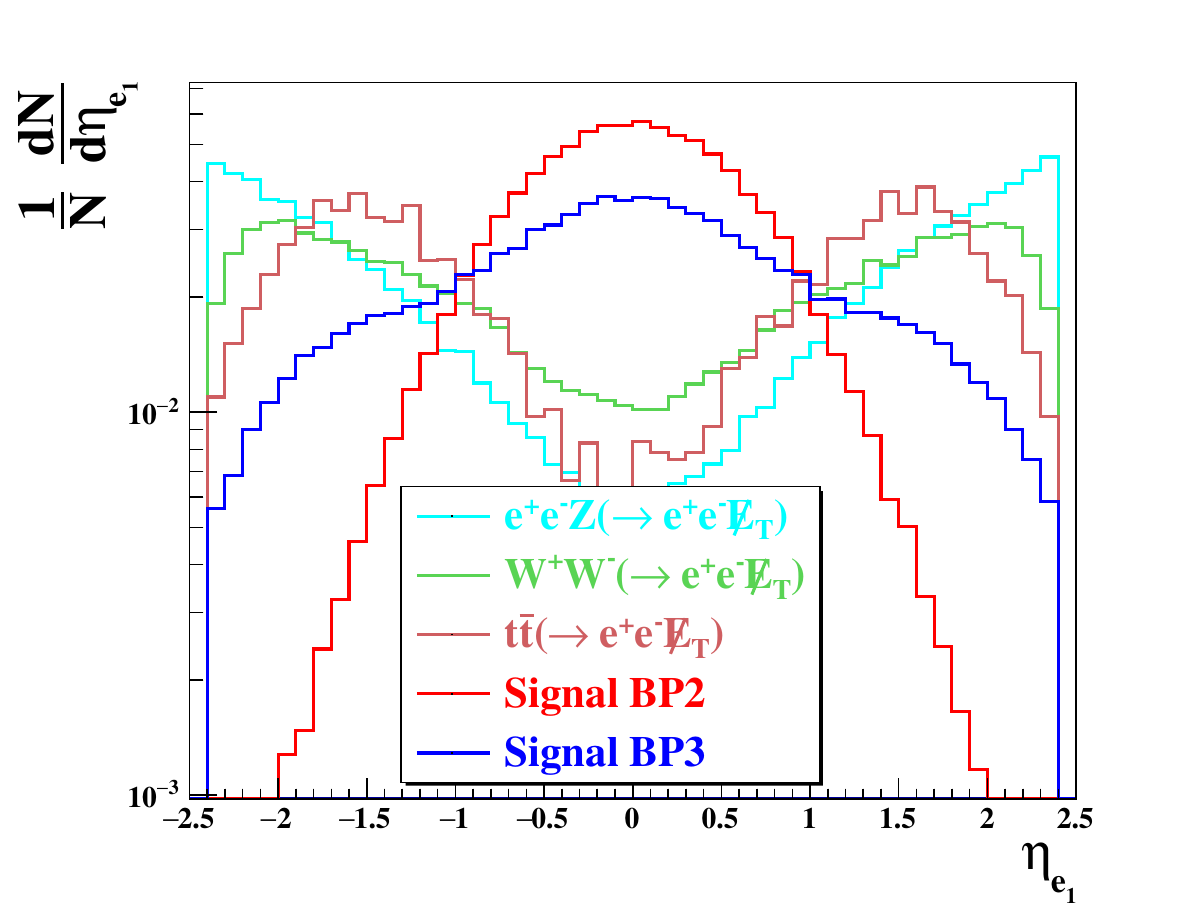}
\includegraphics[angle=0,height=6cm,width=7.5cm]{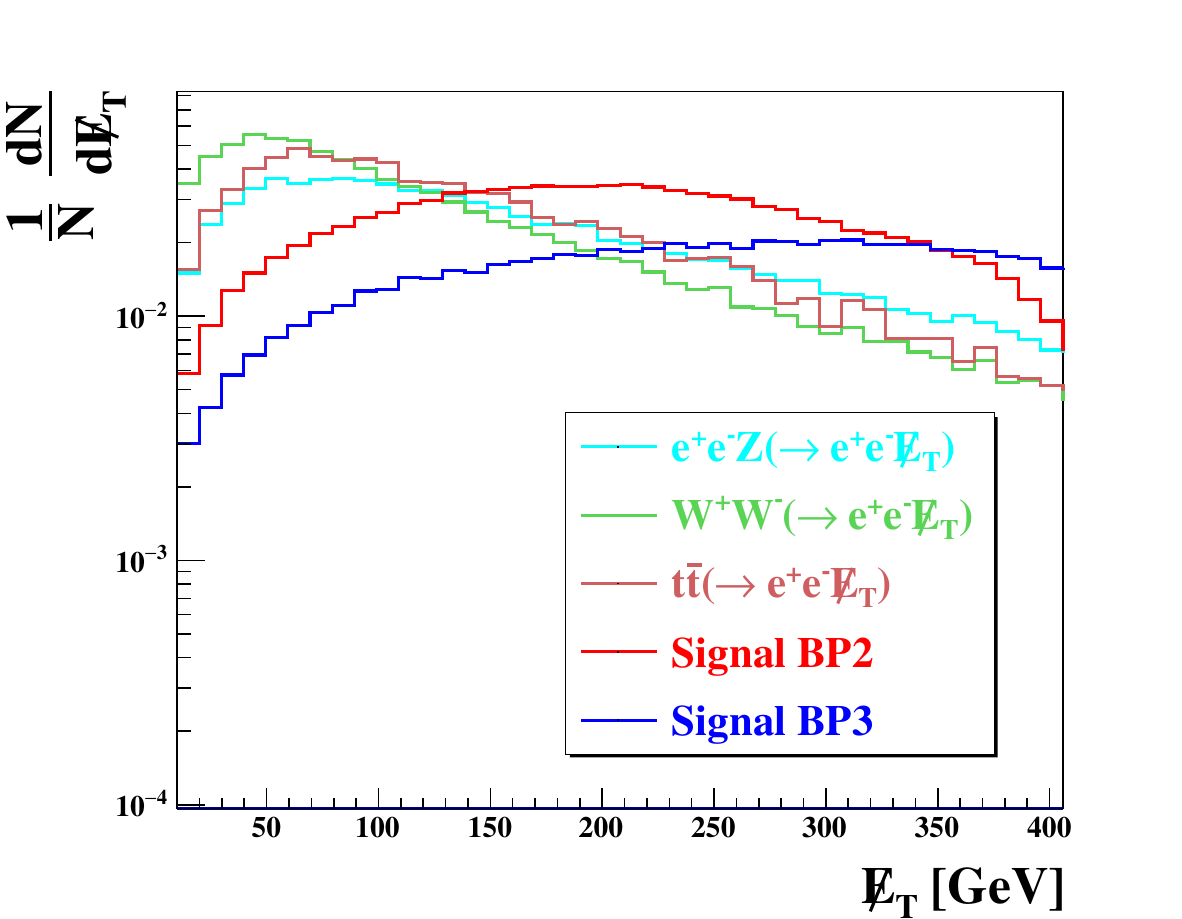}
\caption{Variation of fraction of events for backgrounds and signal with
respect to transverse missing energy and invariant mass of the di-electron
in the left and right panels respectively.}      
\label{histogram-2}
\end{figure}

At the time of generating the events, we have not put any veto to forbid
the processes contains other particles than the leptons and 
missing energy. We can put b-veto
which will reduce $t \bar t$ background. 
In Figs. (\ref{histogram-1}, \ref{histogram-2}), 
we have shown the variation of backgrounds and signals 
about the different kinematical variables namely, transverse momentum of 
the leading electron ($P^{e_1}_{T}$) and second leading 
electron ($P^{e_2}_{T}$), pseudorapidity of the leading electron ($\eta_{e_1}$)
and transverse missing energy ($\cancel{E}_T$). 
From the figures, we can choose the values of the
kinematical variables which will prefer the signal over backgrounds. In this work,
we have used the following kind of cuts on the kinematical variables 
in order to reduce the backgrounds without affecting the signal much. 
The details of the cuts on the kinematical variables are as follows,  

\begin{enumerate}

\item[A0.] We have considered the events which contain
opposite sign di-electron ($e^{+} e^{-}$) and transverse
missing energy ($\cancel{E}_{T}$). We have also put the
minimal cut on the transverse momentum of the electrons
which is $p^{min}_{T, e} \geq 10$ GeV. We have collected
the events which satisfy the pseudorapidity of the electrons
$\eta_{e} < 2.5$. These cuts have been implemented at the time of
the partonic generation of the events.

\item[A1.] We consider events that have opposite sign 
di-electron pair ($e^{+} e^{-}$) in the final state.

\item[A2.] From the left panel of Fig.\,(\ref{histogram-1}), we can see that if 
we put strong cut on the leading electron $P^{e_1}_{T} \geq 130$ GeV 
then we can reduce the $t \bar t$ background. We have chosen
relatively soft cut on the second leading electron
which is $P^{e_2}_{T} \geq 60$ GeV\,.

\item[A3.] To reduce the background which comes from $ZZ$ mode, we
have used $Z-veto$. $Z-veto$ means, we have accepted the the events 
which violate the condition $|m_{ee} - 91.2| < 10$ GeV, where
$m_{ee}$ is the di-electron invariant mass.

\item[A4.] In order to eliminate the $t \bar t$ background, we have 
implemented $b-$veto. This implies we have rejected the events which contains 
$b-$quarks in the final state.

\item[A5.] From the left panel of Fig.\,(\ref{histogram-2}), we can see
that background and signal peak at different values of pseudorapidity.
Therefore, to reduce the background without affecting the signal much
we have considered the events which have pseudorapidity in
the range, $|\eta^{e}| < 1.5$.

\item[A6.] From the right panel of Fig.\,(\ref{histogram-2}),
we can see that if we implement a cut on the transverse
missing energy then background can be reduced significantly.
We have adapted the missing energy cut which is 
$\cancel{E}_{T} > 160$ GeV.
\end{enumerate}

\def\I{i}
\begin{center}
\begin{table}[ht!]
\begin{tabular}{||c|c||}
\hline
\hline
\begin{tabular}{c|c}
    \multicolumn{2}{c}{SM Backgrounds at 1 TeV ILC}\\ 
    \hline
    Channels & Cross section (fb) \\ 
    \hline
    $e^{+} e^{-} Z \,(\rightarrow \nu_{l}\bar{\nu_{l}}) $ & $15.59$ \\ 
    \hline

    $W^{+}(\rightarrow e^{+} \nu_{l})\, W^{-} (\rightarrow e^{-} \bar{\nu_{l}})$
    & $13.24$\\
    \hline
    $t (\rightarrow b e^{+} \nu_{l})\,\bar{t} (\rightarrow \bar{b} e^{-} \bar{\nu_{l}})$
    & $1.61$\\
    \hline
    Total Backgrounds & ~\\ 
\end{tabular}
&
\begin{tabular}{c|c|c|c|c|c}
    \multicolumn{6}{c}{Effective Cross section after applying cuts (fb)}\\   
    \hline 
    A0\,+\,A1 & \,\,~~A2~~\,\, &\,\, ~~A3~~\,\, &\,\, ~~A4~~\,\, &\,\, ~~A5~~ &\,\,~~A6~~ \\
    \hline
    $10.07$ & $6.07$ & $5.88$ & $5.88$ & $3.92$ & $2.43$  \\

    \hline
    $8.88$ & $6.25$ & $6.25$ & $6.25$ & $4.28$ & $1.57$  \\
    \hline
    $0.73$ & $0.20$ & $0.20$ & $0.05$ & $0.04$ & $0.02$ \\
    \hline
    ~ & ~ & ~ & ~ & ~& 4.02 \\
\end{tabular}\\
\hline
\hline
\end{tabular}
\caption{Cut-flow table for the obtained cross sections corresponding to the
{different SM backgrounds. See the text for the details of the cuts
A0-A6. The c.m.energy is  $\sqrt{s}=1 $ TeV, relevant for ILC.}}
\label{tab:1tev}
\end{table}
\end{center}


\vspace{-1.2cm}
\def\I{i}
\begin{center}
\begin{table}[h!]
\begin{tabular}{||c|c||}
\hline
\hline
\begin{tabular}{c|c}
    \multicolumn{2}{c}{SM Backgrounds at $3$ TeV CLIC}\\ 
    \hline
    Channels & Cross section (fb) \\ 
    \hline
    $e^{+} e^{-} Z \,(\rightarrow \nu_{l}\bar{\nu_{l}}) $ & $6.18$ \\ 
    \hline

    $W^{+}(\rightarrow e^{+} \nu_{l})\, W^{-} (\rightarrow e^{-} \bar{\nu_{l}})$
    & $1.44$\\
    \hline
    $t (\rightarrow b e^{+} \nu_{l})\,\bar{t} (\rightarrow \bar{b} e^{-} \bar{\nu_{l}})$
    & $0.19$\\
    \hline
    Total Backgrounds & ~\\ 
\end{tabular}
&
\begin{tabular}{c|c|c|c|c|c}
    \multicolumn{6}{c}{Effective Cross section after applying cuts (fb)}\\   
    \hline 
    A0\,+\,A1 & \,\,~~A2~~\,\, &\,\, ~~A3~~\,\, &\,\, ~~A4~~\,\, &\,\, ~~A5~~ &\,\,~~A6~~ \\
    \hline
    $2.95$ & $2.85$ & $2.83$ & $2.83$ & $1.06$ & $0.82$  \\
    \hline

    $0.74$ & $0.73$ & $0.73$ & $0.73$ & $0.36$ & $0.27$  \\
    \hline
    $0.01$ & $0.004$ & $0.004$ & 0.001 & 0.0006 & 0.0005 \\
    \hline
    ~ & ~ & ~ & ~ & ~& 1.09 \\
\end{tabular}\\
\hline
\hline
\end{tabular}
\caption{Cut-flow table for the obtained cross sections corresponding to the
SM backgrounds. The details of the cuts A0-A6 are mentioned in the text.
We perform the simulation for $3$ TeV CLIC.}
\label{tab:3tev}
\end{table}
\end{center}
In Tables\,(\ref{tab:1tev},\ref{tab:3tev}), we have shown the 
survival of the backgrounds for ILC ($\sqrt{s} = 1$ TeV) and 
CLIC ($\sqrt{s} = 3$ TeV) colliders, respectively. 
In Table\,(\ref{tab:signal}), we have shown the response of the 
signal production cross section after applying different cuts,
A0 to A6. We can see that the cuts are effective in lowering the backgrounds 
and at the same time cuts reduce the signal cross section 
less significantly than the backgrounds. In determining the statistical significance
of signal over background, we have used Eq.\,(\ref{formula-significance}).
in Eq.\,(\ref{formula-significance}), $s$ corresponds
to number of events \footnote{We determine the number of events
by multiplying the cross section with the luminosity.}
for signal after applying all the cuts and $b$ is the number
of background events after applying all the cuts,       
\begin{equation}
\mathcal{S} = \sqrt{2 \times \left[ (s+b) {\rm ln}(1 + \frac{s}{b}) - s \right]}.
\label{formula-significance}
\end{equation}

\def\I{i}
\begin{center}
\begin{table}[!ht]
\begin{tabular}{||c|c|c|c|c|c|c|c|c|c|c|c||}
\hline
\hline
    \multicolumn{4}{||c|}{Signal at $e^{+}e^{-}$ Collider} & \multicolumn{6}{|c|}{Effective CS after cuts (fb)} & \multicolumn{2}{|c||}{Stat Significance ($\mathcal{S}$)}\\
    \hline
     Experiment & Mass (GeV) & $\beta_{e\chi}$ & CS (fb) &{A0+A1} & {A2} & {A3} & {A4} &{A5} & {A6} & {$\mathcal{L} = 1~{\rm fb^{-1}}$}& {$\mathcal{L} = 10$ ${\rm fb^{-1}}$} \\
\hline
     \multirow{2}{*} {1 TeV ILC}& 350.0 & 0.1 & 41.20 & 29.60 & 24.03 & 23.73 & 23.73 & 22.57 & 19.53 & 6.65 & 21.03 \\
     \cline{2-12}
     & 450.0 & 0.1 & 29.56 & 21.30 & 19.83 & 19.60 & 19.60 & 19.50 & 17.38 & 6.11 & 19.32 \\
    \hline
    \hline
     \multirow{2}{*} {3 TeV CLIC} & 600.0 & 0.1 & 5.13 & 2.91 & 2.78 & 2.78 & 2.78 & 2.16 & 2.05 & 1.60 & 5.04 \\    
    \cline{2-12}
      & 700.0 & 0.1 & 5.47 & 3.08 & 2.99 & 2.99 & 2.99 & 2.45 & 2.34  & 1.78 & 5.64 \\

%
%
\hline
\hline
\end{tabular}
\caption{Cut-flow table of signal cross section
at ILC ($\sqrt{s} = 1$ TeV) and CLIC ($\sqrt{s} = 3$ TeV).
We have considered $\beta^{L}_{e\chi} = \beta^{R}_{e\chi} = \beta_{e\chi} = 
0.1$ and different values of $\chi^{\pm}$ mass.}
\label{tab:signal}
\end{table}
\end{center}
The last two columns in Table\,\,\ref{tab:signal} corresponds to the statistical
significance of the signal. For $1$ TeV collider, we can see that the
present model can have more than $6 \sigma$ significance for $1\,\,fb^{-1}$
luminosity. For $3$ TeV collider, we need $10\,\,fb^{-1}$ luminosity
in order to get the $5 \sigma$ statistical significance for the
signal. We can see that the current model can be tested at
the very early run of the ILC and CLIC colliders. 

 \begin{figure}[h!]
\centering
\includegraphics[angle=0,height=6cm,width=7.5cm]{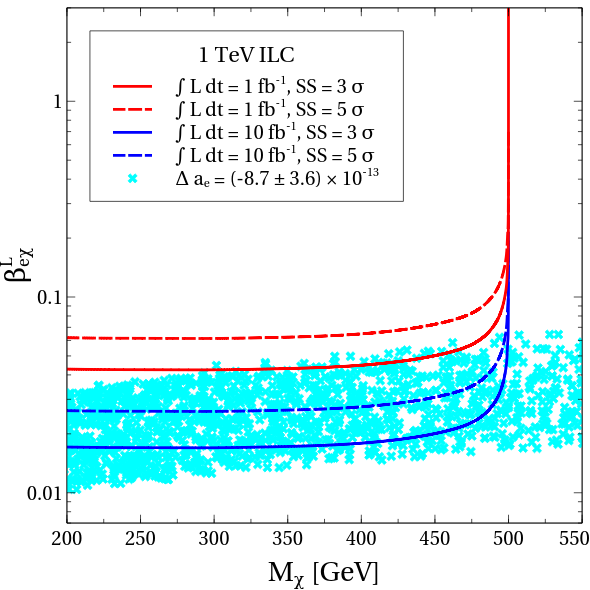}
\includegraphics[angle=0,height=6cm,width=7.5cm]{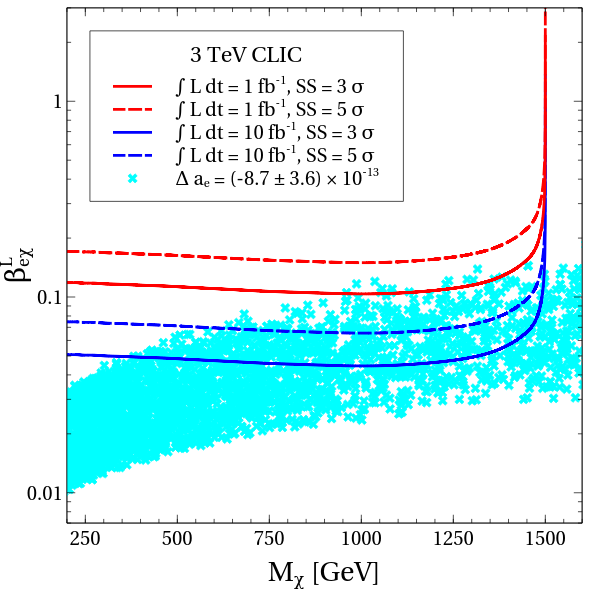}
\caption{Left and right panels show the allowed regions (by cyan colour points) 
after satisfying the $(g-2)_e$
in $\beta^{L}_{e\chi} - M_{\chi}$ plane for 1 TeV and 3 TeV lepton colliders,
respectively. The solid and dashed lines correspond to the variation 
of $3\sigma$ and $5\sigma$ statistical significance lines of 
$e^{+}e^{-}\cancel{E}_{T}$ signal
with the charged singlet fermion mass for the luminosity 1 $fb^{-1}$
and 10 $fb^{-1}$, respectively. The other parameters kept
fixed at $\beta^{R}_{e\chi} = \beta^{L}_{e\chi}$, 
$M_{\phi_4} = 100$ MeV, $\theta_{D} = 0.1$ and $M_{\phi_2}$
has been varied between 1 to 10 TeV.}      
\label{beta-mchi-sigma}
\end{figure}
In Fig.\,(\ref{beta-mchi-sigma}), we have shown the scatter plots in the 
$M_{\phi_4} - \beta^{L}_{e\chi}$ plane after satisfying $(g-2)_{e}$
by the cyan colour points. In the figure, the left panel corresponds to
the 1 TeV ILC collider whereas the right panel is for the 3 TeV CLIC 
collider. The variation among the cyan colour points are due to
the variation of $\phi_2$ mass, $M_{\phi_2}$, which
we have considered in the range $1-10$ TeV. We have kept fixed the other
parameters as mentioned in the caption of the figure. We see from 
both the panel that a sharp correlation between $M_{\chi}$ and
$\beta^{L}_{e\chi}$ after satisfying the
$(g-2)_{e}$ in the range $\Delta a_{e} = (- 8.7 \pm 3.6) \times 10^{-13}$.
Moreover, the parameters which we have varied for $(g-2)_{e}$ also 
affect the production cross section
of $\chi^{+} \chi^{-}$ channel at $e^{+} e^{-}$ collider. We have displayed 
the $1\sigma$ and $3\sigma$ statistical significance lines of the signal,
$e^{+} e^{-} \rightarrow e^{+} e^{-} \cancel{E}_{T}$, by the solid 
and dashed lines, respectively, whereas the red and blue colors on them
imply the $1\,fb^{-1}$ and $10\,fb^{-1}$ luminosity. 
It is easily understood from the figure that if we increased the 
luminosity keeping statistical significance fixed then
we can access the lower values of $\beta^{L}_{e\chi}$  as well. 
As exhibited in the right panel of Fig.\,(\ref{ep-em-CS}),
the production cross section of $\chi^{+}\chi^{-}$
with the charged fermion mass, $M_{\chi}$, is flat except at the kinematical
threshold limit $M_{\chi} \lesssim \frac{\sqrt{s}}{2}$ where 
the production cross section falls abruptly. In Table \ref{tab:signal},
we have shown the signal strength remains after applying different cuts 
$A0 - A6$ for two benchmark points. Therefore, from the table, we
can determine the average value of cut efficiency for the signal
which is the ratio of the signal cross section after applying the
A6 cut and the production cross section without any cuts. We find 
the average value of the cut efficiency for 1 TeV collider is 0.53 
whereas for 3 TeV, it is 0.43. We used these values in finding the 
$1\sigma$ and $3 \sigma$ isocontours 
of statistical significance for the signal in the 
$M_{\chi} - \beta^{L}_{e\chi}$ plane. We see that we need constant 
value of $\beta^{L}_{e\chi}$ for wide range
of $\chi^{\pm}$ mass. Moreover,  
one can notice that we need higher values of 
$\beta^{L}_{e\chi}$ in the kinematical threshold region 
($M_{\chi} \simeq \frac{\sqrt{s}}{2}$, $s$ is {c.o.m} energy)
limit because the production cross section of $\chi^{+} \chi^{-}$
sharply falls there. Additionally, the most appealing thing to 
be noticed here is that we have overlap region in the
$M_{\chi} - \beta^{L}_{e\chi}$ plane which gives us the correct value
of $(g-2)_e$ and the $\geq 3 \sigma$ statistical significance of the
signal over background.

In order to provide an overall picture to the
readers, in Table \ref{tab:BP}, we present four plausible
benchmark points (BP1, BP2, BP3 and BP4) of the present model
and the corresponding numerical values of several physical quantities
such as dark matter relic density, direct detection cross sections,
dark matter self interaction and discrepancy in leptons
anomalous magnetic moment.     
\begin{table}[h!]
\begin{center}
\vskip 0.5cm
\begin{tabular} {|c|c|c|c|c|}
\hline
Parameters/& BP1& BP2 & BP3 & BP4\\
Observables & & & &\\
\hline
\hline
$M_{\phi_4}$ (GeV) & $29.216\times10^{-3}$&
$128.116\times 10^{-3}$ &
$168.756\times 10^{-3}$ & $236.691\times10^{-3}$\\
$M_{\phi_2}$ (GeV) & 7536.32& 3511.42  & 1282.32 & 3123.67\\
$M_{Z_{\mu\tau}}$ (GeV) & $39.405\times10^{-3}$ 
& $15.986\times10^{-3}$ & $124.048 \times 10^{-3}$ &$168.81\times10^{-3}$\\
$g_{Z_{\mu\tau}}$ & $6.626\times10^{-4}$& 
$5.878\times10^{-4}$ & $1.003\times 10^{-3}$ & $1.282\times10^{-3}$\\
$\theta_{\mu\tau}$ & $4.720\times10^{-6}$&$4.720\times10^{-6}$&
$4.720\times10^{-6}$ & $4.720\times10^{-6}$\\
$\theta_{D}$ & $-2.347\times10^{-2}$ &
$-0.815\times10^{-2}$ & $-0.1422\times 10^{-2}$ & $-9.559\times10^{-2}$ \\
$\xi$ & $0.034\times10^{-2}$ & $1.022\times 10^{-2}$ &
$ 0.343\times 10^{-2}$ & $0.413\times10^{-2}$ \\
$\beta^{L}_{e\chi}$ & 0.12 & 0.22&0.70&0.066\\
$\beta^{R}_{e\chi}$ & 0.12 & 0.22&0.70&0.066\\
$\Omega_{\phi_4}h^2$ & 0.1191 & 0.1196 & 0.1199 & 0.1220\\
$\sigma_{\rm SI}$ (cm$^2$)& $5.719\times10^{-49}$& 
$1.317\times10^{-49}$ & $1.966\times 10^{-52}$ & $6.958\times10^{-45}$ \\
$\sigma_{elec}$ (cm$^2$)& $7.073\times10^{-54}$&
$1.057\times10^{-55}$ & $9.819\times 10^{-59}$ & $1.994\times10^{-51}$\\
$\sigma_{\rm SD}$ (cm$^2$)& $2.352\times10^{-54}$&
$5.416\times10^{-55}$ & $8.085\times 10^{-58}$ & $2.861\times10^{-50}$\\
$\frac{\sigma_{\rm self}}{M_{\phi_4}}$ cm$^2/$g & 3.171 & 1.024& 0.808 &  0.529\\
$\Delta{a_{\mu}}$ & $2.551\times10^{-9}$ 
& $2.995\times10^{-9}$ & $2.213\times 10^{-9}$ & $2.502\times10^{-9}$\\
$\Delta{a_{e}}$ & $-9.867\times 10^{-13}$ & 
$-9.620\times 10^{-13}$& $-9.441\times 10^{-13}$ & $-9.661\times 10^{-13}$ \\
\hline
\hline
\end{tabular}
\end{center}
\caption{Plausible four benchmark points (BP1, BP2, BP3 and BP4) with the
numerical values of other physical quantities considered in this work.}
\label{tab:BP}
\end{table} 
\section{Summary and conclusion}
\label{Sec:conclu}
In this work, we have extended the minimal U(1)$_{L_{\mu}-L_{\tau}}$
model by a scalar doublet ($\Phi^\prime_2$), a singlet scalar ($\Phi^\prime_4$)
and a vector like singlet fermion $\chi$ to address the deviations found in experiments
from the theoretical predictions of anomalous magnetic moments
for both muon and electron. All these new fields have specific
${L_{\mu}-L_{\tau}}$ charges as required by the new Yukawa interactions. We have shown that
in the minimal model, considering the present bounds on $\zmt$ from various
experiments particularly from Borexino, it is not possible to address both
these anomalies simultaneously. Apart from the one loop contribution
due to the neutral gauge boson $\zmt$ similar to the minimal
${L_{\mu}-L_{\tau}}$ model, the additional contribution coming
from one loop diagrams involving $\chi$ and $\phi_2$ or $\phi_4$ 
provide the deficit in $(g-2)_e$ as required by the experimental
data when the relevant parameters remain within the
following range  $\beta^{L}_{e\chi}\gtrsim5\times10^{-2}$,
$10^{-3}\,\,{\rm rad}\lesssim|\theta_D|\lesssim0.1$ rad and
$1\,\,{\rm TeV}\leq M_{\phi_2}\leq$ 10 TeV.
Interestingly, in order to achieve this, we also have a
natural SIMP dark matter candidate $\phi_4$, an admixture of $\Phi^\prime_4$
and $\phi^\prime_2$ (neutral part of $\Phi^\prime_2$), the signature
of which can be found as missing energy at the upcoming $e^{+} e^{-}$ linear colliders
like ILC and CLIC. In order to explore the dynamics of dark matter
in detail we have considered all possible theoretical and experimental
constraints arising from dark matter self interaction, perturbativity
and unitarity, spin dependent and spin independent elastic scatterings,
kinetic equilibrium between dark and visible sectors, Higgs invisible
decay branching and the relic density bound. The kinetic equilibrium
of the SIMP dark matter with the SM bath is possible due to
the elastic scatterings with $\nu_{\mu}$ and $\nu_{\tau}$ where
$\zmt$ plays the role of mediator. We have shown that the parameter
space in $\gmt-M_{\zmt}$ plane satisfying $(g-2)$ is fully consistent
with the range of $\gmt$ and $M_{\zmt}$ necessary for maintaining
kinetic equilibrium of dark matter with the SM bath.  
 
The characteristics of the SIMP dark matter is achieved through
the number changing $3 \rightarrow 2$ processes like
$\phi_{4} \phi_{4} \phi_{4} \rightarrow \phi^{\dagger}_{4} \phi_{4}$, 
$\phi^{\dagger}_{4} \phi_{4} \phi_{4} \rightarrow 
\phi^{\dagger}_{4} \phi^{\dagger}_{4}$
which supersede the contribution coming from $2 \rightarrow 2$ processes 
($\phi_{4} \phi_{4} \rightarrow f \bar{f}$) due to the appearance of 
the $\phi^{3}_{4}$ term when $\Phi^\prime_3$ gets a VEV $\vmt$
and breaks the $L_{\mu}-L_{\tau}$ symmetry. Therefore, the symmetry
breaking scale is involved in the freeze-out dynamics of dark matter
and thereby determining the final abundance of $\phi_4$.   
We have found that we need an MeV scale dark matter with
$M_{\phi_4}\lesssim200$ MeV to satisfy the unitarity bound
which at the same time being consistent with
the self interaction limit $0.1\,\,{\rm cm^2/g}\leq\frac{\sigma_{\rm self}}{M_{\phi_4}}
\leq 10\,\,{\rm cm^{2}}/{\rm g}$ and the relic density
bound $0.117\leq \Omega_{\phi_4}h^2\leq0.123$ considered in this work.
Moreover, we have also searched for the collider signature of
the charged fermion ($\chi^{\pm}$) at the  
$e^{+} e^{-}$ linear colliders. For the present model, at $e^{+}e^{-}$
collider, we have an additional $t$-channel diagram, in comparison to the 
hadron collider, mediated by the MeV scale SIMP dark matter, which enhances
$\chi^{+} \chi^{-}$ pair production cross section. The produced
$\chi^{\pm}$ can decay as $\chi^{\pm} \rightarrow e^{\pm} \phi_4$. Therefore,
we have studied an opposite sign di-electron and missing energy
 ($e^{+} e^{-} \cancel{E}_{T}$) signal at the final state. 
After investigating the relevant backgrounds which can mimic the 
present signal and performing the cut based analysis of signal and
backgrounds we find that TeV scale $\chi^{\pm}$ can be detected 
at the early run of $e^{+} e^{-}$ collider with $\geq 3 \sigma$
statistical significance for luminosity as low as 10 $fb^{-1}$.
We have also discussed the compatible region in the $\beta^{L}_{e\chi} - M_{\chi}$
parameter space which can simultaneously explain the $(g-2)_{e}$
and also demands $\geq 3 \sigma$ statistical significance for the signal.
Therefore, upon conclusion, our model can accommodate both $(g-2)_{e,\mu}$,
neutrino masses and mixings, a natural SIMP dark matter and also an interesting
collider imprint of the dark sector including the charged fermion
$\chi^{\pm}$ at the upcoming $e^{+}e^{-}$ linear colliders. 
\section*{Acknowledgement}
The authors would like to thank Jinsu Kim for the useful discussion.
The research of A.B. was supported by Basic Science Research Program
through the National Research Foundation of Korea(NRF) funded by the
Ministry of Education through the Center for Quantum Spacetime (CQUeST)
of Sogang University (NRF-2020R1A6A1A03047877).
This work used the Scientific Compute Cluster at GWDG,
the joint data center of Max Planck Society for the Advancement
of Science (MPG) and University of G\"{o}ttingen.
\appendix
\section*{Appendix}
\section{$(g-2)_{e}$ in the extended model}
\label{Sec:g-2_ext}
{The one loop Feynman diagrams for $(g-2)_e$ where
we have contributions from the dark sector scalars including dark matter $\phi_4$
are shown in Fig.\,\ref{Fig:Feyn_dia_g-2_scalar}.
\begin{figure}[h!]
\includegraphics[height=4cm,width=15cm,angle=0]{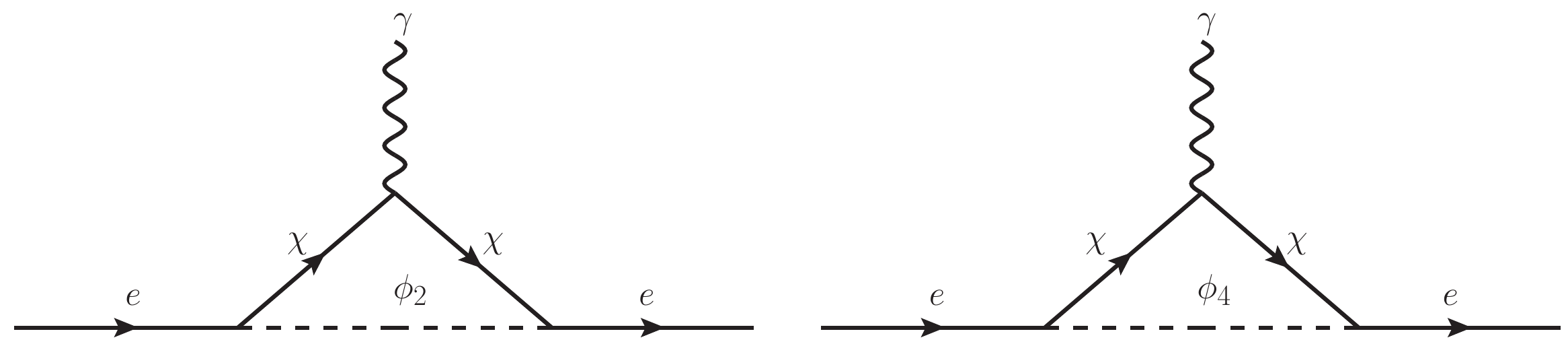}
\caption{Feynman diagrams contributing to the anomalous magnetic moment.}
\label{Fig:Feyn_dia_g-2_scalar}
\end{figure}
The expression of $\Delta{a}_e$ due to these scalar mediated diagrams
is given by \cite{Leveille:1977rc}
\begin{eqnarray}
\Delta{a}^{scalar}_e&=&-\dfrac{Q^\chi_{em}\,m_e}{8\pi^2}
\left[(g^s_{\phi_2})^2\,\mathcal{I}(m_e,\,M_\chi,\,M_{\phi_2})
+(g^p_{\phi_2})^2\,\mathcal{I}(m_e,\,-M_\chi,\,M_{\phi_2})
\right.\nn \\ && \left. 
(g^s_{\phi_4})^2\,\mathcal{I}(m_e,\,M_\chi,\,M_{\phi_4})
+(g^p_{\phi_4})^2\,\mathcal{I}(m_e,\,-M_\chi,\,M_{\phi_4})
 \right]\,,
\end{eqnarray}
where the loop integral function $\mathcal{I}(m_1,m_2,m_3)$ is defined as
\begin{eqnarray}
\mathcal{I}(m_1,\,m_2,\,m_3) = \int_{0}^{1}dx \dfrac{x^2-x^3+\frac{m_2}{m_1}x^2}
{m_1^2x^2+(m_2^2-m_1^2)x+m^2_3(1-x)} \,
\label{loop_int_scalar}
\end{eqnarray}
and the associated coefficients $g^{(s,\,p)}_{({\phi_2},\,{\phi_4})}$
have the following expressions
\begin{eqnarray}
g^s_{\phi_2}&=&-\dfrac{1}{2}\left(\beta^{R}_{e\chi}\cos\theta_D-\beta_{e\chi}^L\sin\theta_D\right),\nn\\
g^p_{\phi_2}&=&-\dfrac{1}{2}\left(\beta^{R}_{e\chi}\cos\theta_D+\beta_{e\chi}^L\sin\theta_D\right),\nn \\
g^s_{\phi_4}&=&-\dfrac{1}{2}\left(\beta^{R}_{e\chi}\sin\theta_D+\beta_{e\chi}^L\cos\theta_D\right),\nn\\
g^p_{\phi_4}&=&-\dfrac{1}{2}\left(\beta^{R}_{e\chi}\sin\theta_D-\beta_{e\chi}^L\cos\theta_D\right).
\end{eqnarray}
These are scalar and pseudo scalar couplings of $\phi_i$ with $e$ and $\chi$
i.e.\,\,we have an interaction like $\overline{e}\left(g^{s}_{\phi_i}+\gamma_5\,g^{p}_{\phi_i}\right)\chi\phi_i$. 
The total BSM contribution in $(g-2)_e$ in the extended model is
$\Delta{a}_e=\Delta{a}^{scalar}_e + \Delta{a}^{\zmt}_e$, where $\Delta{a}^{\zmt}_e$, the
contribution due to $\zmt$, is given in Eq.\,\,(\ref{del_al}). However, since there
is no new Yukawa coupling for muon, the BSM effect in the anomalous magnetic moment
of muon is solely due to $\zmt$ as also in the minimal $L_{\mu}-L_{\tau}$ model. }
\section{Necessary Vertex factors}
\label{App:Vertices}
In this appendix we have listed the necessary vertex factors.
\begin{eqnarray}
&&\phi_4\phi_4\phi_4: -3\sqrt{2}\,\xi \vmt \cos^3\theta_{D}\,,\\
&&\phi^\dagger_4\phi^\dagger_4\phi^\dagger_4: -3\sqrt{2}\,\xi \vmt \cos^3\theta_{D}\,,\\
&& \phi_4\phi_4\phi^\dagger_4\phi^\dagger_4: -4\left(\lambda_4\,\cos^4\theta_{D}+
\lambda_2\,\sin^4\theta_{D} + \lambda_{24} \sin^2 \theta_{D}\cos^2\theta_{D}
\right)\,,\\
&& h_1\phi^\dagger_2\phi_4: -\dfrac{1}{2}\cos \theta\left(\sqrt{2}\,\mu \cos 2\theta_D +
v(\lambda_{12}+\lambda^\prime_{12}-\lambda_{14})\sin 2\theta_{D}\right) \nn \\
&&~~~~~~~~~~~~~~~~~~+v_{\mu\tau} (\lambda_{23} - \lambda_{34}) \cos \theta_D \sin \theta_D \sin \theta\,,\\
&& h_1\phi_2\phi^\dagger_4: -\dfrac{1}{2}\cos \theta\left(\sqrt{2}\,\mu \cos 2\theta_D +
v(\lambda_{12}+\lambda^\prime_{12}-\lambda_{14})\sin 2\theta_{D}\right) \nn \\
&&~~~~~~~~~~~~~~~~~~+v_{\mu\tau} (\lambda_{23} - \lambda_{34}) \cos \theta_D \sin \theta_D \sin \theta\,, \\
&& h_3 \phi^\dagger_2\phi_2:
-\vmt \cos\theta \left(\lambda_{23} \cos^2 \theta_D + \lambda_{34} \sin^2 \theta_D \right)
-\sin \theta \Big(-\sqrt{2} \mu \sin \theta_D \cos \theta_D \nn \\ &&
~~~~~~~~~~~~~~~~~~ + v \cos^2 \theta_D (\lambda_{12} +\lambda^\prime_{12})
+\lambda_{14} v \sin^2\theta_D\Big)\,,
\\
&& h_3 \phi^\dagger_4\phi_4: 
-\vmt \cos\theta \left(\lambda_{34} \cos^2 \theta_D + \lambda_{23} \sin^2 \theta_D \right)
-\sin \theta \Big(+\sqrt{2} \mu \sin \theta_D \cos \theta_D \nn \\ &&
~~~~~~~~~~~~~~~~~~ + v \sin^2 \theta_D (\lambda_{12} +\lambda^\prime_{12})
+\lambda_{14} v \cos^2\theta_D\Big)\,,
\\
&& h_3\phi_4\phi_4\phi_4: -3\sqrt{2}\,\xi\cos^3\theta_{D}\cos\theta\,,\\
&& h_3\phi^\dagger_4\phi^\dagger_4\phi^\dagger_4: -3\sqrt{2}\,\xi\cos^3\theta_{D}\cos\theta\,, \\
&& h_1\phi_4\phi_4\phi_4: 3\sqrt{2}\,\xi\cos^3\theta_{D}\sin\theta\,,\\
&& h_1\phi^\dagger_4\phi^\dagger_4\phi^\dagger_4: 3\sqrt{2}\,\xi\cos^3\theta_{D}\sin\theta\,,\\
&& \phi_2 \phi_4\phi_4: 3 \sqrt{2}\,\vmt\,\xi \cos^2\theta_D \sin\theta_D\,, \\
&& \phi^\dagger_2 \phi^\dagger_4\phi^\dagger_4: 3 \sqrt{2}\,\vmt\,\xi \cos^2\theta_D \sin\theta_D\,,
\\
&& h\phi^\dagger_4\phi_4: 
-\sin\theta \Big(v \lambda_{14} \cos^2\theta_D + 
\sqrt{2} \mu \cos \theta_D \sin \theta_D + 
    v (\lambda_{12} + \lambda_{12}^\prime) \sin^2\theta_D\Big) + \nn \\ && 
 \vmt \cos \theta \Big(\lambda_{34} \cos^2\theta_D + \lambda_{23}
 \sin^2\theta_D\Big) \\
&& \phi^\dagger_4(p_1)\phi_4(p_2) Z: \dfrac{g_2}{2\,\cos\theta_W}
\sin^2\theta_D(p_1-p_2)^\alpha\,, \\
&& \phi^\dagger_4(p_1)\phi_4(p_2)\zmt: \dfrac{\gmt}{3} (p_1-p_2)^\alpha\,.
\end{eqnarray}
\bibliographystyle{JHEP}
\bibliography{draftV2.bib} 
\end{document}